\documentclass[twocolumn,aps,superscriptaddress]{revtex4}

\usepackage{amssymb}
\usepackage{amsmath}
\usepackage{graphicx}
\usepackage[normalem]{ulem}
\usepackage{appendix}

\begin{document}

\title{Bayesian inference of finite-nuclei observables based on the KIDS model}

\author{Jun Xu\footnote{xujun@zjlab.org.cn}}
\affiliation{Shanghai Advanced Research Institute, Chinese Academy of Sciences,
Shanghai 201210, China}
\affiliation{Shanghai Institute of Applied Physics, Chinese Academy
of Sciences, Shanghai 201800, China}
\author{Panagiota Papakonstantinou\footnote{ppapakon@ibs.re.kr}}
\affiliation{Rare Isotope Science Project, Institute for Basic Science, Daejeon 34000, Korea}

\date{\today}

\begin{abstract}

Bayesian analyses on both isoscalar and isovector nuclear interaction parameters are carried out based on the Korea-IBS-Daegu-SKKU (KIDS) model under the constraints of nuclear structure data of $^{208}$Pb and $^{120}$Sn. Under the constraint of the neutron-skin thickness, it is found that incorporating the curvature parameter $K_{sym}$ of nuclear symmetry energy as an independent variable significantly broadens the posterior probability distribution function (PDF) of the slope parameter $L$, and affects the related correlations. Typically, the anticorrelation between $L$ and the symmetry energy at saturation density disappears, while a positive correlation between $L$ and $K_{sym}$ is observed. Under the constraint of the isoscalar giant monopole resonance (ISGMR), incorporating the skewness parameter as an independent variable also significantly broadens the posterior PDF of the nuclear matter incompressibility $K_0$. Even with the broad uncertainties of higher-order parameters of the equation of state (EOS), robust constraints of $L<90$ MeV and $K_0<270$ MeV are obtained. Our results suggest some compatibility between the ISGMR data of $^{208}$Pb and $^{120}$Sn but not of the isovector observables especially the neutron-skin thickness.

\end{abstract}

\maketitle

\section{Introduction}

Understanding properties of nuclear interactions is one of the main goals of nuclear physics. These properties are mostly characterized by the nuclear matter EOS and the in-medium nucleon effective mass, containing isoscalar and isovector parts. The knowledge of the density dependence of the symmetric nuclear matter (SNM) EOS and the nuclear symmetry energy as well as the information of the nucleon effective mass is important in understanding nuclear systems from nuclear structures, nuclear reactions, and nuclear astrophysics. So far difficulties in understanding the accurate knowledge of the nuclear matter EOS and the nucleon effective mass appear in two aspects. On one hand, nuclear systems are mostly neutron-rich systems, so observables are thus affected by both isoscalar and isovector nuclear interactions, and they can be sensitive to not only the nuclear matter EOS but also the nucleon effective mass, hampering us from constraining accurately each individual physics quantity. On the other hand, different observables are sensitive to nuclear matter properties at different density regions, calling for better knowledge of detailed EOS parameters and a nuclear interaction model with a more flexible energy-density functional (EDF).

Thanks to the available experimental data sets in the multimessage era of nuclear physics, the Bayesian analysis serves as a suitable tool to constrain multiple physics quantities from multiple observables, and it has several advantages over the traditional $\chi^2$ fitting in revealing the relevant model parameters~\cite{Ken01}. Besides obtaining the posterior PDF of an individual physics quantity, the Bayesian analysis can also reveal the correlation between model parameters under the constraint of the experimental data. One has to keep in mind that these correlations between model parameters are generally built with given experimental data based on a particular nuclear interaction model, and the Bayesian analysis serves as a useful tool with which to reveal that correlation. The resulting posterior PDFs and correlations can be model dependent.

Observables from nuclear structure and collective response are reliable probes of the nuclear matter EOS up to saturation density and more generally of the nuclear interaction in both isoscalar and isovector channels. The excitation energy of the ISGMR, a breathing oscillation mode of a nucleus, serves as a good probe of the incompressibility $K_0$ of normal nuclear matter~\cite{Bla80,You99,Gar18,Col14,Shl06,Pie10,Kha12,Mar18}. However, it has been observed that within a given theoretical model, the ISGMR data always favor a smaller $K_0$ value for Sn isotopes than for heavy nuclei, leading to the question of why Sn is so soft (soft Tin puzzle)~\cite{Pie10,Pie07,Gar07}. The isoscalar giant quadrupole resonance (ISGQR), an oscillation mode with quadrupole deformation of a nucleus, has been found to be much affected
by the isoscalar nucleon effective mass $m_s^\star$~\cite{Boh75,Boh79,Bla80,Klu09,Roc13a,Zha16,Kon17,Bon18,Xu20a}. The neutron-skin thickness $\Delta r_{np}$ is the difference in root-mean-square neutron and proton radii, and its values for heavy nuclei are among the most robust probes of the nuclear symmetry energy at subsaturation densities~\cite{Bro00,Typ01,Chuck01,Fur02,Tod05,Cen09,Zha13,India1,India2,Xu20b,New21}. The recent PREXII data of $\Delta r_{np} = 0.283 \pm 0.071$ fm for $^{208}$Pb from parity violating electron-nucleus scatterings~\cite{PREXII} favors a rather stiff nuclear symmetry energy~\cite{Ree21}, inconsistent with the old $\Delta r_{np}$ data for Sn isotopes from proton elastic scattering experiments~\cite{Ter08} as well as the CREX data for $^{48}$Ca to be announced~\cite{Hor21}. The IVGDR is an oscillation mode in which neutrons and protons move collectively relative to each other, and is a good probe of the nuclear symmetry energy~\cite{Rei99,Tri08,Rei10,Pie12,Vre12,Roc13b,Col14,Zha14,Roc15,zhangzhen15,zhenghua16,Geb16,Li21} and the isovector nucleon effective mass $m_v^\star$~\cite{Zha16,Kon17,Xu20a}. The stiff nuclear symmetry energy from the PREXII data is also inconsistent with the IVGDR data~\cite{Pie21}, leading to another puzzle (PREXII puzzle).

In previous studies~\cite{Xu20,Xu21,Xu21b}, the posterior PDFs of the nucleon effective mass, the value and the slope parameter $L$ of symmetry energy at saturation density, and the nuclear matter incompressibility as well as its isospin dependence were extracted by employing the Bayesian analysis based on the standard Skyrme-Hartree-Fock (SHF) model. In the present study, we employ a similar analysis method but based on a more flexible KIDS model, and try to address both puzzles mentioned above. With more parameters in the KIDS model, the higher-order EOS parameters, e.g., the curvature parameter $K_{sym}$ of symmetry energy and the skewness EOS parameter $Q_0$ of SNM at saturation density, can be varied as an independent model parameter. This may have important effects on constraining the symmetry energy as well as the incompressibility of nuclear matter. In the present work, we investigate the posterior PDFs of physics quantities step by step by incorporating more and more available nuclear structure data of $^{208}$Pb and $^{120}$Sn. As we will see, incorporating $K_{sym}$ as an independent variable may largely affect the constraint on $L$ and related correlations, while incorporating $Q_0$ may largely affect the constraint on both the isoscalar and the isovector part of the incompressibility. Under the constraints of the neutron-skin thickness, the IVGDR, and the ISGMR, substantial overlaps of the PDFs from the data of $^{208}$Pb and $^{120}$Sn are observed for $L$ and for $K_0$. We have also compared predictions of $^{120}$Sn observables using the posterior PDFs from the constraints of $^{208}$Pb data to the corresponding experimental data, and vice versa, in order to quantify the ``PREXII puzzle" and the ``soft Tin puzzle".

This manuscript is structured as follows. In Sec. II, we present the theoretical framework, namely the KIDS EDF and the standard SHF model, elements of random-phase approximation, and the Bayesian analysis method as well as the parameters and data used in the analysis. In Sec. III, we present and discuss our results. First, in Sec. IIIA, a simple sensitivity study provides an overview of how strongly the different variables can affect the predictions for the observables of interest. Next, in Sec. IIIB, we incorporate the constraints from the neutron-skin thickness and examine the posterior PDFs in both the SHF and KIDS models. We demonstrate the compatibility of the various data by extracting predictions for isovector quantities based on the PREXII measurement and comparing with existing data. Next, in Sec. IIIC, we incorporate all constraints from isovector observables and examine the resulting PDFs in both the SHF and KIDS models, showing, among other things, that the strong constraints on $K_{sym}$ obtained within SHF disappear within KIDS. Finally, in Sec. IIID, we incorporate the additional constraint from the ISGMR data and discuss the posterior PDFs in the SHF and KIDS model, and compare predictions of observables with the experimental data. In Sec. IV we summarize our findings.

\section{Theoretical framework}

In the present study, the effective nuclear interaction is taken from the KIDS model, and the standard Skyrme interaction is also compared as a reference~\cite{MSL0}. For both EDFs of the KIDS and the standard SHF model, coefficients can be explicitly expressed in terms of macroscopic quantities but with different numbers of independent variables. The standard Hartree-Fock method is used to obtain the ground-state properties of spherical nuclei of interest, and the random-phase approximation method is used to study nucleus resonances of different types. The Bayesian analysis method is used in evaluating the posterior probability distribution function of physics parameters as well as their correlations by comparing properties of ground-state nuclei and nucleus resonances with the available experimental data.

\subsection{The effective nuclear interaction}

The KIDS framework for the nuclear matter EOS and EDF offers the possibility to explore systematically the effect of EOS and other interaction parameters on predictions for a variety of observables. The EOS of homogeneous nuclear matter is expressed as an expansion in the cubic root of the density, which is physically well motivated~\cite{Pap18}. Although up to four terms have been found optimal for a wide range of densities, the expansion can be extended to accommodate any set of EOS parameters~\cite{Gil19a} so that they can be varied independently of each other. Any set of EOS parameters characterizing, typically, the saturation point and the density dependence of the symmetry energy, is readily transposed to a nuclear EDF for finite nuclei in the highly convenient form of an extended Skyrme functional, with the additional freedom of choosing the values for the effective mass and other interaction parameters which are not active in homogeneous nuclear matter~\cite{Gil19a,Gil19}. Studies of symmetry-energy parameters based on astronomical observations and bulk nuclear properties were publicized recently~\cite{Gil21,Gil21a}, and a pilot study of the neutron-skin thickness was also conducted~\cite{Pap21}.

The effective interaction for the KIDS model can be considered as an extension of that for the standard SHF model, and the interaction form between two nucleons at the positions $\vec{r}_1$ and $\vec{r}_2$ can be expressed as
\begin{eqnarray}\label{SHFI}
v_{KIDS}(\vec{r}_1,\vec{r}_2) &=& (t_0+y_0P_\sigma)\delta(\vec{r}) \notag \\
&+& \frac{1}{2} (t_1+y_1P_\sigma)[{\vec{k}'^2}\delta(\vec{r})+\delta(\vec{r})\vec{k}^2] \notag\\
&+&(t_2+y_2P_\sigma)\vec{k}' \cdot \delta(\vec{r})\vec{k} \notag\\
&+&\frac{1}{6}\sum_{i=1}^{i_{max}} (t_{3i}+y_{3i}P_\sigma)\rho^{i/3}(\vec{R})\delta(\vec{r}) \notag\\
&+& i W_0(\vec{\sigma}_1+\vec{\sigma_2})[\vec{k}' \times \delta(\vec{r})\vec{k}].
\end{eqnarray}
In the above, $\vec{r}=\vec{r}_1-\vec{r}_2$ and $\vec{R}=(\vec{r}_1+\vec{r}_2)/2$ are respectively the relative and central coordinates of the two nucleons, $\vec{k}=(\nabla_1-\nabla_2)/2i$ is the relative momentum operator and $\vec{k}'$ is its complex conjugate acting on the left, and $P_\sigma=(1+\vec{\sigma}_1 \cdot \vec{\sigma}_2)/2$ is the spin exchange operator. Note that the above effective interaction is the same as the standard Skyrme interaction except for the density-dependent term, which in the standard SHF model is written as
\begin{equation}\label{vshf}
v_{SHF}^{(\rho)} = \frac{1}{6}(t_3+y_3P_\sigma)\rho^\alpha(\vec{R})\delta(\vec{r}),
\end{equation}
with $t_3$, $y_3$, and $\alpha$ being the parameters. The density-dependent term in Eq.~(\ref{SHFI}) is now a summation of terms from power $1/3$ to $i_{max}/3$, and we take $i_{max}=3$ in the present study.

The default values of parameters in the KIDS model are determined as follows. Except for the spin-orbit coupling constant, which we fix at $W_0=133$ MeVfm$^5$~\cite{MSL0}, the other 12 parameters $t_0$, $t_1$, $t_2$, $t_{31}$, $t_{32}$, $t_{33}$, $y_0$, $y_1$, $y_2$, $y_{31}$, $y_{32}$, and $y_{33}$ in the KIDS model can be solved inversely from the macroscopic quantities, i.e., the saturation density $\rho_0$, the binding energy $E_0$, the incompressibility $K_0$, and the skewness EOS parameter $Q_0$ of SNM at $\rho_0$, the isoscalar and isovector nucleon effective mass $m_s^\star$ and $m_v^\star$ in normal nuclear matter, the symmetry energy $E_{sym}^0$ and its slope parameter $L$, curvature parameter $K_{sym}$, and skewness parameter $Q_{sym}$ at $\rho_0$, and the isoscalar and isovector density gradient coefficient $G_S$ and $G_V$. For the detailed expressions, we refer the reader to Appendix~\ref{app}. The isoscalar nucleon effective mass $m_s^\star=0.82m$, with $m$ being the bare nucleon mass, presently chosen so as to reproduce both the excitation energies of isoscalar giant quadruple resonance in $^{208}$Pb and $^{120}$Sn, to be shown later. $G_S=132$ MeVfm$^5$, $G_V=5$ MeVfm$^5$, and $m_v^\star=0.7m$ are chosen as the default values as those in the MSL0 model~\cite{MSL0}, while the default values of other EOS parameters are taken to be those from the KIDS-P4 parameterization that reproduces the APR EOS rather well~\cite{Gil19}. In the present study using the Bayesian analysis, we only vary the quantities, to which the experimental observables are most sensitive after doing the sensitivity analysis, within their prior ranges, while other quantities are fixed at their default values, as shown in Table~\ref{T1}.

In the case of the standard SHF model, which we will use for comparison, we proceed in a similar manner, but there are fewer parameters that can be explored freely. Specifically, the higher-order EOS parameters, i.e., $Q_0$ and $K_{sym}$, are not varied independently because they can be fully expressed in terms of other parameters. Indeed, based on the standard SHF EDF~\cite{MSL0}, $Q_0$ is related to lower-order EOS parameters through the relation
\begin{eqnarray}\label{Q0shf}
   Q_0 &=& - 3(3\alpha+1)\frac{3}{5}\left(\frac{3\pi^2}{2}\right)^{2/3}\frac{\hbar^2}{2m}\rho_0^{2/3} \notag\\
   &+& 45(\alpha + 1)E_0    +  (3\alpha+2)K_0.
\end{eqnarray}
$K_{sym}$ is related to lower-order EOS parameters through the relation~\cite{Gil21}
\begin{eqnarray}\label{Ksymshf}
K_{sym} &=& (2-3\alpha) \frac{2}{3}\left(\frac{3\pi^2}{2}\right)^{2/3}\frac{\hbar^2}{2m}\rho_0^{2/3} \notag\\
&\times& \left[-3 \left(\frac{m}{m_v^\star}-1\right)+4\left(\frac{m}{m_s^\star}-1\right)\right] \notag\\
&-& 3 (1+\alpha) (3E_{sym}^0-L) \notag\\
&+& (1+3\alpha) \frac{1}{3} \left(\frac{3\pi^2}{2}\right)^{2/3}\frac{\hbar^2}{2m} \rho_0^{2/3}.
\end{eqnarray}
$\alpha$ in the above formulas is the exponential coefficient in Eq.~(\ref{vshf}), and it depends in a non-linear way on $\rho_0$, $E_0$, $K_0$, and $m_s^\star$ via the relation
\begin{eqnarray}\label{alpha}
&&9(1+\alpha )E_0 + K_0  \notag\\
&=&\left[(1+3\alpha )   +
      2(2-3\alpha ) \left(\frac{m}{m_s^{\star}}-1\right)  \right] \notag\\
&\times& \frac{3}{5} \left( \frac{3\pi^2}{2}\right)^{2/3}
         \frac{\hbar^2}{2m}
           \rho_0^{2/3}.
\end{eqnarray}
With a fixed $\alpha$, one sees that $Q_0$ is linear in $K_0$ and $K_{sym}$ is linear in $3E_{sym}^0-L$ without any constraints in the standard SHF model. We will return to this point later.

\begin{table}\small
  \caption{Default values of macroscopic quantities as well as their prior ranges in the KIDS model used in the present study.}
    \begin{tabular}{|c|c|c|}
  \hline
   & default value & prior range \\
   \hline
    $\rho_0$ (fm$^{-3}$) & 0.16 & -\\
    $E_0$ (MeV) & $-$16 & -\\
    $K_0$ (MeV) & 240 & 200 $\sim$ 300 \\
    $Q_0$ (MeV) & $-$373 &  $-$800 $\sim$ 400\\
    $E_{sym}$ (MeV) & 33 & 25 $\sim$ 35 \\
    $L$ (MeV) & 49 & 0 $\sim$ 120 \\
    $K_{sym}$ (MeV) & $-$156 & $-$400 $\sim$ 100 \\
    $Q_{sym}$ (MeV) & 583 & 0 $\sim$ 1000 \\
    $m_s^\star/m$ & 0.82 & -\\
    $m_v^\star/m$ & 0.7 & 0.5 $\sim$ 1 \\
    $G_S$ (MeVfm$^5$) & 132 & -\\
    $G_V$ (MeVfm$^5$) & 5 & -\\
    $W_0$ (MeVfm$^5$) & 133 & -\\
   \hline
    \end{tabular}
  \label{T1}
\end{table}

The potential energy per nucleon in the KIDS EDF, which formally corresponds to the effective interaction [Eq.~(\ref{SHFI})] employed in the Hartree-Fock approximation, has the form
\begin{eqnarray}\label{SHFpot}
E_{pot} &=& \frac{3}{8}t_0 \rho - \frac{1}{8} (t_0+2y_0) \frac{\rho_3^2}{\rho} \nonumber\\
&+& \sum_{i=1}^{i_{max}} \left[\frac{1}{16} t_{3i} \rho^{1+i/3} - \frac{1}{48} (t_{3i}+2y_{3i})\rho^{-1+i/3}\rho_3^2 \right] \nonumber\\
&+& \frac{1}{16} (3t_1+5t_2+4y_2) \rho \tau \nonumber\\
&-& \frac{1}{16}(t_1+2y_2-t_2-2y_2) \rho_3 \tau_3 \nonumber\\
&+& \frac{1}{64} (-9t_1+5t_2+4y_2) \rho \nabla^2\rho \nonumber\\
&+& \frac{1}{64} (3t_1+6y_1+t_2+2y_2) \rho_3 \nabla^2\rho_3 \nonumber\\
&-& \frac{3}{4} W_0 \rho \nabla \cdot \vec{J} - \frac{1}{4} W_0 \rho_3 \nabla \cdot \vec{J}_3,
\end{eqnarray}
where $\rho=\rho_n+\rho_p$, $\tau=\tau_n+\tau_p$, and $\vec{J}=\vec{J}_n+\vec{J}_p$ are the isoscalar nucleon number density, the kinetic density, and the spin-current density, respectively, and $\rho_3=\rho_n-\rho_p$, $\tau_3=\tau_n-\tau_p$, and $\vec{J}_3=\vec{J}_n-\vec{J}_p$ are the corresponding isovector densities, respectively. For the standard SHF EDF, we refer the reader to Ref.~\cite{MSL0} for comparison. Here we assume that the nuclei investigated in the present study are spherical and consider only time-even terms in the EDF. Using the variational principle, one obtains the single-nucleon Hamlitonian and the Schr\"odinger equation. Solving the Schr\"odinger equation leads to the eigen-energies and wave functions of constituent nucleons, based on which the binding energy, the charge radius, and the neutron-skin thickness can be obtained from this standard procedure~\cite{Vau72}.

\subsection{Giant resonances from RPA method}

The nuclear response to external fields is studied by applying the random phase approximation (RPA) and using the Hartree-Fock basis obtained from the above EDF. For the present study, we use the open source routine of Ref.~\cite{Col13}, after modifying to incorporate the extended density dependence in Eq.~(\ref{SHFI}). The operators for exciting the IVGDR, ISGMR, and ISGQR are chosen respectively as
\begin{equation}
\hat{F}_{\rm IVGDR} = \frac{N}{A} \sum_{i=1}^Z r_i Y_{\rm 1M}(\hat{r}_i) - \frac{Z}{A} \sum_{i=1}^N r_i Y_{\rm 1M}(\hat{r}_i), \label{QIVGDR}
\end{equation}
\begin{equation}
\hat{F}_{\rm ISGMR} = \sum_{i=1}^A r_i^2 Y_{00}(\hat{r}_i),
\end{equation}
\begin{equation}
\hat{F}_{\rm ISGQR} = \sum_{i=1}^A r_i^2 Y_{2M}(\hat{r}_i),
\end{equation}
where $N$, $Z$, and $A$ are respectively the neutron, proton, and nucleon numbers in a nucleus, $r_i$ is the coordinate of the $i$th nucleon with respect to the center-of-mass of the nucleus, and $Y_{\rm 00}(\hat{r}_i)$, $Y_{\rm 1M}(\hat{r}_i)$, and $Y_{2M}(\hat{r}_i)$ are the spherical harmonics with the magnetic quantum number $M$ degenerate in spherical nuclei. Using the RPA method~\cite{Col13}, the strength function
\begin{equation}
S(E) = \sum_\nu |\langle \nu|| \hat{F}  || \tilde{0} \rangle |^2 \delta(E-E_\nu)
\end{equation}
of a nucleus resonance in a given channel can be obtained, where the square of the reduced matrix element $|\langle \nu|| \hat{F}  || \tilde{0} \rangle |$ represents the transition probability from the ground state $| \tilde{0} \rangle $ to the excited state $| \nu \rangle$ under the action of the external field $\hat{F}$. The moments of the strength function for the corresponding resonance type can then be calculated from
\begin{equation}
m_k = \int_0^\infty dE E^k S(E).
\end{equation}
For the IVGDR, the centroid energy $E_{-1}$ and the electric polarizability $\alpha_D$ can be obtained from the moments of the strength function through the relation
\begin{eqnarray}
E_{-1} &=& \sqrt{m_1/m_{-1}}, \\
\alpha_D &=& \frac{8\pi e^2}{9} m_{-1}.
\end{eqnarray}
For the ISGMR, the RPA results of the excitation energy
\begin{equation}
E_{ISGMR}=m_1/m_0
\end{equation}
are compared with the corresponding experimental data. For the ISGQR, we compare the peak values of the strength function directly to the corresponding experimental data.

\subsection{Bayesian analysis}

We employ the Bayesian analysis method to obtain the PDFs of model parameters from the experimental data, and the calculation method can be formally expressed as the Bayes' theorem
\begin{equation}
P(M|D) = \frac{P(D|M)P(M)}{\int P(D|M)P(M)dM},
\end{equation}
where $P(M|D)$ is the posterior probability for the model $M$ given the data set $D$, $P(D|M)$ is the likelihood function or the conditional probability for a given theoretical model $M$ to predict correctly the data $D$, and $P(M)$ denotes the prior probability of the model $M$ before being confronted with the data. The denominator of the right-hand side of the above equation is the normalization constant. For the prior PDFs, we choose the model parameters $p_1=E_{sym}^0$ uniformly within $25 \sim 35$ MeV, $p_2=L$ uniformly within $0 \sim 120$ MeV, $p_3=m_v^\star/m$ uniformly within $0.5 \sim 1$, and $p_4=K_0$ uniformly within $200 \sim 300$ MeV. In order to study the isospin dependence of the incompressibility, we also choose $p_5=K_{sym}$ uniformly within $-400 \sim 100$ MeV based
on analyses of terrestrial nuclear experiments and EDFs~\cite{Tew17,Zha17}. Although it is not the purpose to constrain $Q_{sym}$, we vary $p_6=Q_{sym}$ uniformly within $0 \sim 1000$ MeV in order to take into account the uncertainties of higher-order EOS parameters and thus obtain a conservative constraint on other quantities. In addition, $p_7=Q_0$ is varied uniformly within $-800 \sim 400$ MeV~\cite{Tew17,Zha17} in the most complete calculation with the ISGMR data incorporated. The theoretical results of $d^{th}_1=\Delta r_{np}$, $d^{th}_2=E_{-1}$, $d^{th}_3=\alpha_D$, and $d^{th}_4=E_{ISGMR}$ from the SHF-RPA method are compared with the experimental data $d^{exp}_{1 \sim 4}$, and a likelihood function is used to quantify how well these model parameters reproduce the corresponding experimental data
\begin{eqnarray}
&&P[D(d_1,d_2,d_3,d_4)|M(p_1,p_2,p_3,p_4,p_5,p_6,p_7)] \notag\\
&=& \Pi_{i=1}^4 \Bigg \{ \frac{1}{2\pi \sigma_i} \exp\left[-\frac{(d^{th}_i-d^{exp}_i)^2}{2\sigma_i^2}\right] \Bigg\}, \label{llh}
\end{eqnarray}
where $\sigma_{i}$ is the $1\sigma$ error of the data $d_i^{exp}$. The calculation of the posterior PDFs is based on the Markov-Chain Monte Carlo (MCMC) approach using the Metropolis-Hastings algorithm~\cite{Met53,Has70}. Since the MCMC process does not start from an equilibrium distribution, initial samples in the so-called burn-in period have to be thrown away. After the average of each model parameter becomes stable, the posterior PDF of a single model parameter $p_i$ can be calculated from
\begin{equation}\label{1dpdf}
P(p_i|D) = \frac{\int P(D|M) P(M) \Pi_{j\ne i} dp_j}{\int P(D|M) P(M) \Pi_{j} dp_j},
\end{equation}
while the correlated PDF of two model parameters $p_i$ and $p_j$ can be calculated from
\begin{equation}\label{2dpdf}
P[(p_i,p_j)|D] = \frac{\int P(D|M) P(M) \Pi_{k\ne i,j} dp_{k}}{\int P(D|M) P(M) \Pi_{k} dp_k}.
\end{equation}
In the present study, we incorporate the experimental data and the corresponding sensitive model parameters step by step, so that we can understand where the correlation between model parameters as well as their posterior PDFs come from.


\begin{widetext}
\begin{table*}\small
  \caption{Experimental data of the neutron-skin thickness $\Delta r_{np}$, the centroid energy $E_{-1}$ of the IVGDR and electric polarizability $\alpha_D$, the excitation energy $E_{ISGMR}$ of the ISGMR, the average energy per nucleon $E_b$, and the charge radius $R_c$ in $^{208}$Pb and $^{120}$Sn used for the Bayesian analysis. For $^{208}$Pb, both the $E_{ISGMR}$ data by TAMU and RCNP are used.}
    \begin{tabular}{|c|c|c|c|c||c|c|}
   \hline
           & $\Delta r_{np}$ (fm) & $E_{-1}$ (MeV)& $\alpha_D$ (fm$^3$) &  $E_{ISGMR}$ (MeV) & $E_b$ (MeV) & $R_c$ (fm) \\
   \hline
   \hline
    $^{208}$Pb & $0.283 \pm 0.071$ & $13.46 \pm 0.10$ & $19.6 \pm 0.6$ & $14.17 \pm 0.28$ $\&$ $13.9 \pm 0.1$ & $-7.867452\pm 3\%$  & $5.5010 \pm 3\%$ \\
    $^{120}$Sn & $0.150 \pm 0.017$ & $15.38 \pm 0.10$ & $8.59 \pm 0.37$ & $15.7 \pm 0.1$ & $-8.504548\pm 3\%$  & $4.6543 \pm 3\%$  \\
   \hline
    \end{tabular}
  \label{T2}
\end{table*}
\end{widetext}

Details of the experimental data for $^{208}$Pb and $^{120}$Sn used in the present study are shown in Table.~\ref{T2}. For the neutron-skin thickness, we adopt the latest PREXII data of $\Delta r_{np} = 0.283 \pm 0.071$ fm for $^{208}$Pb from parity violating electron-nucleus scatterings~\cite{PREXII}, and the predicted values of $\Delta r_{np}=0.150 \pm 0.017$ fm for $^{120}$Sn from $L(\rho^\star=0.10~{\rm fm}^{-3})=43.7 \pm 5.3$ MeV extracted in Ref.~\cite{Xu20b}, with the latter deduced from the neutron-skin thickness of Sn isotopes from proton elastic scattering experiments~\cite{Ter08}. For $^{208}$Pb, the experimental results of the centroid energy $E_{-1}=13.46$ MeV of the IVGDR from photoneutron scatterings~\cite{IVGDRe}, and the electric polarizability $\alpha_D=19.6 \pm 0.6$ fm$^3$ from polarized proton inelastic scatterings~\cite{Tam11} and with the quasi-deuteron excitation contribution subtracted~\cite{Roc15}, are used in the Bayesian analysis. For $^{120}$Sn, we use the experimental data of $E_{-1}=15.38$ MeV of the IVGDR from photoneutron scatterings~\cite{IVGDRe}, and $\alpha_D=8.59 \pm 0.37$ fm$^3$ from combining the proton inelastic scattering and photoabsorption data~\cite{Has15} and with the quasi-deuteron excitation contribution subtracted~\cite{Roc15}, overlaping with $\alpha_D=8.08 \pm 0.60$ fm$^3$ from the latest data extracted through proton inelastic scatterings~\cite{Bas20a,Bas20b}. The $1\sigma$ error of $E_{-1}$ for both $^{208}$Pb and $^{120}$Sn is chosen to be $0.1$ MeV representing the scale of its uncertainty so far~\cite{IVGDRe}. For the excitation energy of the ISGMR from inelastic scatterings of $\alpha$ particles, we use $E_{ISGMR}=15.7 \pm 0.1$ MeV for $^{120}$Sn by the RCNP, Osaka University~\cite{SnRCNP}, and for $^{208}$Pb we use both $E_{ISGMR}=14.17 \pm 0.28$ MeV by the TAMU~\cite{You99} and $E_{ISGMR}=13.9 \pm 0.1$ MeV by the RCNP~\cite{PbRCNP}. Besides comparing with the experimental data of $E_{-1}$, $\alpha_D$, $\Delta r_{np}$, and $E_{ISGMR}$, we have also used a strong constraint that the theoretical calculation should reproduce the binding energy and charge radius of the corresponding nucleus within $3\%$, an uncertainty range for reasonable SHF parameterization as shown in Ref.~\cite{MSL0}, otherwise the likelihood function [Eq.~(\ref{llh})] is set to 0. This condition guarantees that we are exploring a reasonable space of model parameters, and the experimental data of the binding energies and charge radii of $^{208}$Pb and $^{120}$Sn are taken from Refs.~\cite{Aud03,Ang04}. A more precise description of these data is possible, but for each set of EOS parameters it would require fits of the density gradient and spin-orbit parameters to the properties of several more nuclei (to avoid over-fitting), which is beyond the scope of the present study.

\section{Results and discussions}

In the present study, we first fix the value of the isoscalar nucleon effective mass by reproducing the excitation energies of the ISGQR in $^{208}$Pb and $^{120}$Sn. Next, we investigate the sensitivity of involved observables to the physics quantities of interest. The Bayesian analysis is then carried out step by step, by incorporating more observables and physics quantities in the analysis. Results from the KIDS model are compared with those from the standard SHF model, in order to understand the difference from previous studies~\cite{Xu20,Xu21,Xu21b} as well as the model dependence.

\subsection{Sensitivity study}

\begin{figure}[ht]
\includegraphics[scale=0.4]{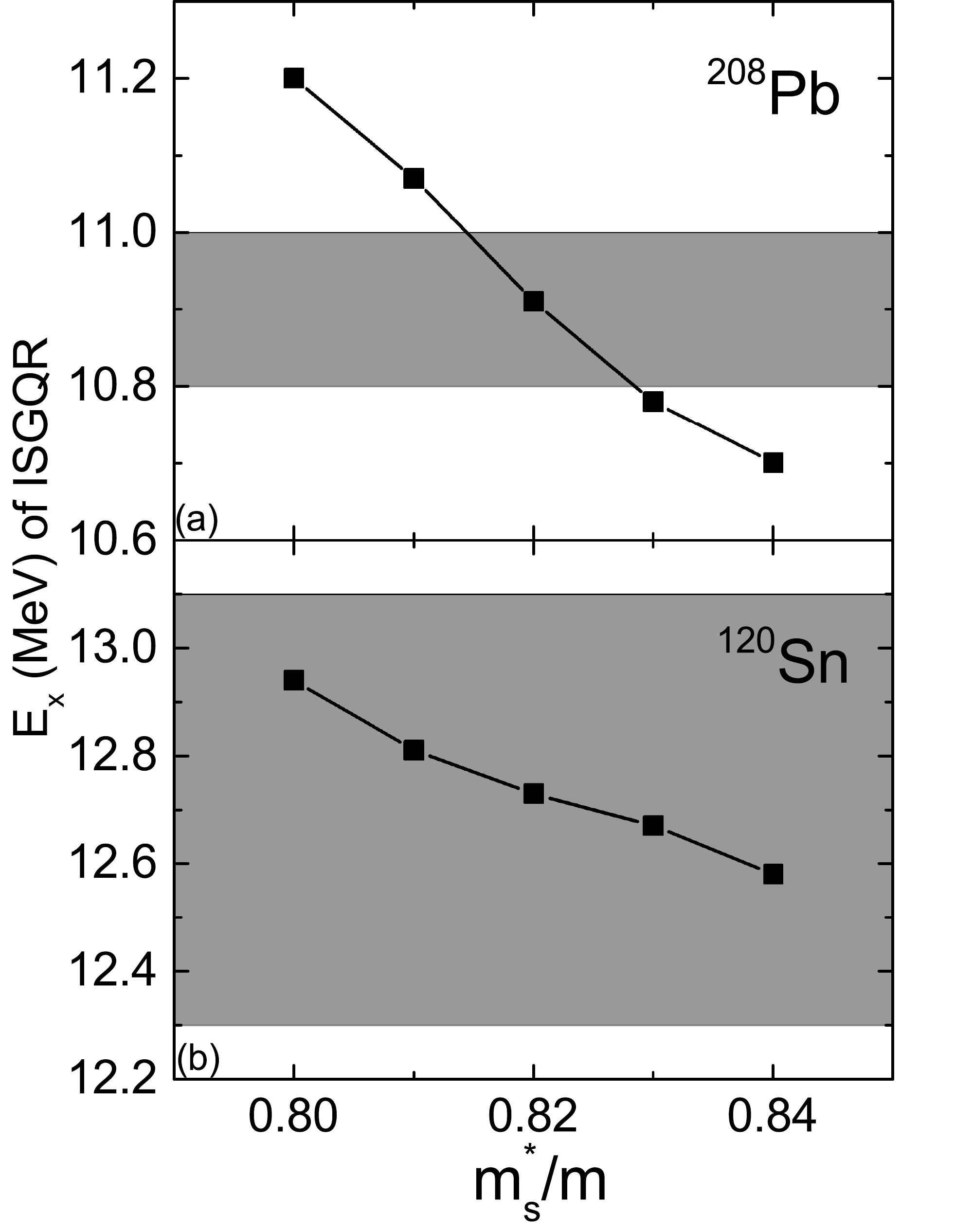}
	\caption{ The excitation energy of ISGQR in $^{208}$Pb (upper) and $^{120}$Sn (lower) from SHF-RPA calculations using default parameters in Table~\ref{T1} but varying $m_s^\star$. The experimental data shown by bands are compared. } \label{fig1}
\end{figure}

\begin{figure*}[ht]
\includegraphics[scale=0.8]{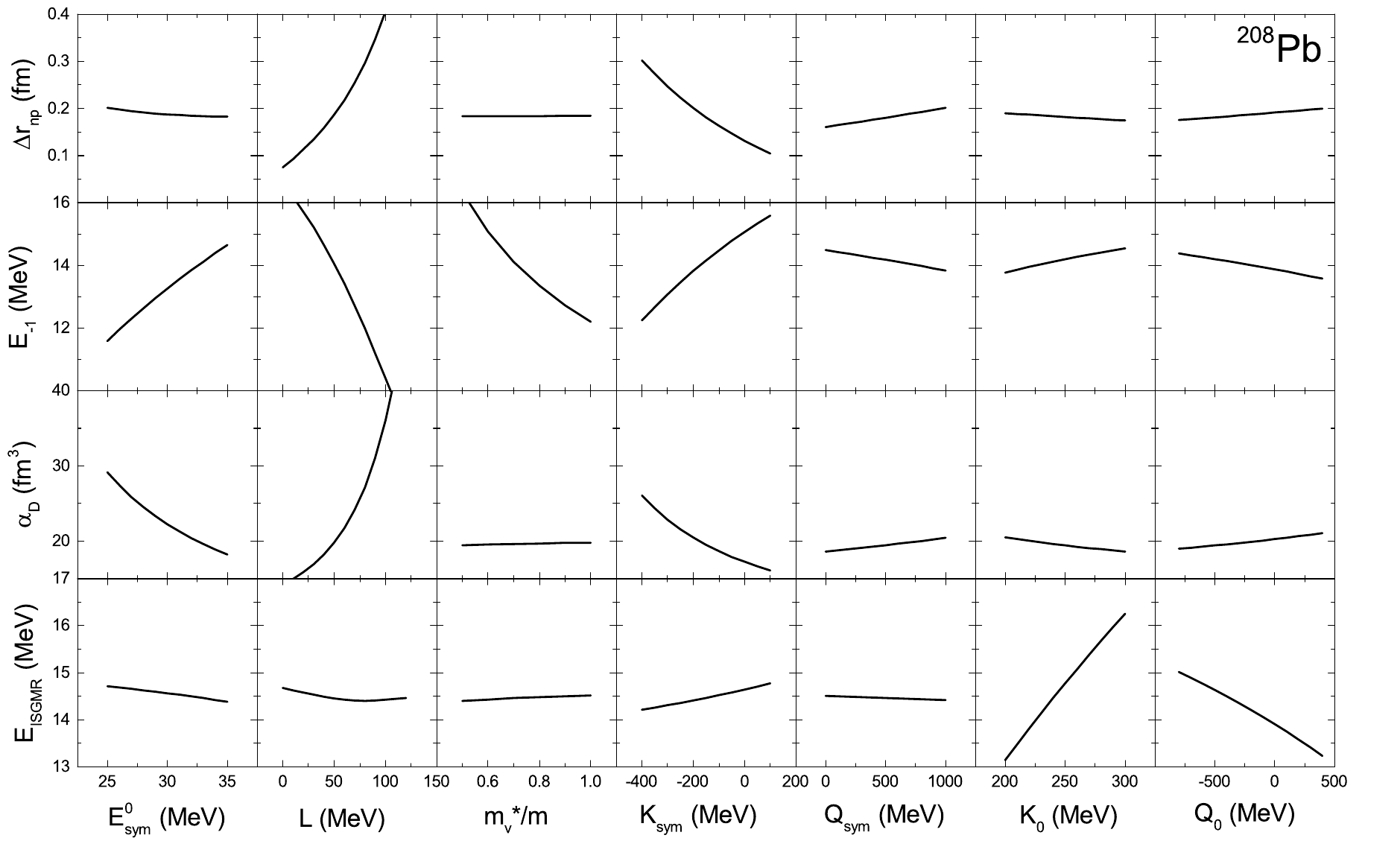}
	\caption{Sensitivity of $\Delta r_{np}$, $E_{-1}$, $\alpha_D$, and $E_{ISGMR}$ in $^{208}$Pb to $E_{sym}^0$, $L$, $m_v^\star$, $K_{sym}$, $Q_{Sym}$, $K_0$, $Q_0$ by changing one quantity a time within its prior range with other quantities fixed at their default values in Table~\ref{T1}. } \label{fig2}
\end{figure*}

\begin{figure*}[ht]
\includegraphics[scale=0.8]{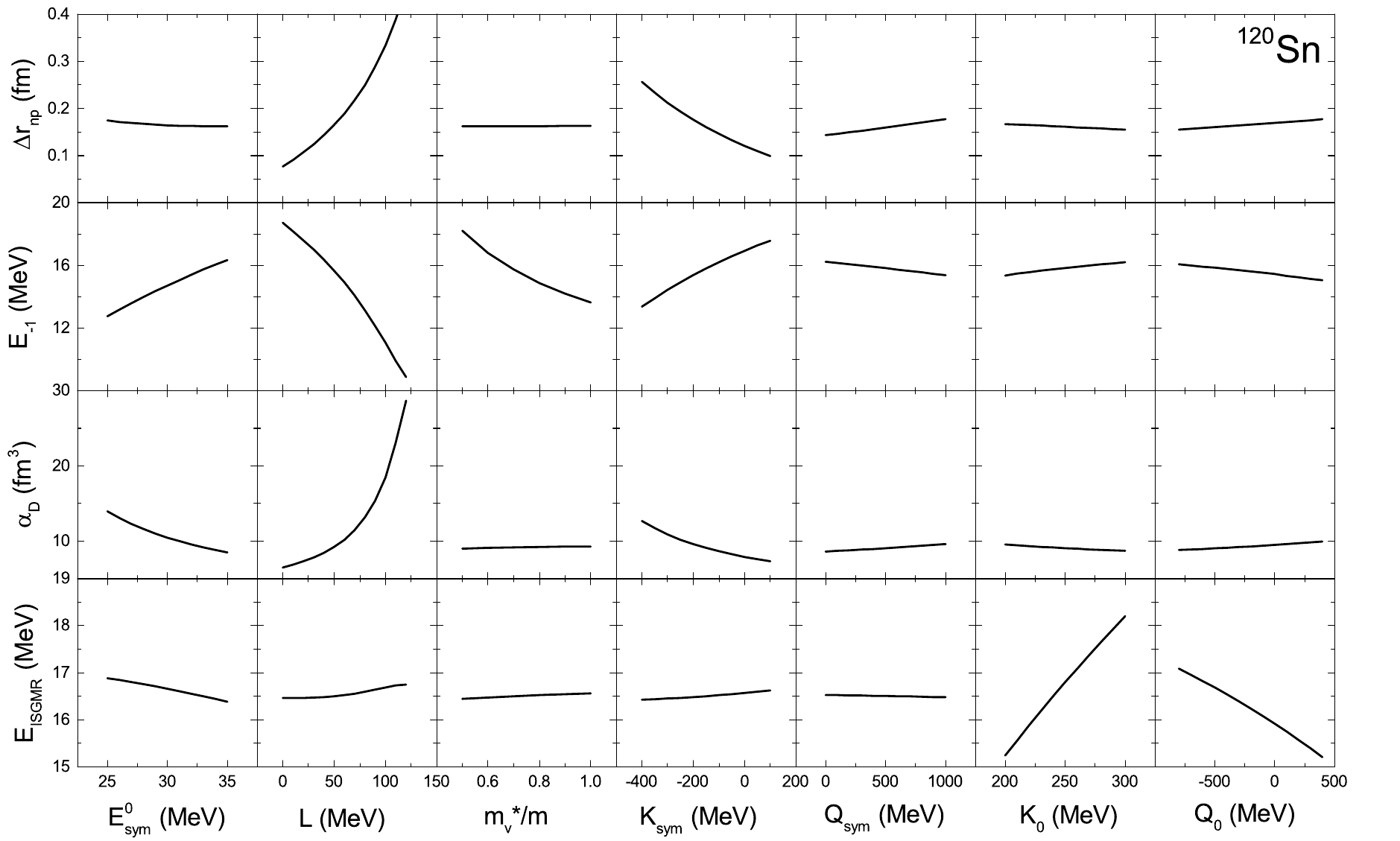}
	\caption{Same as Fig.~\ref{fig2} but for $^{120}$Sn.} \label{fig3}
\end{figure*}

We first show how we determine the isoscalar nucleon effective mass $m_s^\star$ from the ISGQR in $^{208}$Pb and $^{120}$Sn, which is less sensitive to other physics quantities of interest here. As shown in Fig.~\ref{fig1}, the excitation energies $E_x$ of the ISGQR in both $^{208}$Pb and $^{120}$Sn from SHF-RPA calculations based on the KIDS model are seen to decrease with increasing $m_s^\star/m$, and the $E_x$ is seen to be smaller in $^{208}$Pb compared with that in $^{120}$Sn. This is consistent with the intuitive picture that the oscillation frequency becomes smaller in a heavier system with a larger nucleon number or nucleon effective mass. The bands represent the experimental data of $E_x=10.9 \pm 0.1$ MeV in $^{208}$Pb~\cite{ISGQRex1,ISGQRex2,ISGQRex3,You81,Roc13a} and $E_x=12.7 \pm 0.4$ MeV in $^{120}$Sn~\cite{You81} from $\alpha$ inelastic scatterings, respectively shown in panels (a) and (b). It is seen that $m_s^\star/m=0.82$ reproduces the $E_x$ values of the ISGQR in $^{208}$Pb rather well, while the large range of $E_x$ for $^{120}$Sn covers $m_s^\star/m=0.80 \sim 0.84$. We thus fix $m_s^\star/m=0.82$ in the present study.

We further carry out a sensitivity study, by showing the dependence of $\Delta r_{np}$, $E_{-1}$, $\alpha_D$, and $E_{ISGMR}$ in $^{208}$Pb to $E_{sym}^0$, $L$, $m_v^\star$, $K_{sym}$, $Q_{sym}$, $K_0$, and $Q_0$ within the prior ranges of these physics quantities as in Table.~\ref{T1}. In the SHF-RPA calculation based on the KIDS model, we change one quantity at a time while others are fixed at their default values in Table.~\ref{T1}. Results for $^{208}$Pb are shown in Fig.~\ref{fig2} while those for $^{120}$Sn are shown in Fig.~\ref{fig3}, and the sensitivities for the observables of interest are similar in the two systems. We note that this is an illustration of sensitivities free from the experimental data. $\Delta r_{np}$ is seen to be most sensitive to $L$ and moderately sensitive to $K_{sym}$. $E_{-1}$ is seen to be most sensitive to $L$ and moderately sensitive to $E_{sym}^0$, $m_v^\star$, and $K_{sym}$. $\alpha_D$ is seen to be most sensitive $L$ and moderately sensitive to $E_{sym}^0$ and $K_{sym}$. $E_{ISGMR}$ is seen to be most sensitive to $K_0$, moderately sensitive to $Q_0$, and slightly sensitive to $K_{sym}$. It is seen that none of the observables here is sensitive to $Q_{sym}$, varying which does not affect much the conclusion in the present study. Varying other physics variables within their empirical uncertainty ranges, such as $G_S$, $G_V$, and $W_0$, leads to effects on the nuclear structure observables at most comparable to that from $Q_{sym}$. The sensitivity study justifies the validity of the Bayesian analysis by choosing proper physics variables. We also note that the apparent sensitivities of observables to higher-order EOS parameters, such as $K_{sym}$ and $Q_0$, are due to their large prior ranges in Table.~\ref{T1}.

\subsection{Bayesian inference on $\Delta r_{np}$}

\begin{figure}[ht]
\includegraphics[scale=0.21]{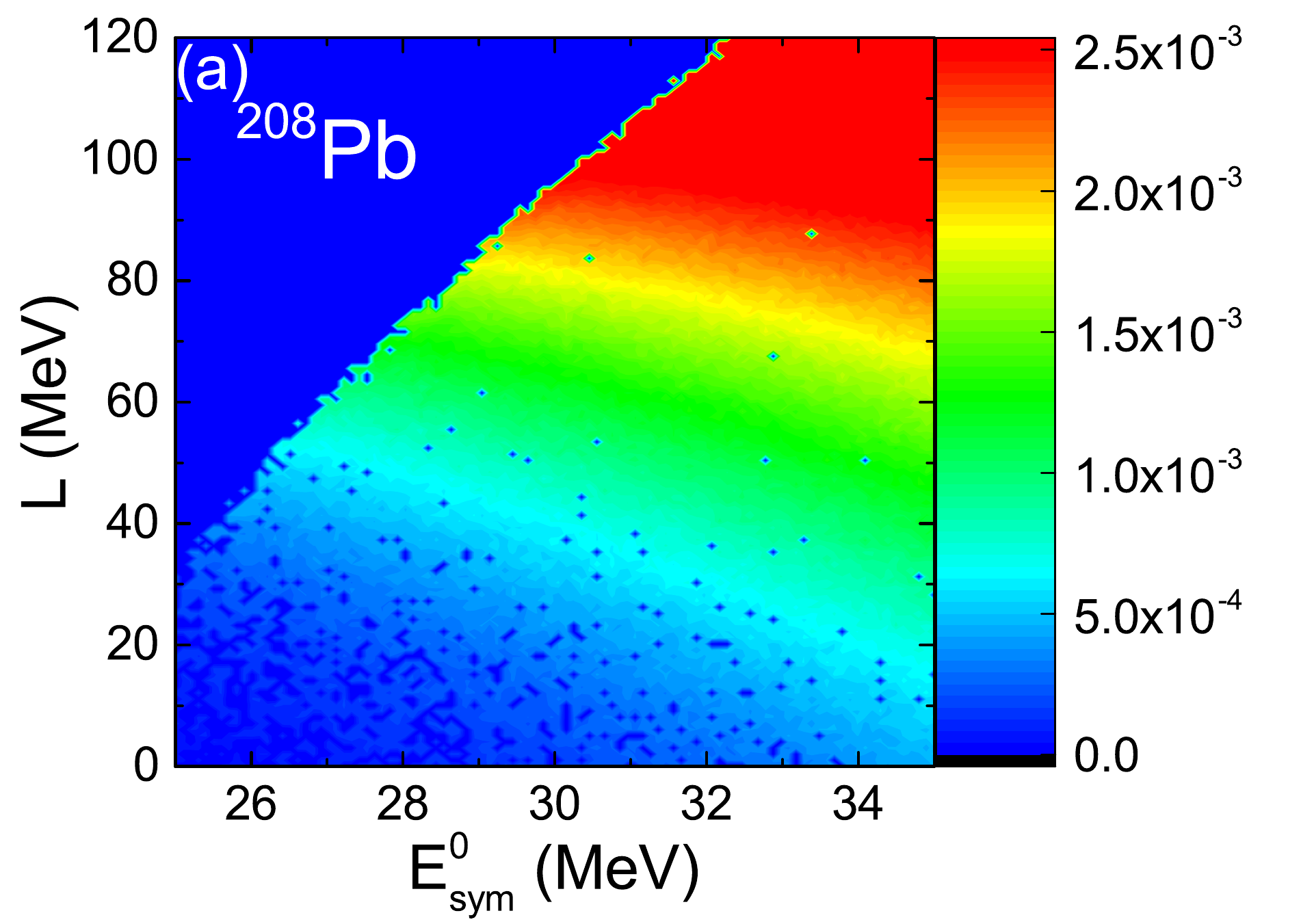} \includegraphics[scale=0.21]{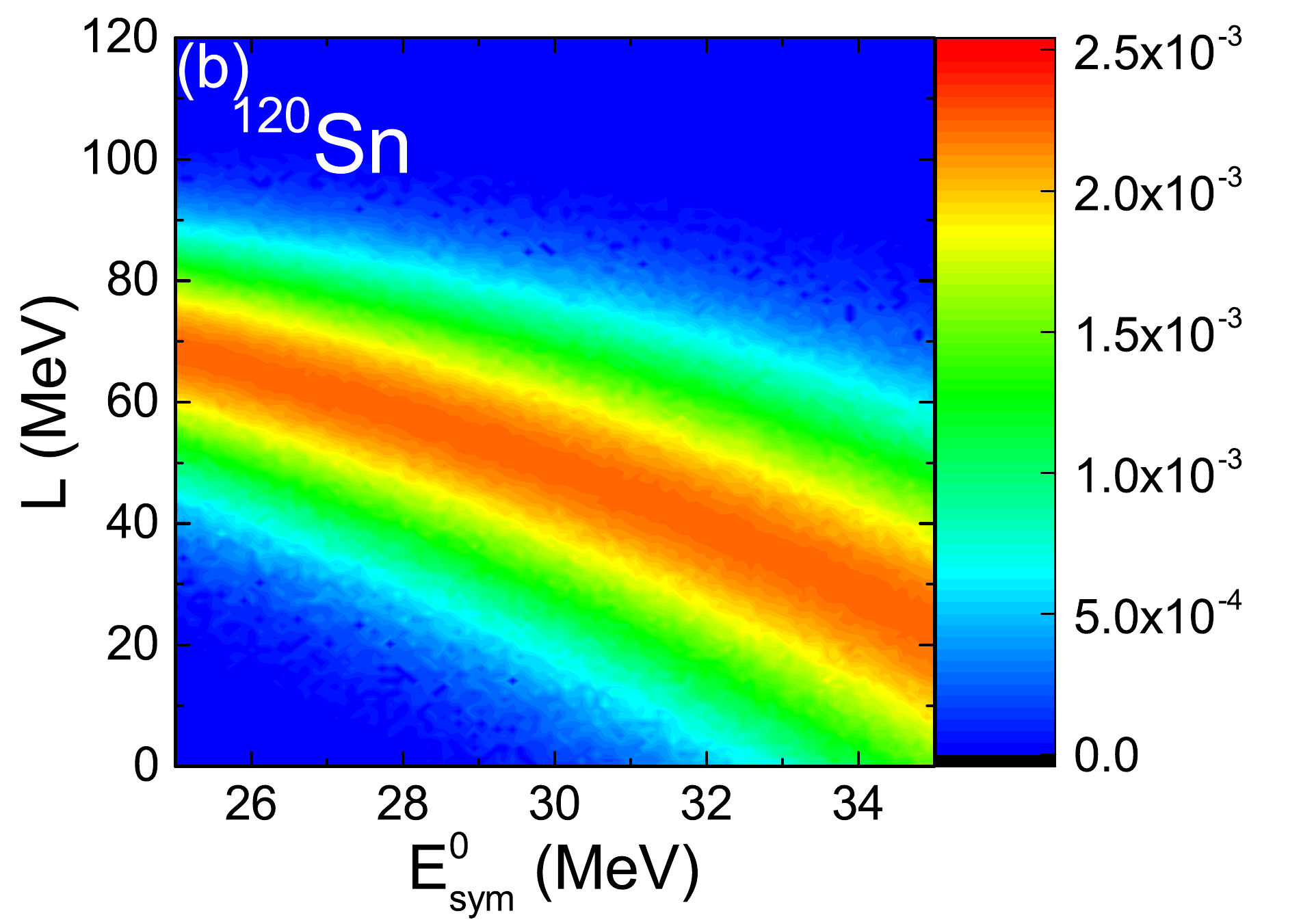} \\
\includegraphics[scale=0.21]{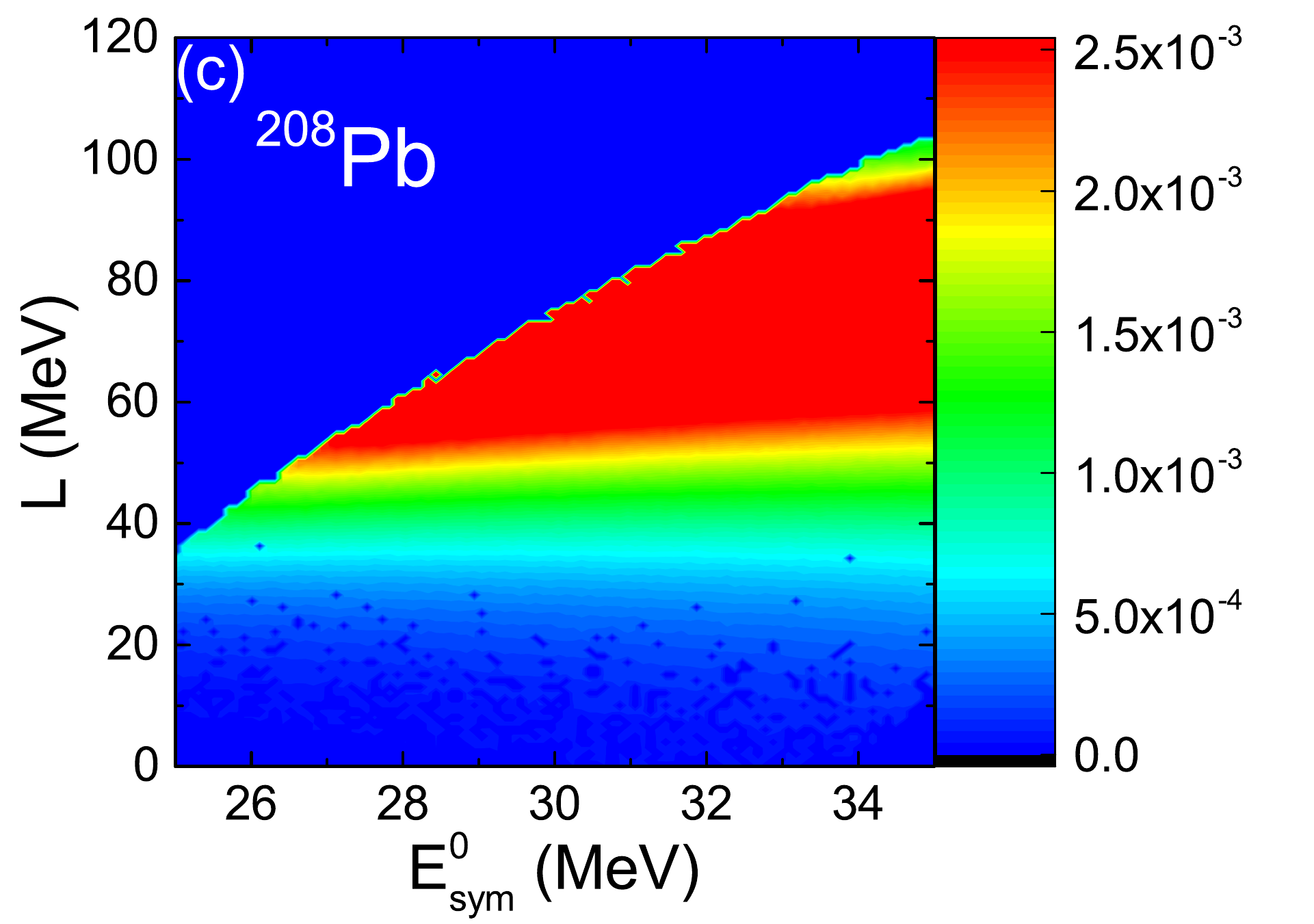} \includegraphics[scale=0.21]{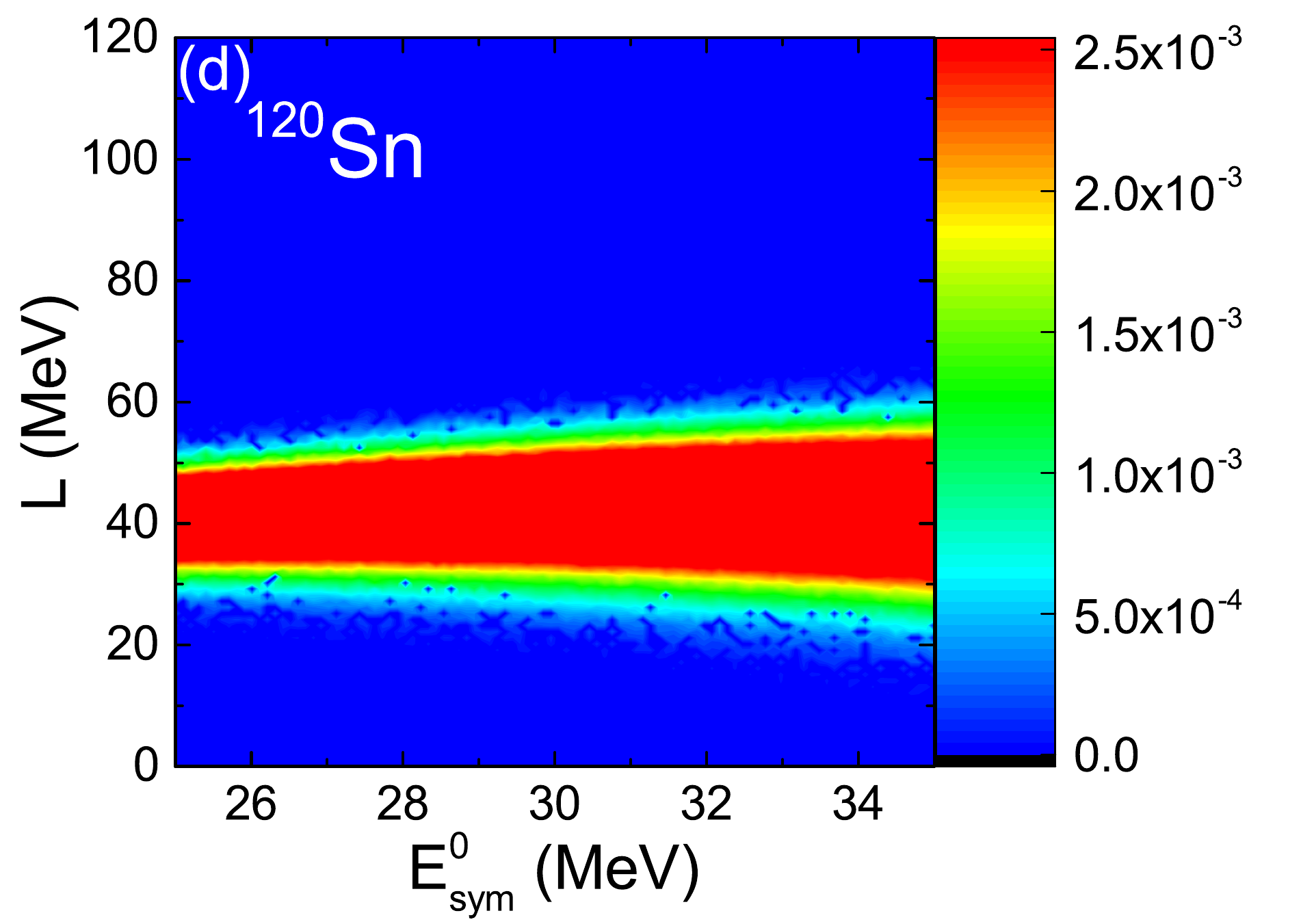} \\
\includegraphics[scale=0.21]{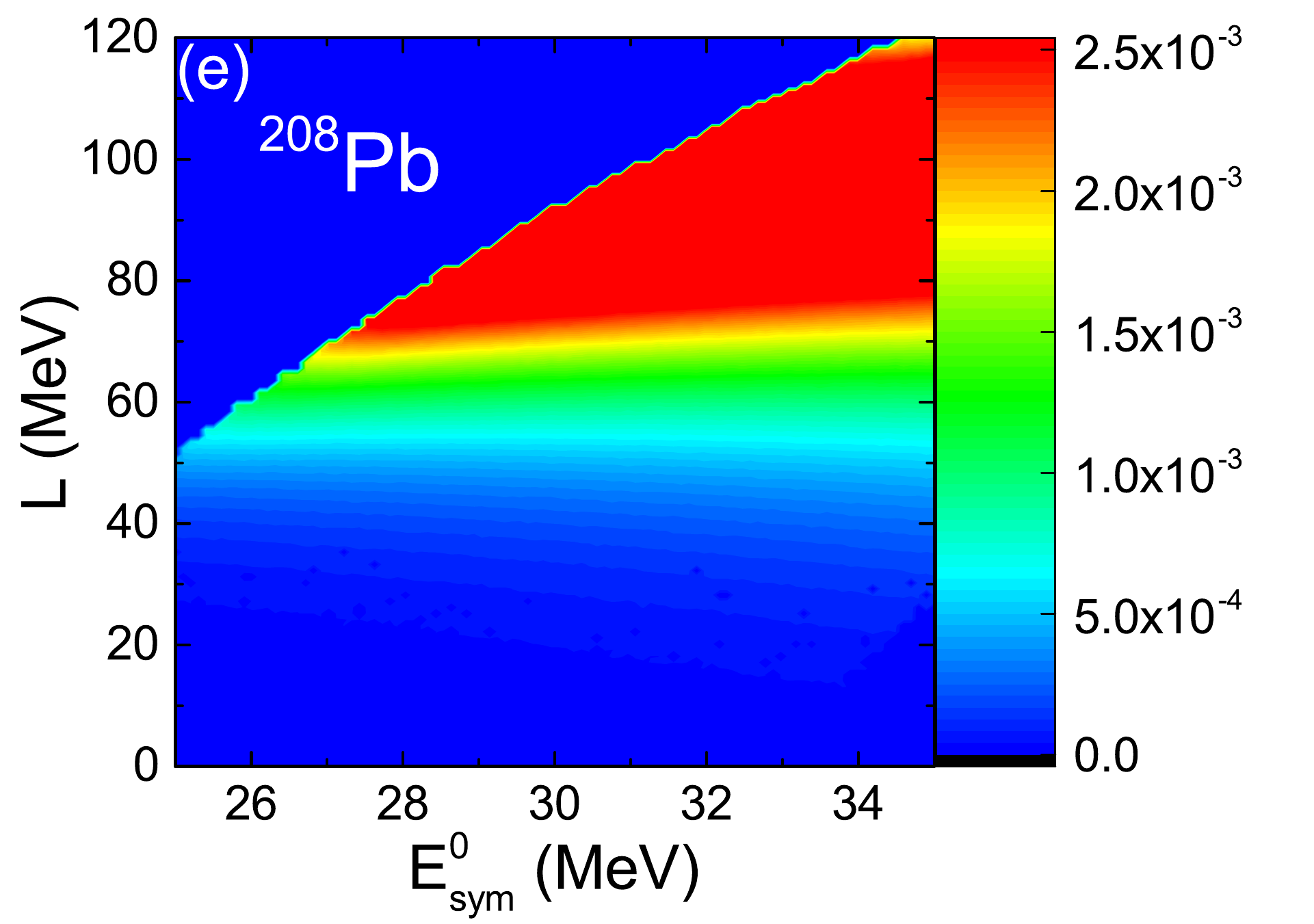} \includegraphics[scale=0.21]{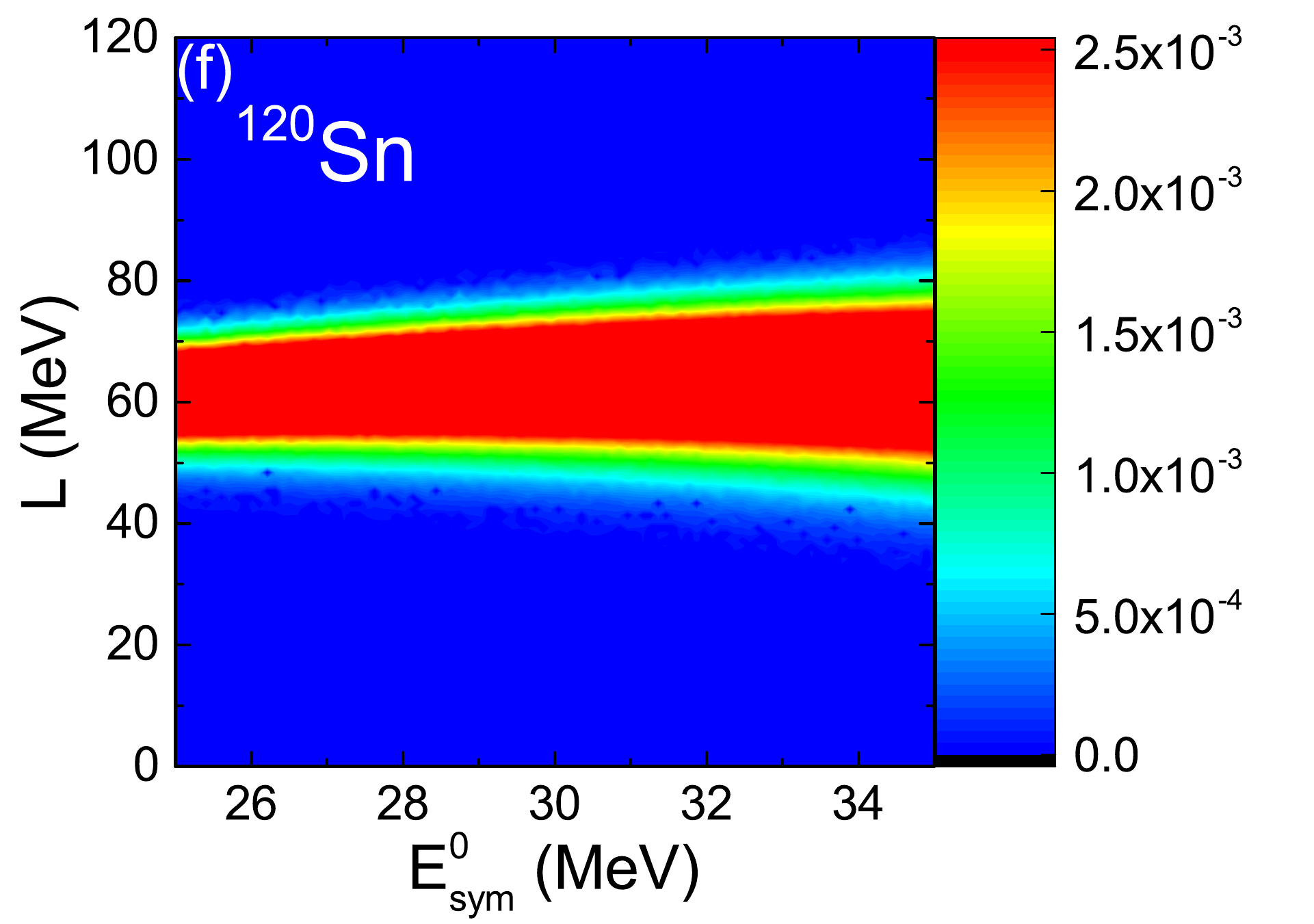} \\
	\caption{Upper: Posterior correlated PDFs between $L$ and $E_{sym}^0$ under the constraint of $\Delta r_{np}$ based on the standard SHF model; Middle: Posterior correlated PDFs of $L$ and $E_{sym}^0$ under the constraint of $\Delta r_{np}$ based on the KIDS model using $K_{sym}=-156$ MeV; Lower: Posterior correlated PDFs of $L$ and $E_{sym}^0$ under the constraint of $\Delta r_{np}$ based on the KIDS model using $K_{sym}=0$ MeV. Results are from only varying $L$, $E_{sym}^0$, and $m_v^\star$, and those in left (right) panels are for $^{208}$Pb ($^{120}$Sn).} \label{fig4}
\end{figure}

We first apply the constraint of only the neutron-skin thickness $\Delta r_{np}$ in Table.~\ref{T2}. As in Ref.~\cite{Xu20b}, we vary $m^\star_v$, $L$, and $E_{sym}^0$ and investigate their posterior PDFs under the constraint of $\Delta r_{np}$ based on the Bayesian analysis. Figure~\ref{fig4} displays the correlated PDFs between $L$ and $E_{sym}^0$ calculated from Eq.~(\ref{2dpdf}) in different scenarios, while the posterior correlated PDFs between $m_v^\star$ and $L$ or $E_{sym}^0$ are trivial, see, e.g., Ref.~\cite{Xu20b}, where similar results were obtained. Fig.~\ref{fig4}(a) and Fig.~\ref{fig4}(b) respectively for $^{208}$Pb and $^{120}$Sn are based on the standard SHF model~\cite{MSL0}. One sees in Fig.~\ref{fig4}(b) that the anticorrelation between $L$ and $E_{sym}^0$ from the $\Delta r_{np}$ in $^{120}$Sn is almost identical to that in Fig.~1(c) of Ref.~\cite{Xu20b}. Compared to the analysis for $^{120}$Sn, the $\Delta r_{np}$ in $^{208}$Pb favors a larger $L$ and with a larger error bar, so the anticorrelation band is shifted and not so obvious due to the limited prior ranges of $L$ and $E_{sym}^0$, as shown in Fig.~\ref{fig4}(a). In the KIDS model, the value of $K_{sym}$ is decoupled from $L$ and $E_{sym}$, different from the case in the standard SHF model, and we will vary it later on. Fig.~\ref{fig4}(c) and Fig.~\ref{fig4}(d) display the correlated PDFs between $L$ and $E_{sym}^0$ with a fixed $K_{sym}=-156$ MeV, and Fig.~\ref{fig4}(e) and Fig.~\ref{fig4}(f) display the similar results but with a fixed $K_{sym}=0$ MeV. The anticorrelations between $L$ and $E_{sym}^0$ is no longer seen, i.e., $\Delta r_{np}$ constrains only $L$ regardless of $E_{sym}^0$, when $K_{sym}$ is fixed as an independent variable rather than coupled to the lower-order parameters. This is different from the intuitive derivation in the appendix of Ref.~\cite{Xu20b}, where the density dependence of the symmetry energy is parameterized as $E_{sym}(\rho)=E_{sym}^0(\rho/\rho_0)^\gamma$. Since the $L$ and $K_{sym}$ are now decoupled in the KIDS model, $E_{sym}(\rho)$ can no longer be parameterized in a density power form with a single $\gamma$ factor. It is also seen that the correlated PDFs depend on the value of $K_{sym}$. For the results of $^{208}$Pb in all scenarios, the regions of too large $L$ and too small $E_{sym}^0$ are ruled out by the rigourous constraint of reproducing the binding energy in the ground state.

\begin{figure}[ht]
\includegraphics[scale=0.21]{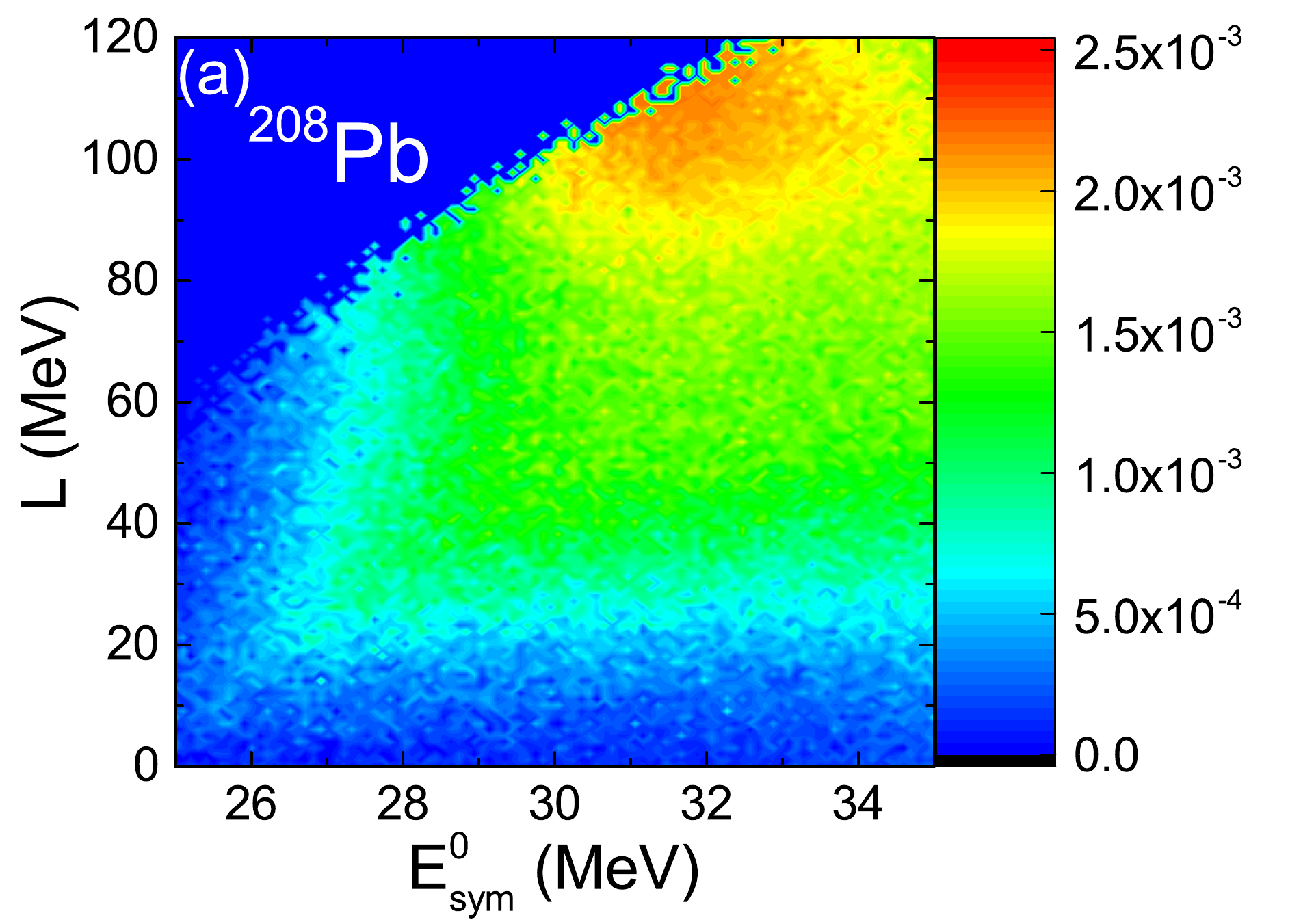} \includegraphics[scale=0.21]{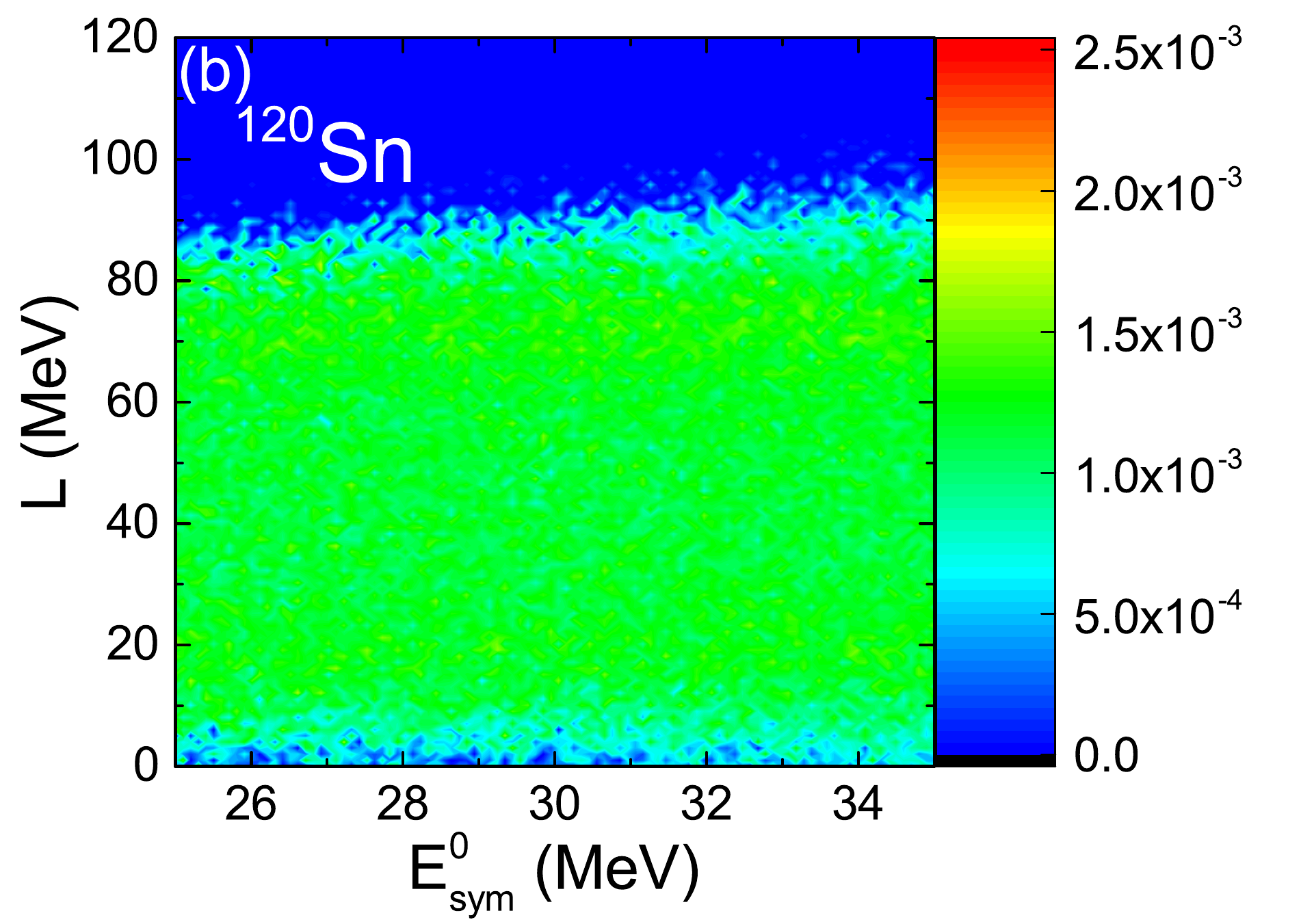} \\
\includegraphics[scale=0.21]{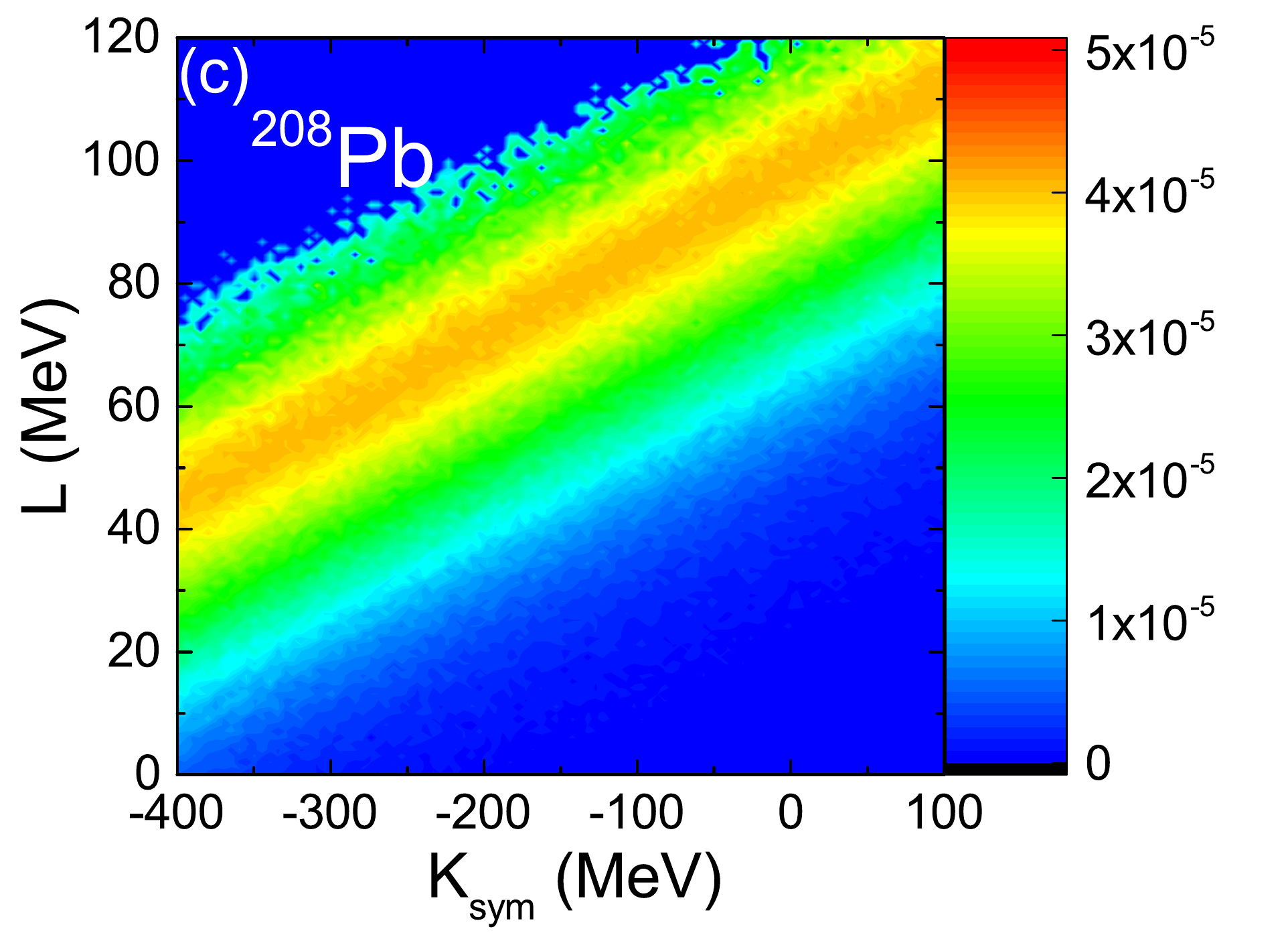} \includegraphics[scale=0.21]{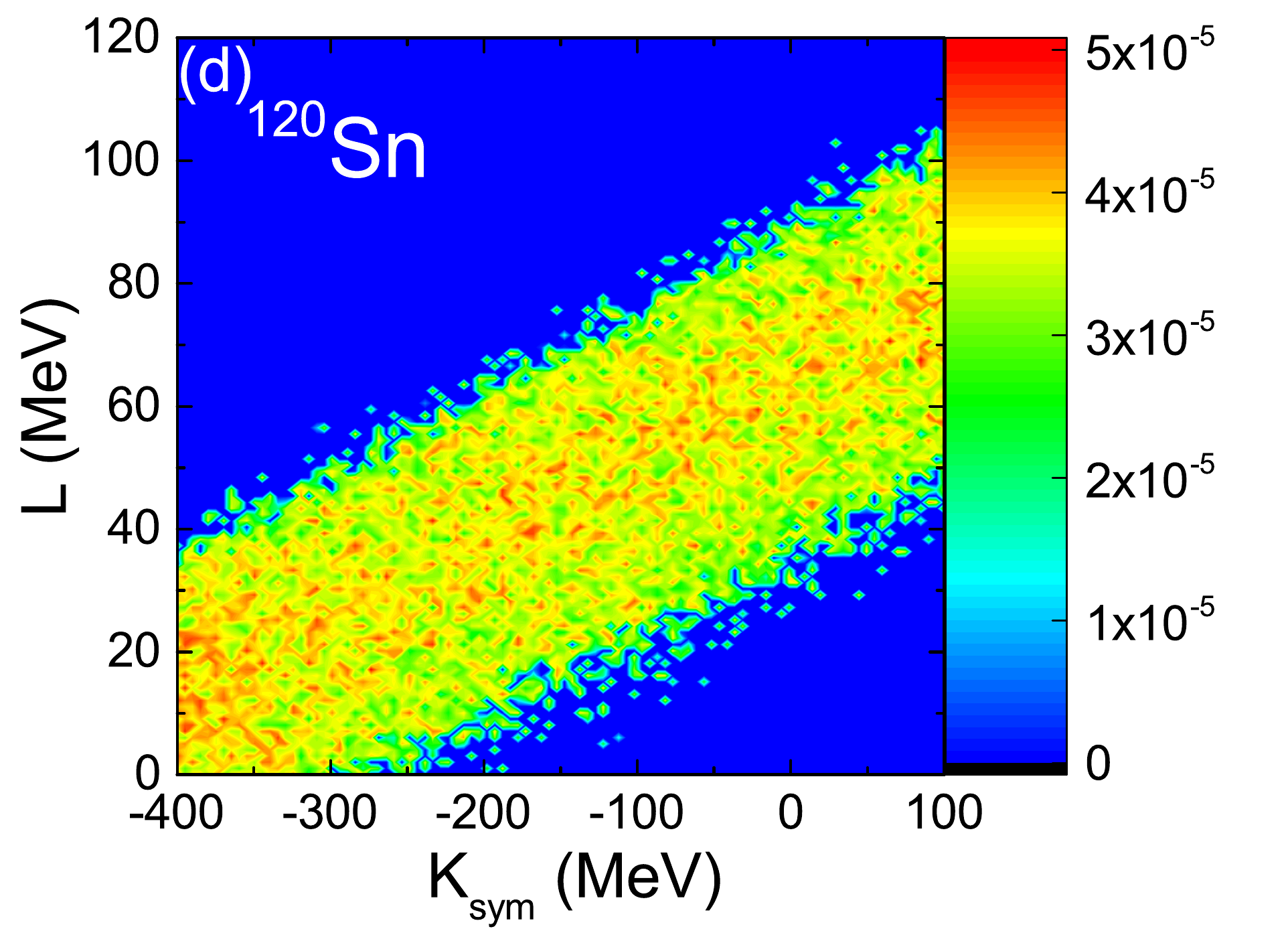} \\
	\caption{Upper: Posterior correlated PDFs between $L$ and $E_{sym}^0$ under the constraint of $\Delta r_{np}$ based on the KIDS model; Lower: Posterior correlated PDFs between $L$ and $K_{sym}$ under the constraint of $\Delta r_{np}$ based the KIDS model. Results are from only changing $L$, $E_{sym}^0$, $m_v^\star$, $K_{sym}$, and $Q_{sym}$, and those in left (right) panels are for $^{208}$Pb ($^{120}$Sn).} \label{fig5}
\end{figure}

It is not surprising that the posterior PDFs under the constraint of $\Delta r_{np}$ depend on the value of $K_{sym}$, since from the sensitivity study (Figs.~\ref{fig2} and \ref{fig3}) $\Delta r_{np}$ is moderately sensitive to $K_{sym}$. We thus vary $m^\star_v$, $L$, $E_{sym}^0$, $K_{sym}$, and also $Q_{sym}$, and the posterior correlated PDFs of interest under the constraint of $\Delta r_{np}$ from the Bayesian analysis are shown in Fig.~\ref{fig5}. One sees from Fig.~\ref{fig5}(a) and Fig.~\ref{fig5}(b) that the posterior correlated PDFs between $L$ and $E_{sym}^0$ are now smeared out, but can be basically regarded as superpositions of the correlated PDFs at different fixed $K_{sym}$. Interestingly, although there is no strong correlation between $L$ and $E_{sym}^0$, we observe a positive correlation between $L$ and $K_{sym}$, as found in Ref.~\cite{New21} based on a different extension of the SHF model. The latter is completely understandable from the positive (negative) correlation between $L$ ($K_{sym}$) and $\Delta r_{np}$ in Figs.~\ref{fig2} and \ref{fig3}, so both $L$ and $K_{sym}$ should increase or decrease to get a similar $\Delta r_{np}$. The latter can be further intuitively understood since both $L$ and $K_{sym}$ characterize the density dependence of symmetry energy so their effects compensate for each other. As shown in Refs.~\cite{Zha13,Xu20b}, the $\Delta r_{np}$ is dominated by the slope parameter of the symmetry energy at about $\frac{2}{3}\rho_0$ based on the standard SHF model, while such sensitivity needs further investigations once $K_{sym}$ becomes an variable independent of $L$. A similar slope of the $L-K_{sym}$ correlation is observed for $^{208}$Pb and $^{120}$Sn, though the intercept values are different, due to overall larger $L$ values favored by the $\Delta r_{np}$ in $^{208}$Pb than in $^{120}$Sn. Under the constraint of $\Delta r_{np}$, it is seen that an accurate constraint on $L$ requires the accurate knowledge of $K_{sym}$.

\begin{figure}[ht]
\includegraphics[scale=0.23]{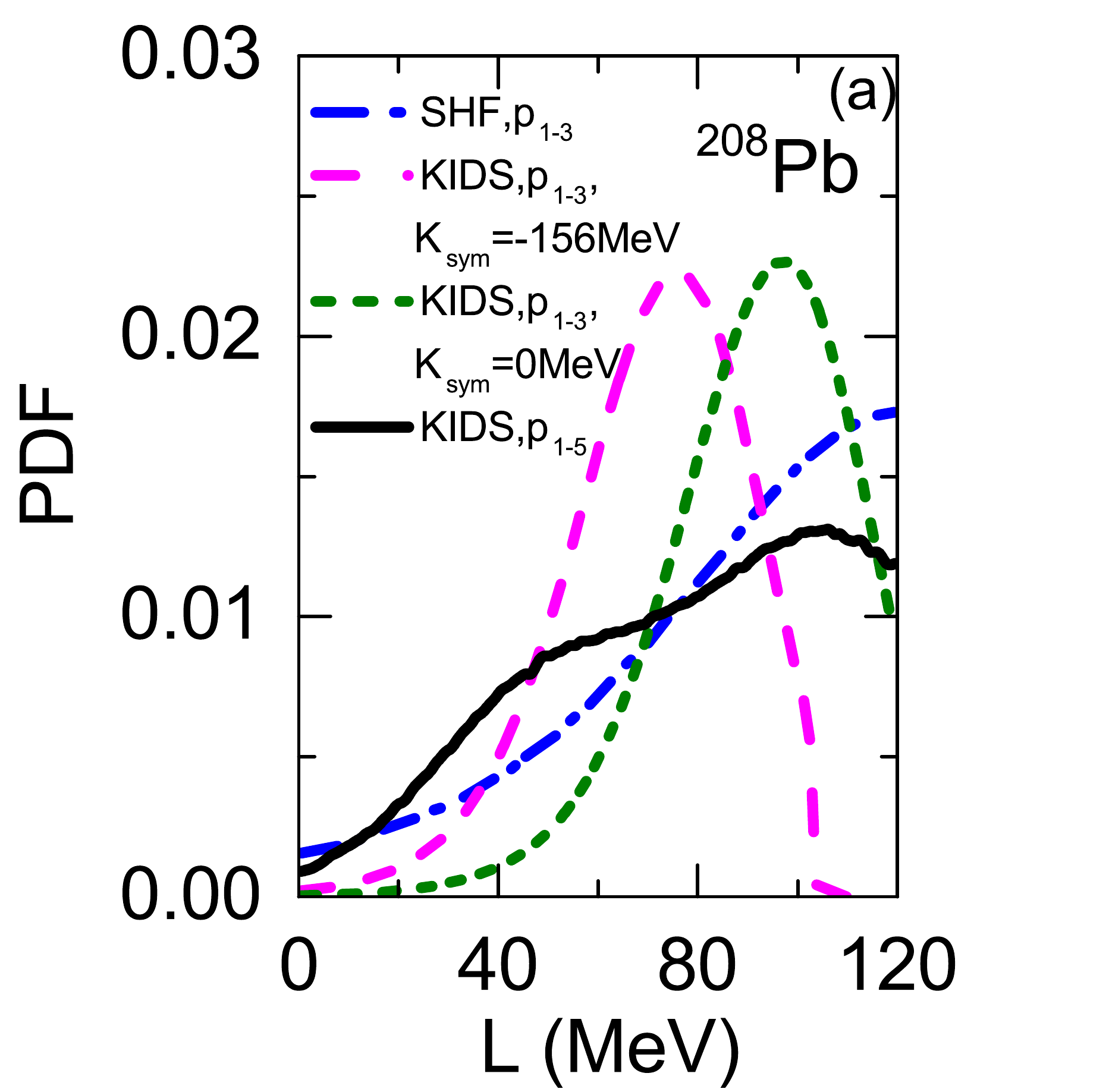} \includegraphics[scale=0.23]{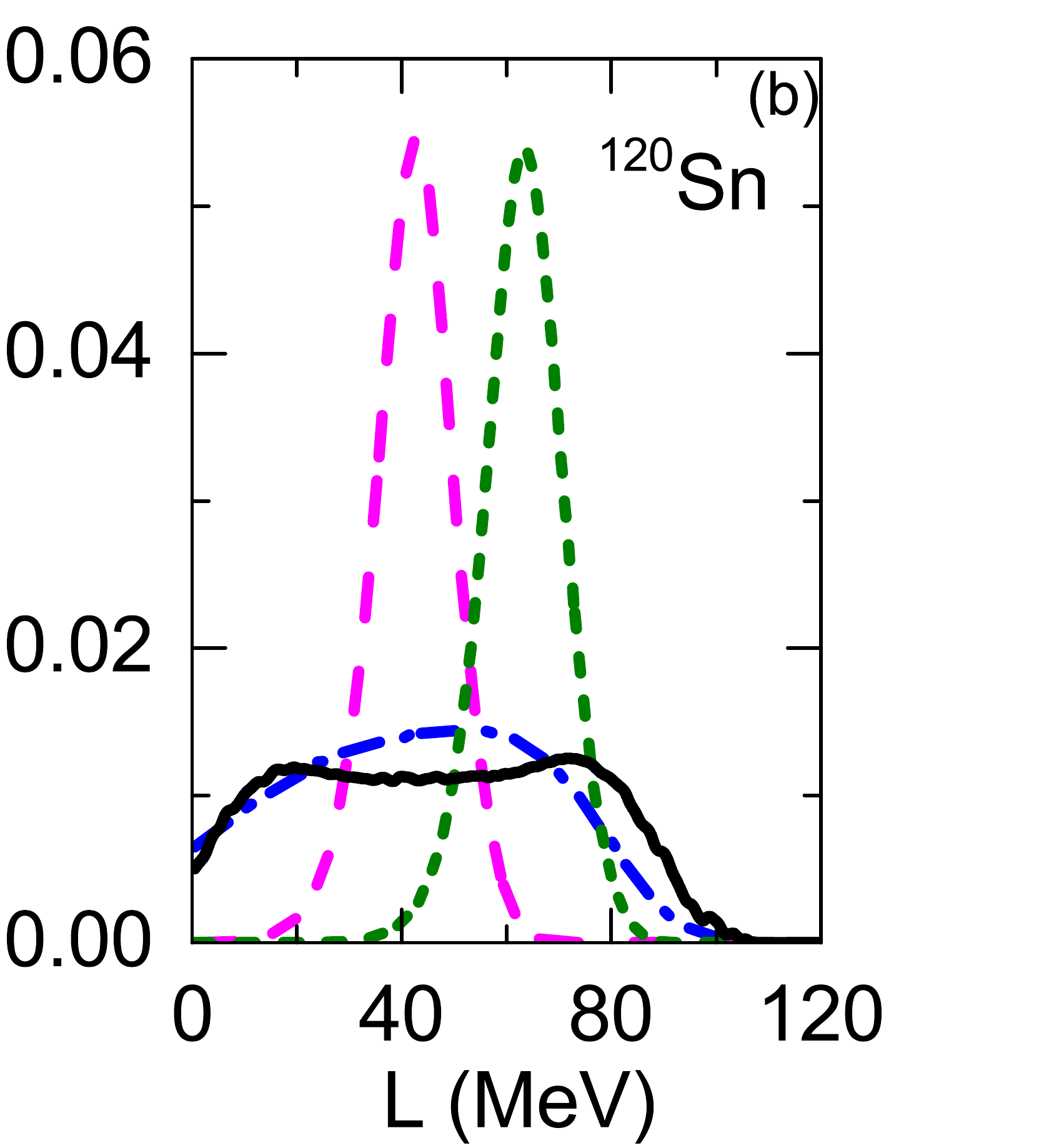}
	\caption{Posterior PDFs of $L$ under the constraint of $\Delta r_{np}$ from the four scenarios in Figs.~\ref{fig4} and \ref{fig5}. Results in the left (right) panel are for $^{208}$Pb ($^{120}$Sn).} \label{fig6}
\end{figure}

Figure~\ref{fig6} compares the posterior PDFs of the slope parameter $L$ of symmetry energy from the four scenarios analyzed in Figs.~\ref{fig4} and \ref{fig5}. Compared with the correlated PDFs, the PDFs of $L$ are basically from integrating other variables according to Eq.~(\ref{1dpdf}). It is seen that the results depend on the EDF as well as the chosen independent variables. From a fixed $K_{sym}=-156$ to 0 MeV, the maximum {\it a posteriori} (MAP) value of $L$ changes from 75 to 97 MeV under the constraint of the $\Delta r_{np}$ in $^{208}$Pb, and changes from 43 to 63 MeV under the constraint of the $\Delta r_{np}$ in $^{120}$Sn. By incorporating $K_{sym}$ and $Q_{sym}$ as independent variables, the posterior PDFs of $L$ become much broader. However, the resulting PDFs of $L$ are similar to those from the Bayesian analysis based on the standard SHF model, where $K_{sym}$ can be determined by $L$, $E_{sym}$, and other quantities. Although we do not show here, we note that none of the scenarios is able to constrain $m_v^\star$, $E_{sym}^0$, $K_{sym}$, and $Q_{sym}$ from only the constraint of $\Delta r_{np}$.

\begin{figure}[ht]
\includegraphics[scale=0.35]{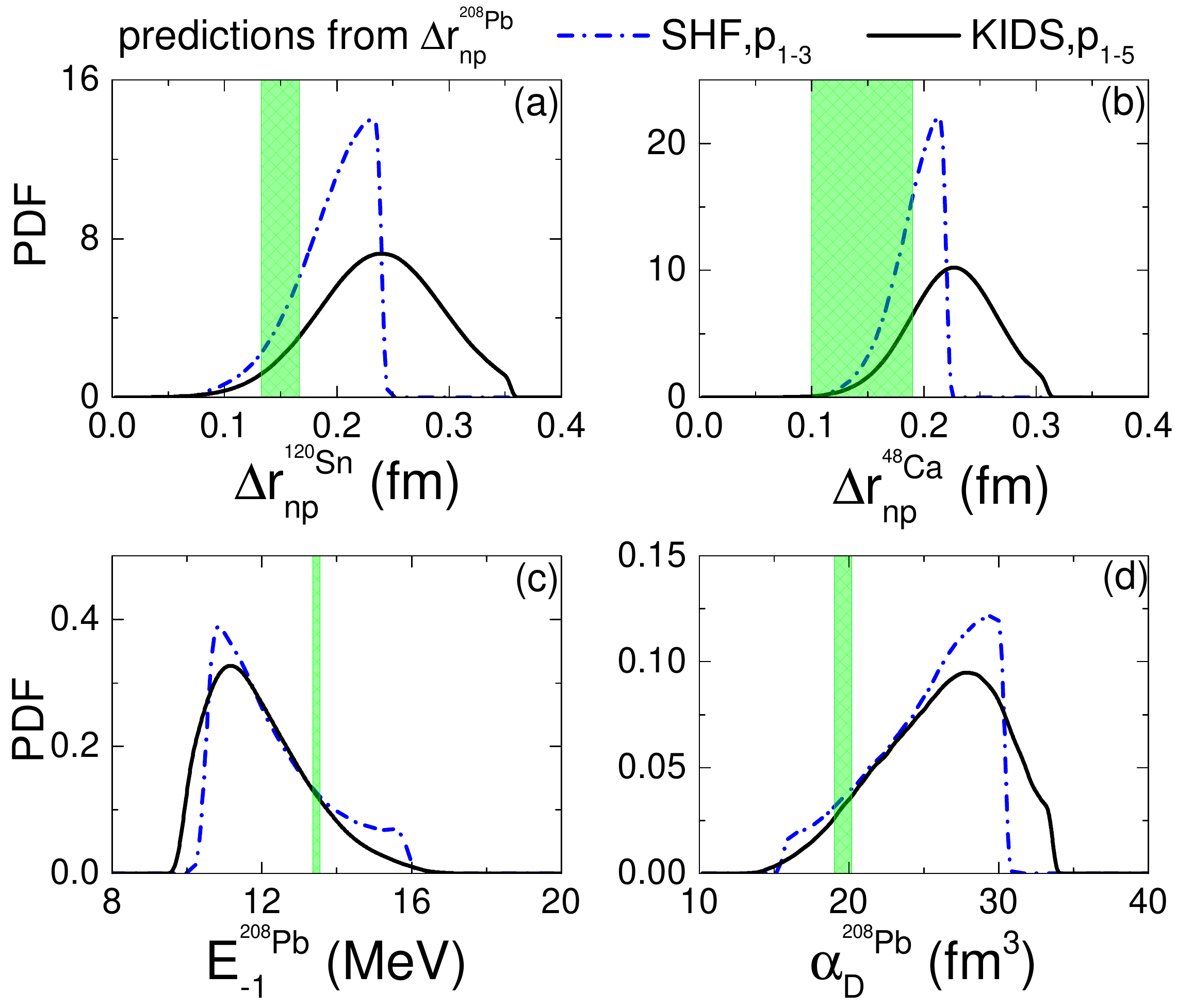}
	\caption{Predictions of $\Delta r_{np}$ in $^{120}$Sn (a), $\Delta r_{np}$ in $^{48}$Ca (b), $E_{-1}$ of $^{208}$Pb (c), and $\alpha_D$ of $^{208}$Pb (d), from the posterior PDFs of physics quantities under the constraint of the $\Delta r_{np}$ in $^{208}$Pb based on the standard SHF and KIDS model, with vertical bands being the available experimental data for comparison. } \label{fig7}
\end{figure}

Using the posterior PDFs of the physics quantities, especially those of $L$, $K_{sym}$, and $E_{sym}^0$, constrained by the $\Delta r_{np}$ in $^{208}$Pb obtained above, we display in Fig.~\ref{fig7} what we can predict on the $\Delta r_{np}$ in $^{120}$Sn and $^{48}$Ca as well as the IVGDR results for $^{208}$Pb, where the available experimental data are shown with vertical bands for comparison. Although the mean value of $\Delta r_{np}$ in $^{208}$Pb favors a large $L$~\cite{Ree21}, its large error bar leads to a diffusive PDF of $L$ as shown in Fig.~\ref{fig6}, giving wide predictions of all observables mentioned above. Generally, there are overlaps compared to $\Delta r_{np}=0.150 \pm 0.017$ fm in $^{120}$Sn from Ref.~\cite{Xu20b}, $\Delta r_{np}=0.10 \sim 0.19$ fm in $^{48}$Ca estimated from Ref.~\cite{Hor21}, $E_{-1}=13.46 \pm 0.10$ MeV from Ref.~\cite{IVGDRe}, and $\alpha_D=19.6 \pm 0.6$ fm$^3$ from Refs.~\cite{Tam11,Roc15}. With more independent variables, especially $K_{sym}$, the KIDS model gives slightly wider predictions than the standard SHF model. We will demonstrate the predictions from the posterior PDFs under the constraints of more experimental data later.

\subsection{Bayesian inference on $\Delta r_{np}$, $E_{-1}$, and $\alpha_D$}

\begin{figure}[ht]
\includegraphics[scale=0.1]{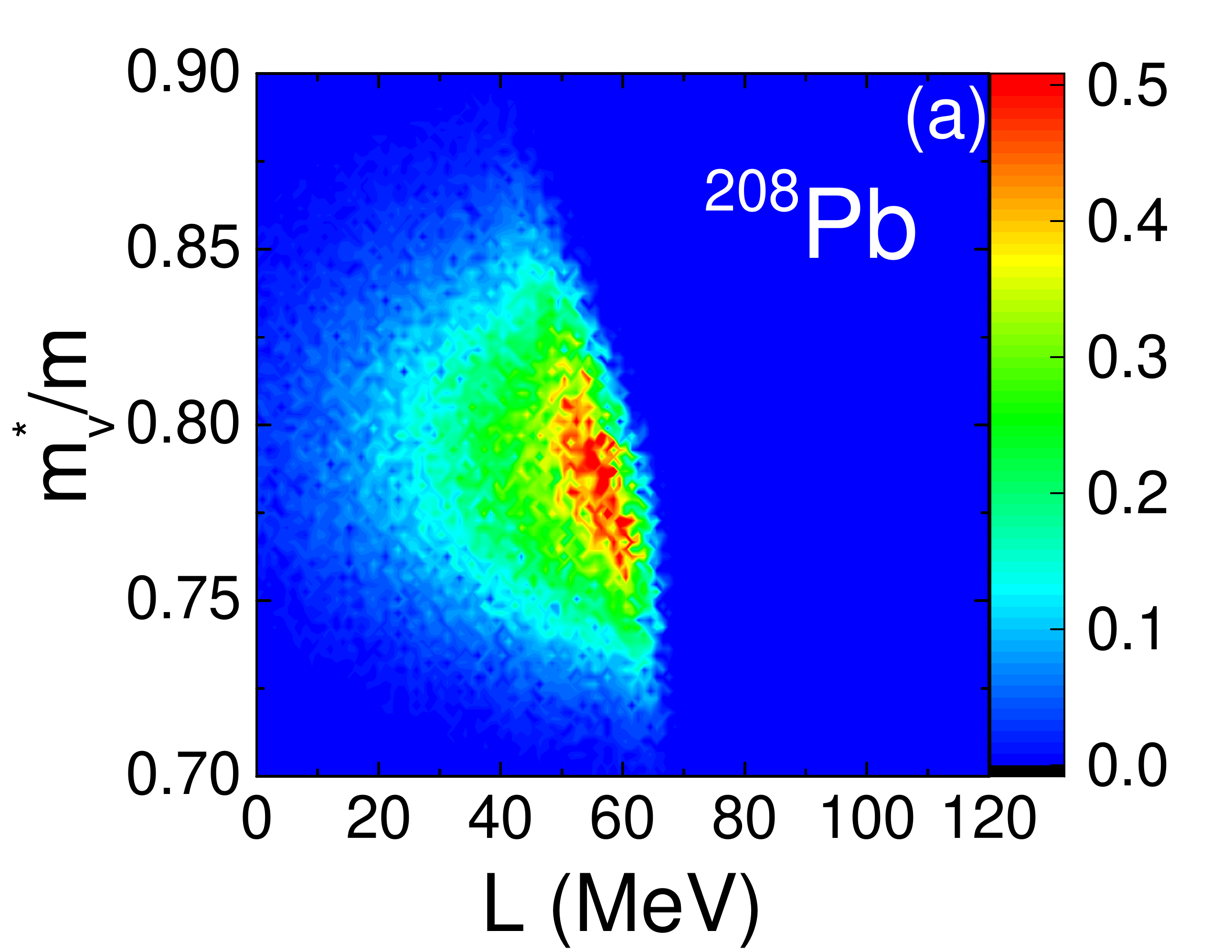}\includegraphics[scale=0.1]{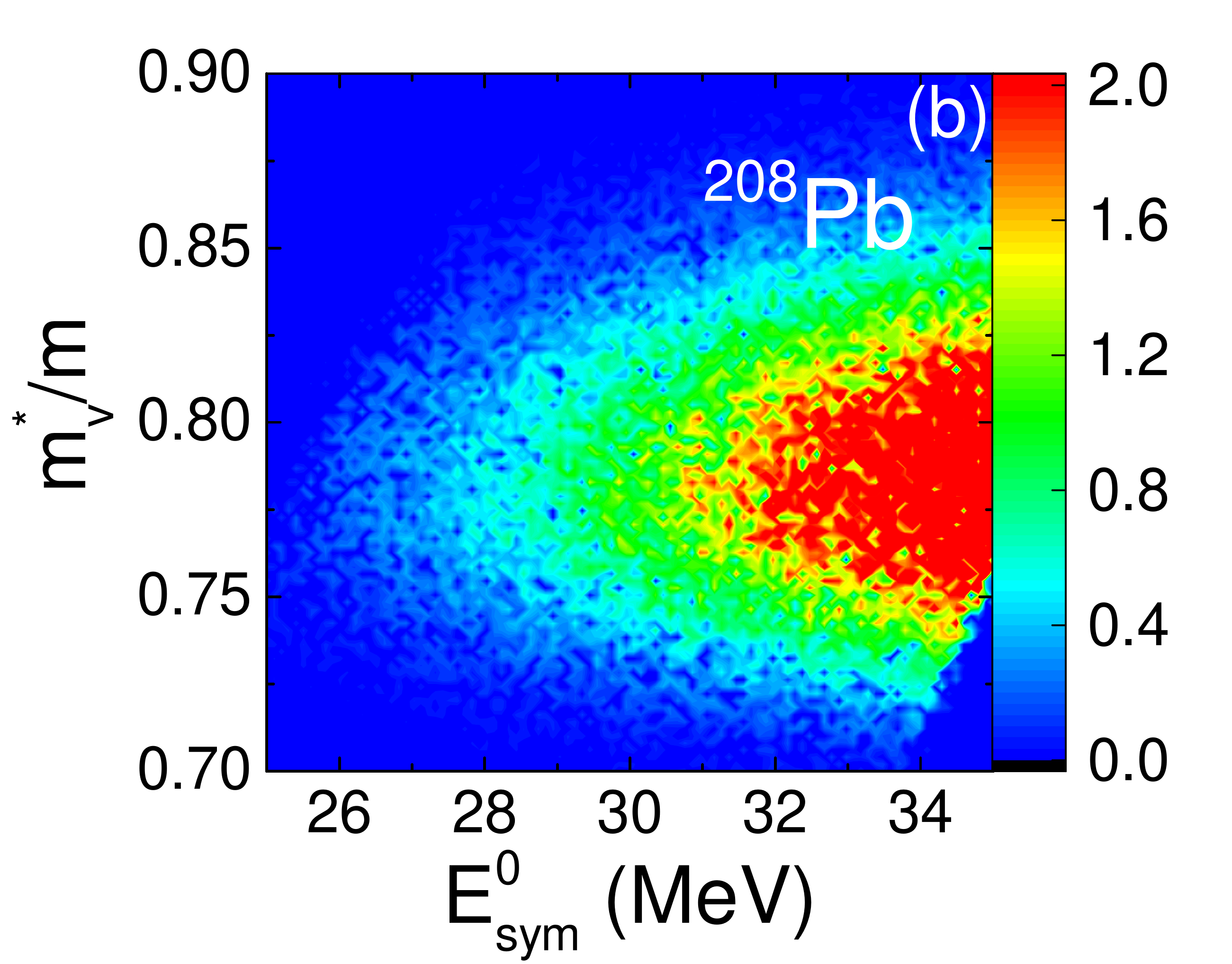}\includegraphics[scale=0.1]{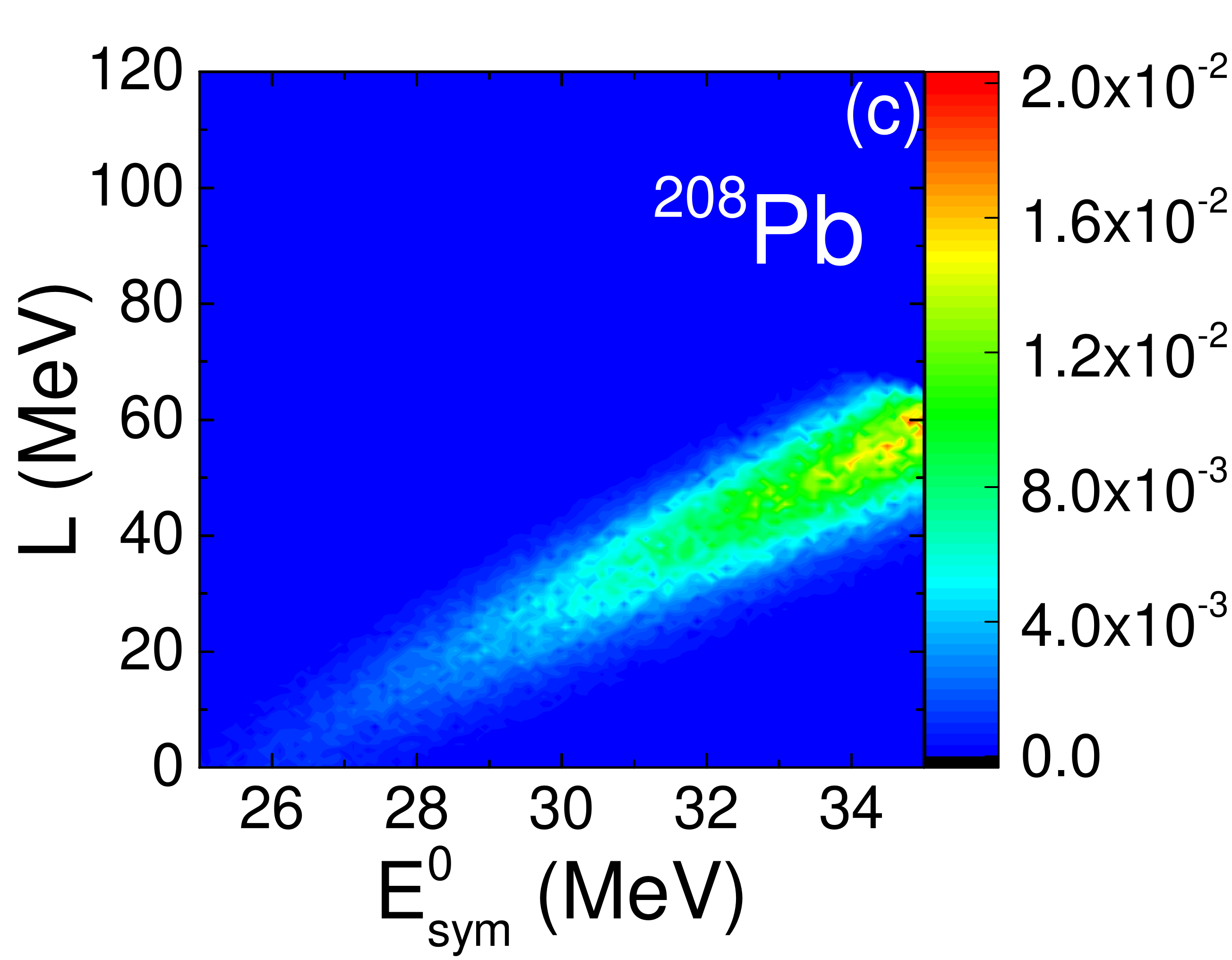}\\
\includegraphics[scale=0.1]{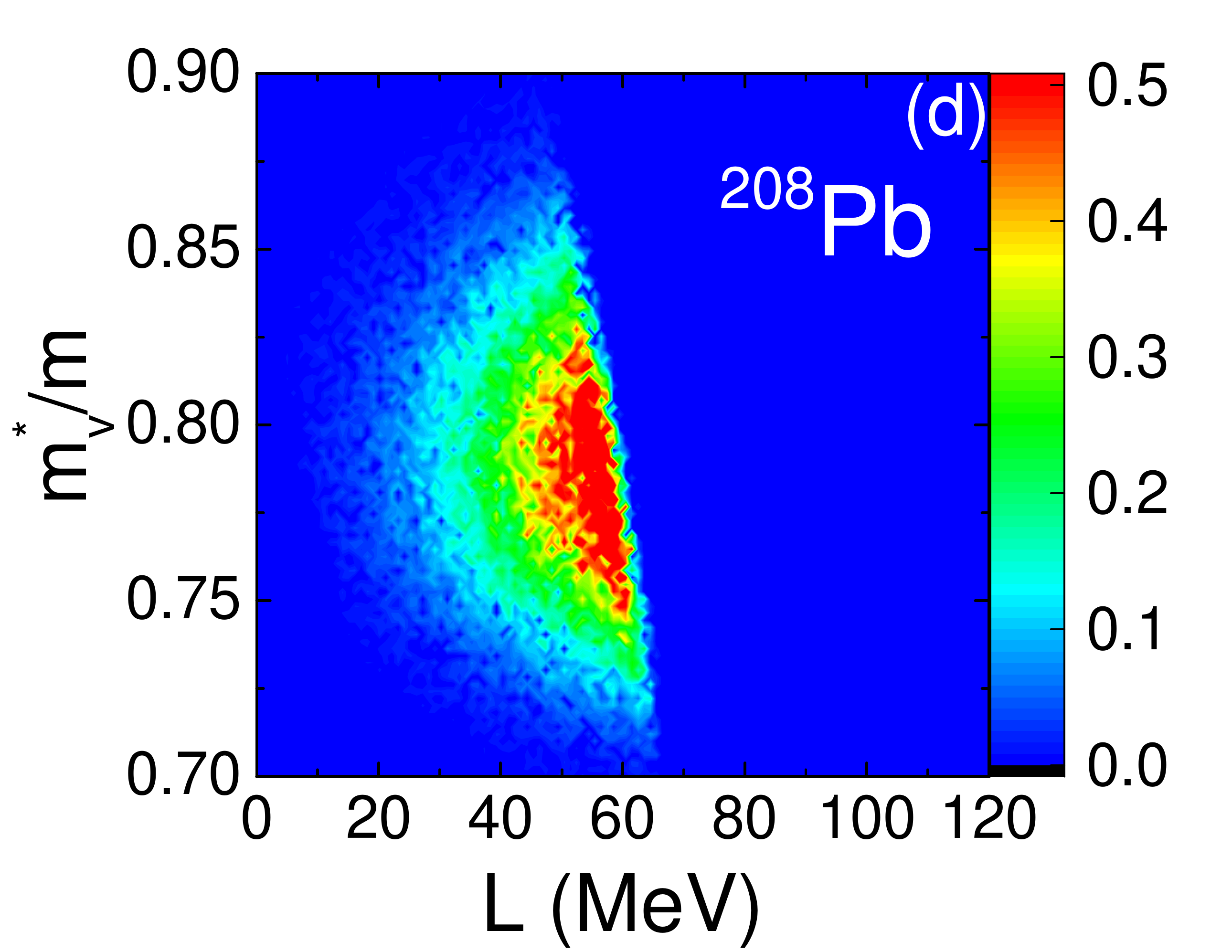}\includegraphics[scale=0.1]{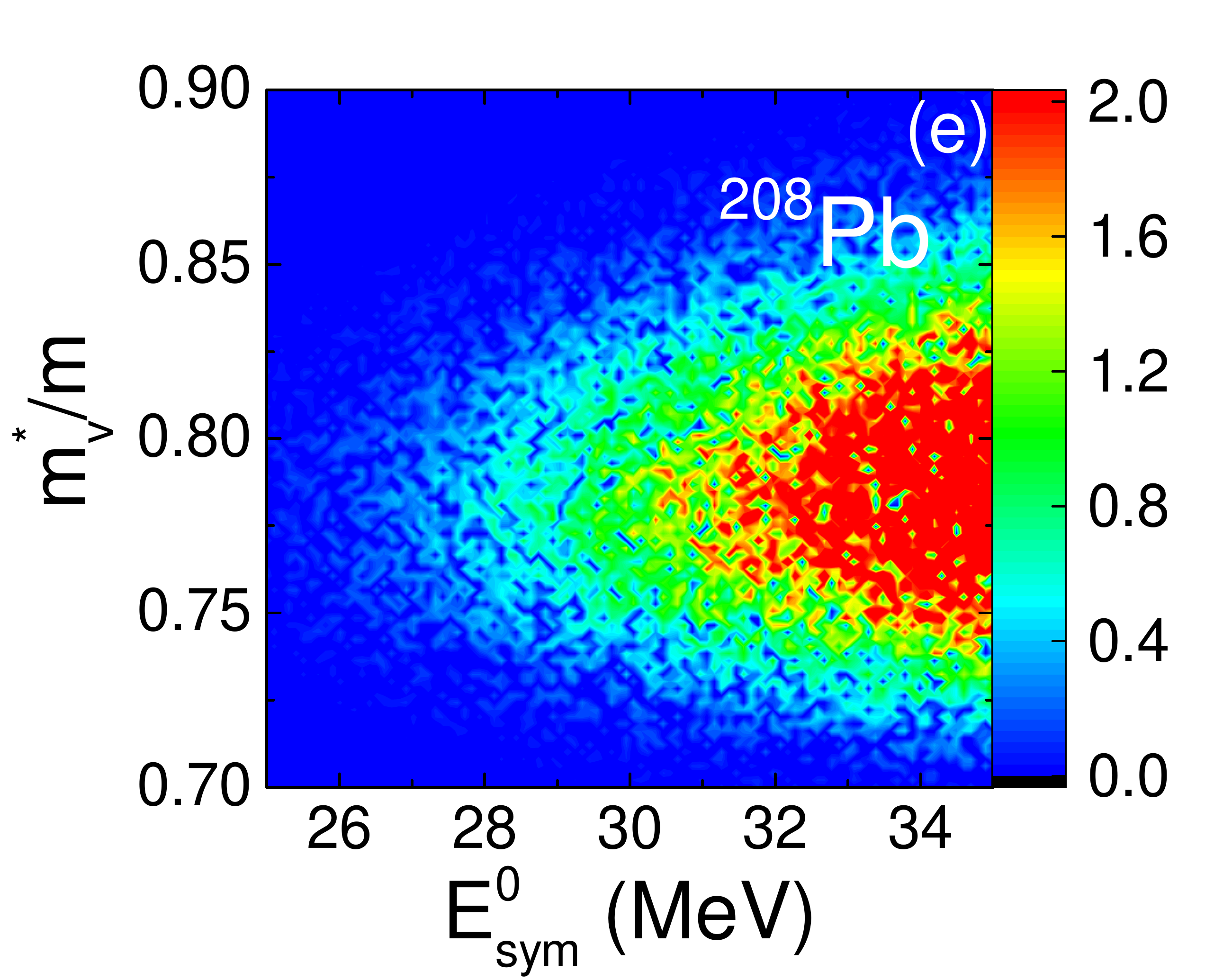}\includegraphics[scale=0.1]{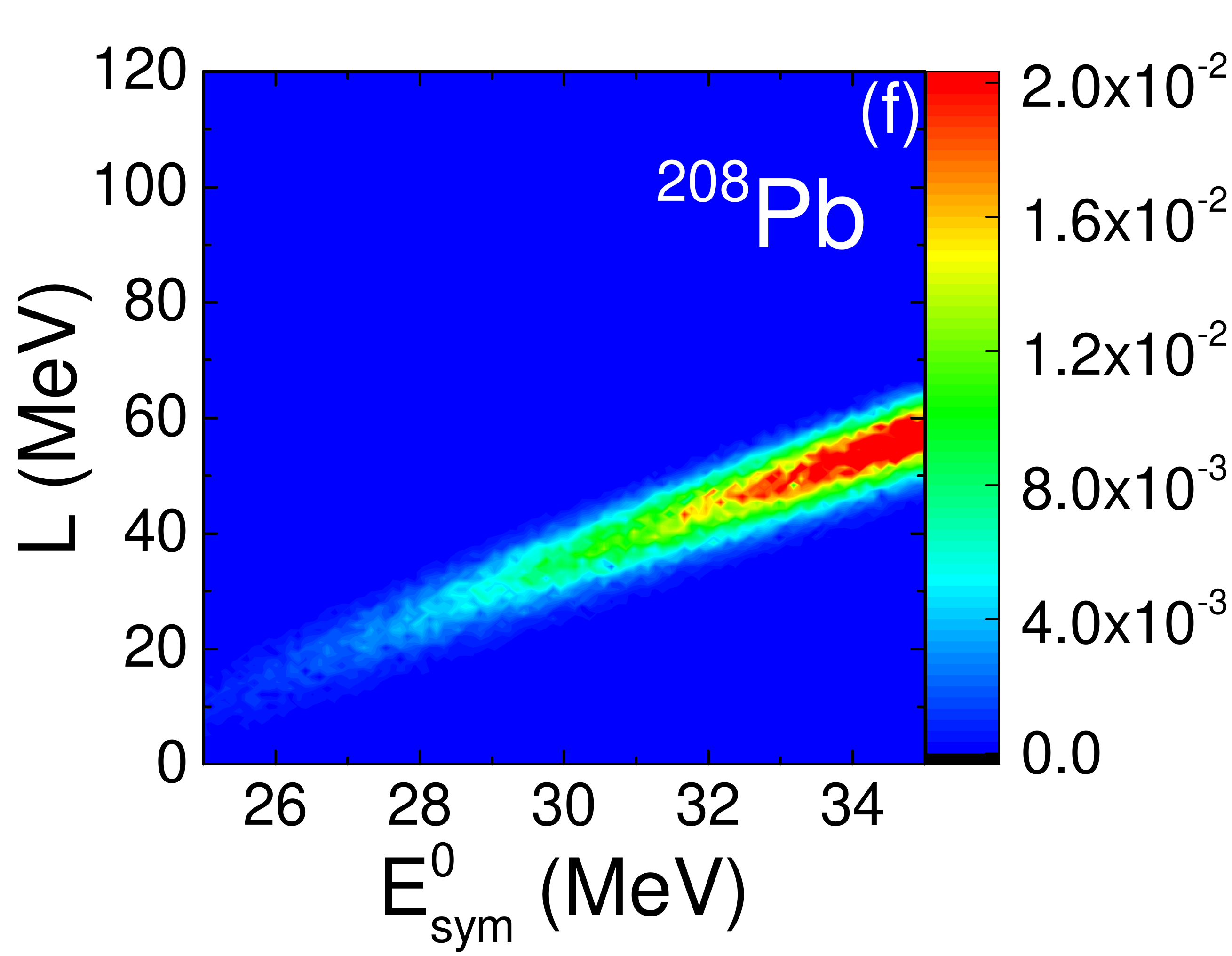}\\
\includegraphics[scale=0.1]{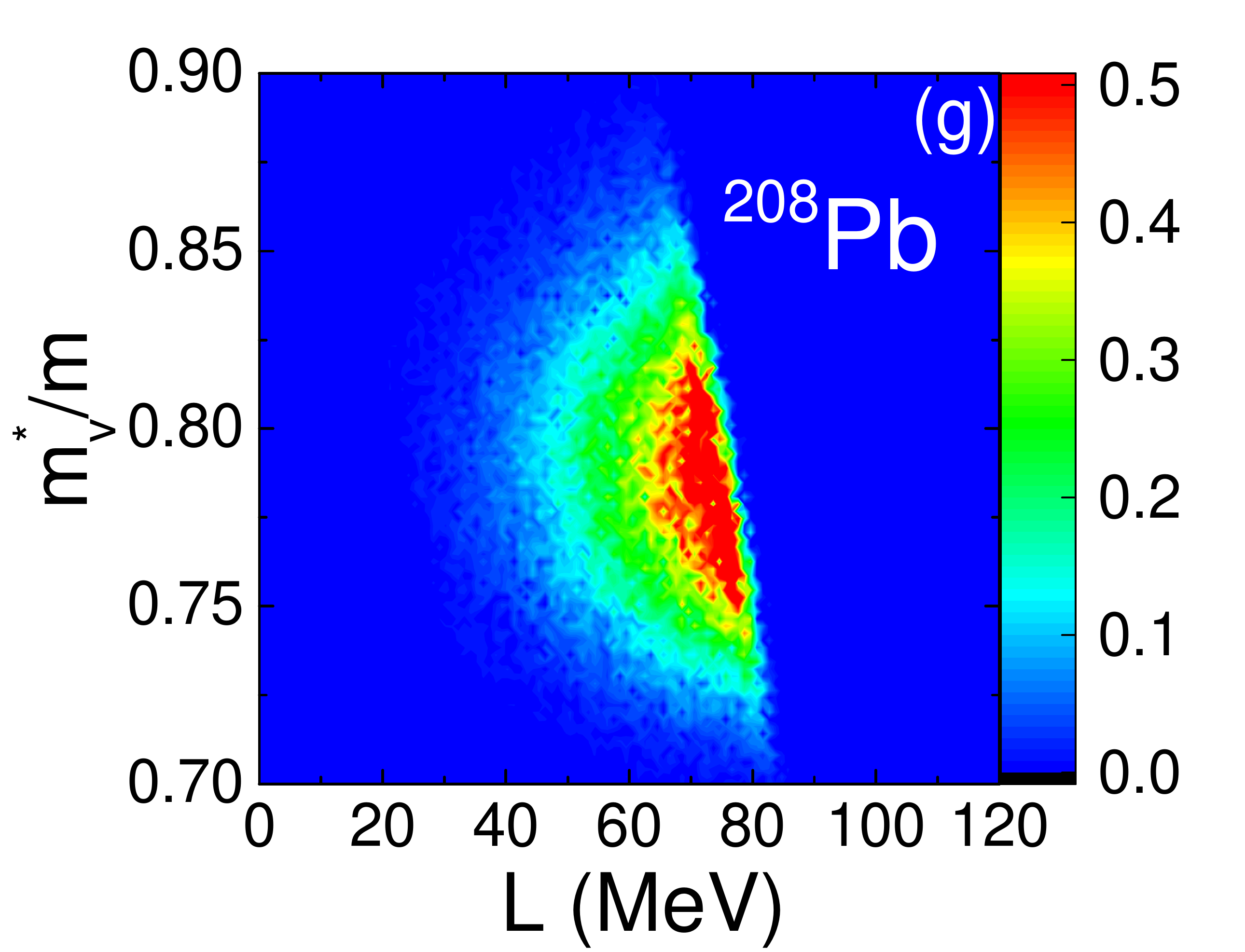}\includegraphics[scale=0.1]{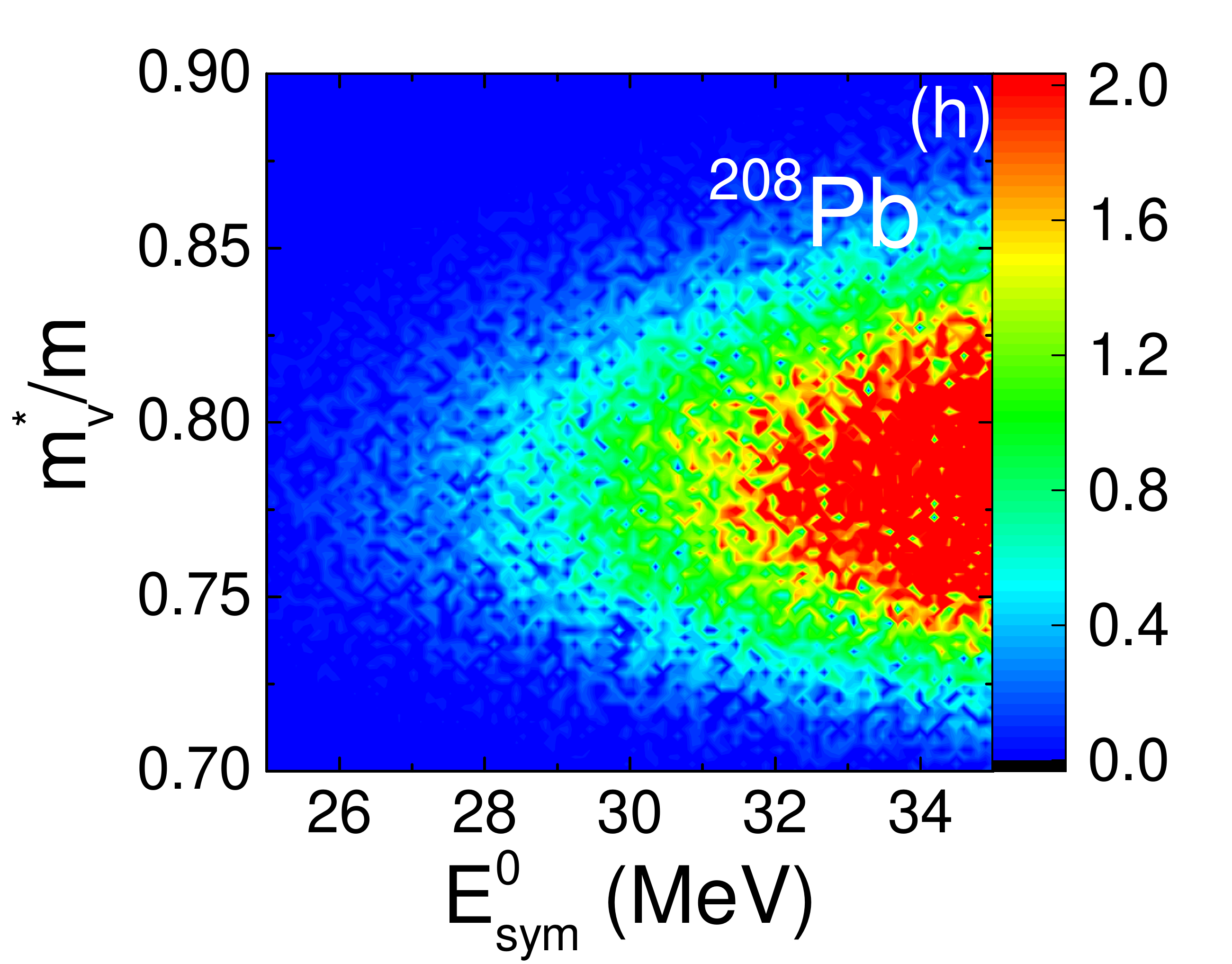}\includegraphics[scale=0.1]{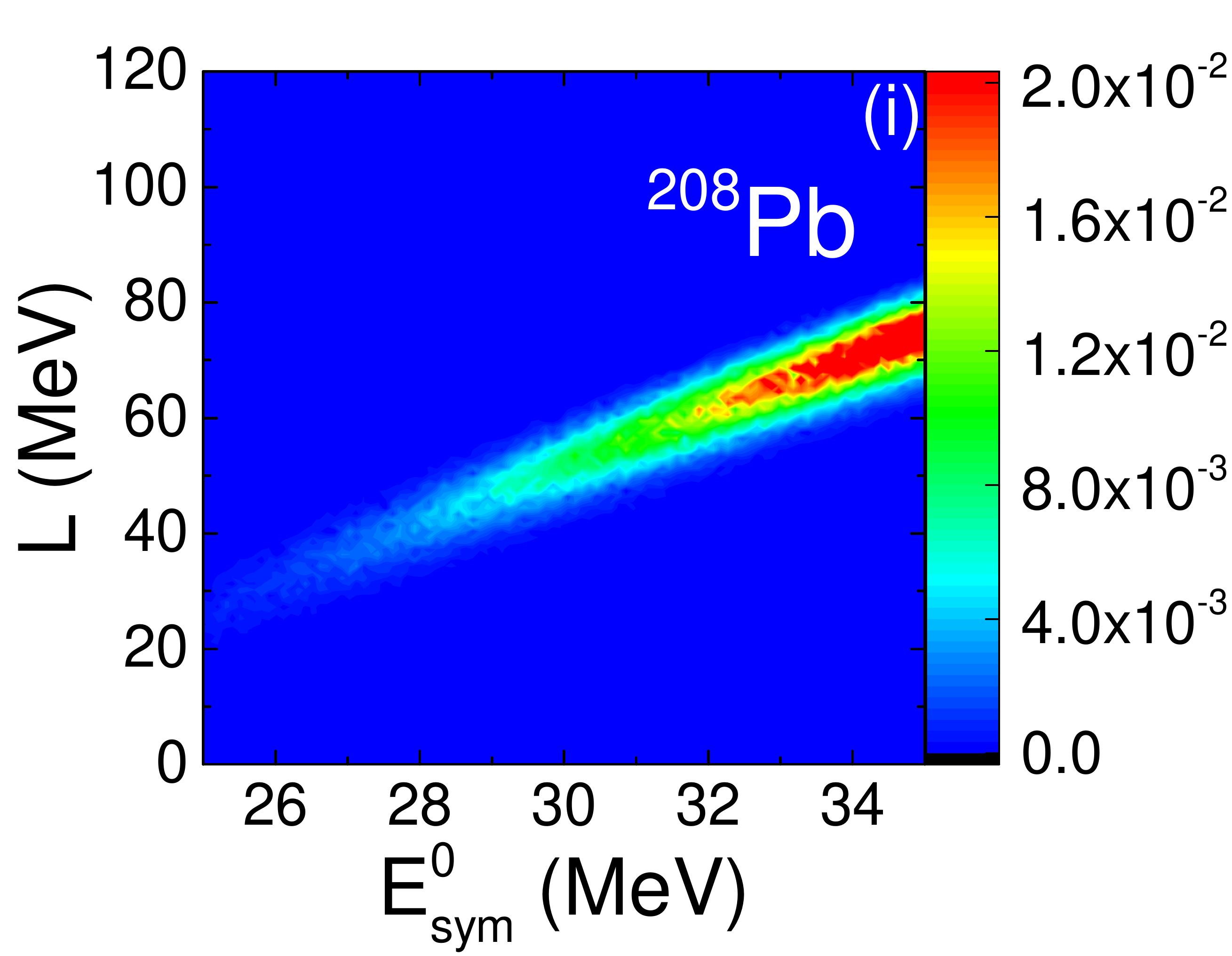}\\
\includegraphics[scale=0.1]{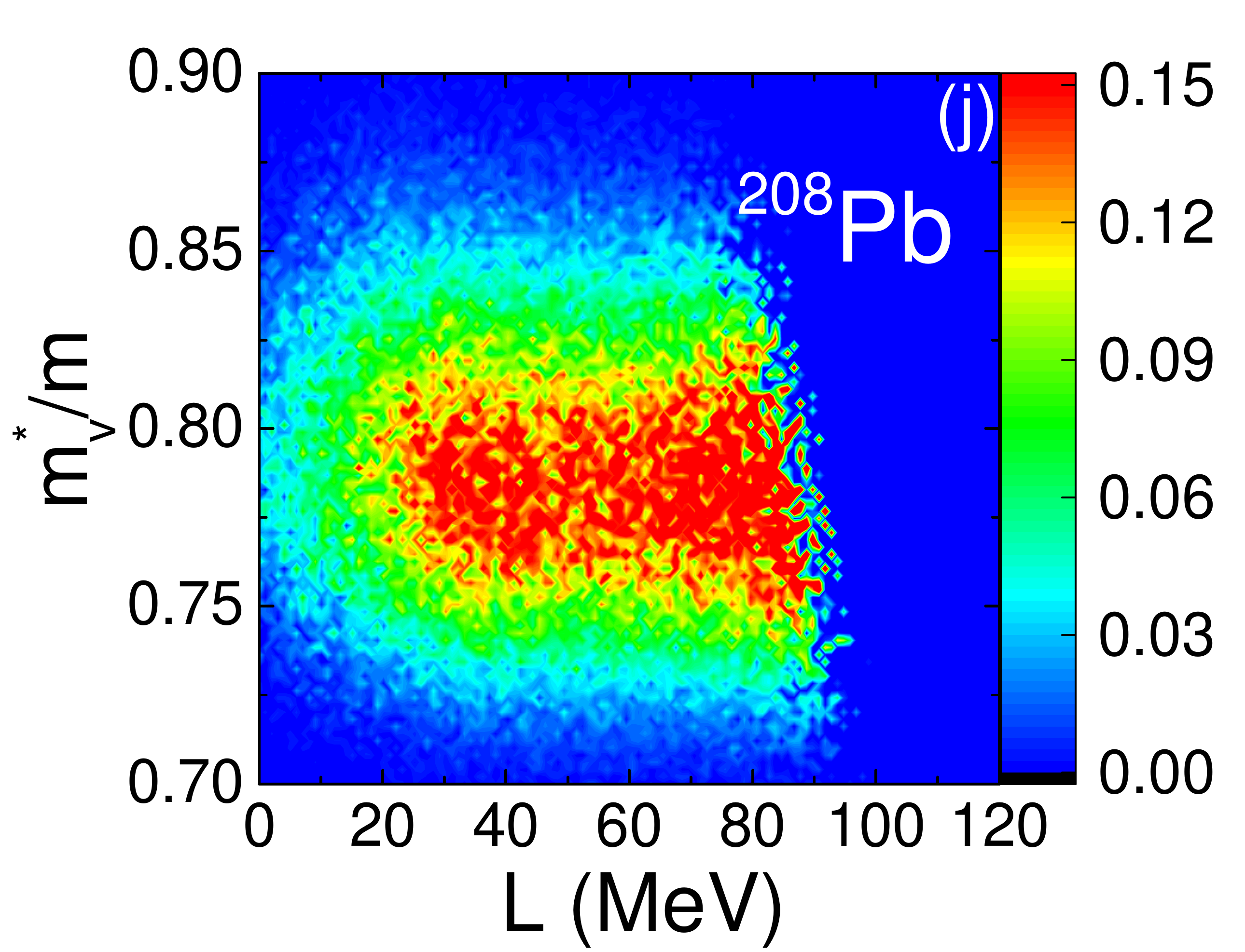}\includegraphics[scale=0.1]{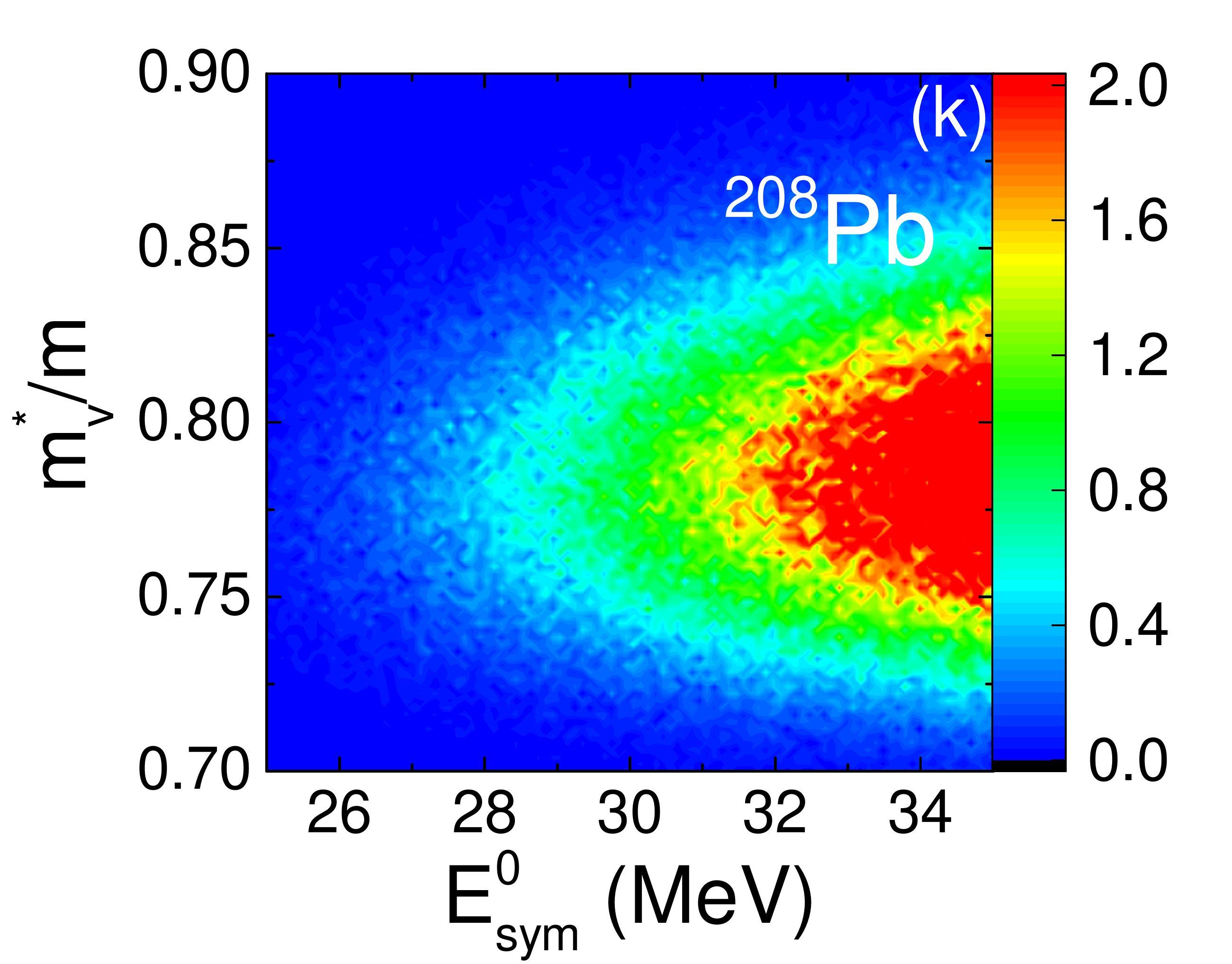}\includegraphics[scale=0.1]{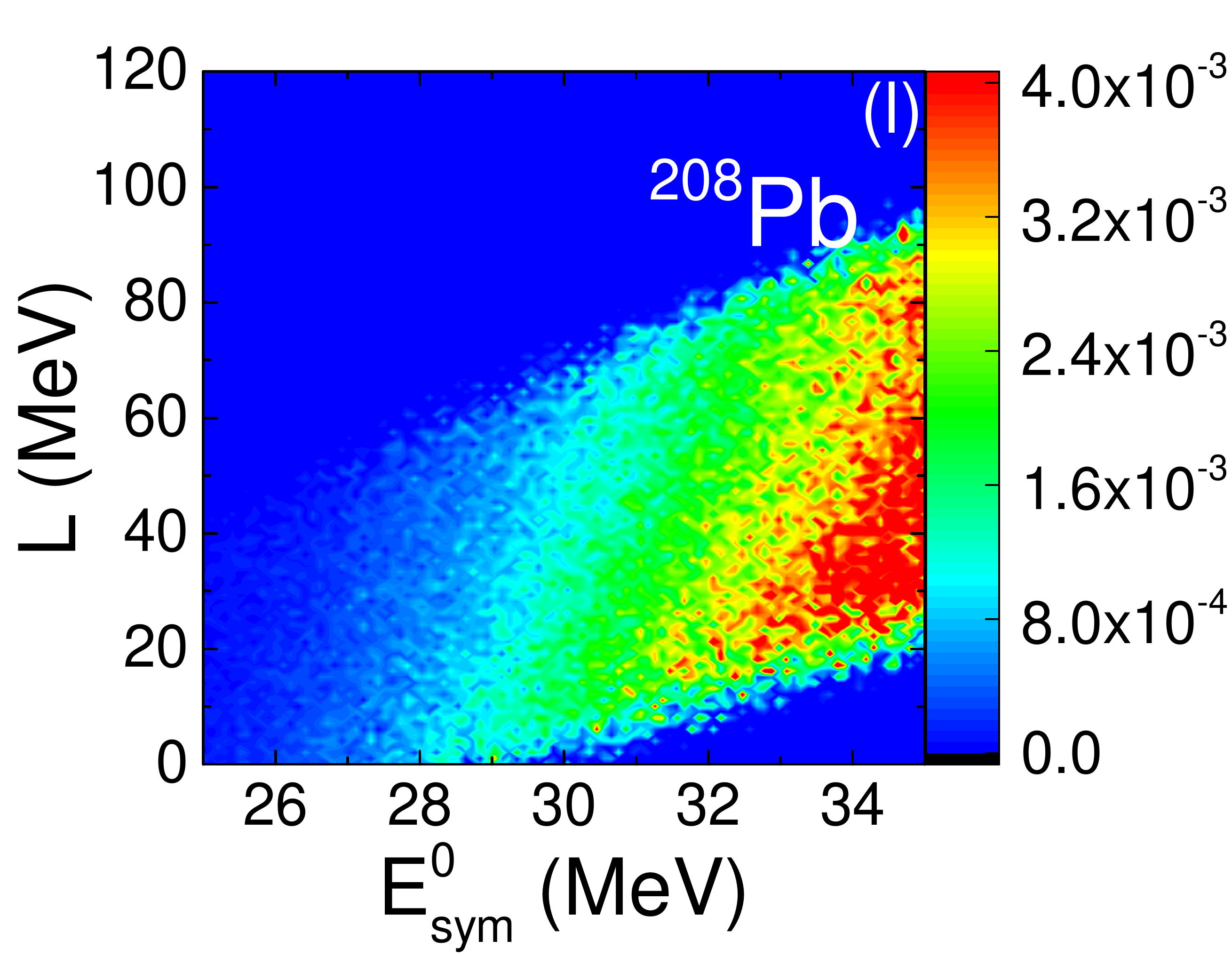}
	\caption{First row: Posterior correlated PDFs between $m_v^\star$, $L$, and $E_{sym}^0$ under the constraints of $\Delta r_{np}$, $E_{-1}$, and $\alpha_D$ in $^{208}$Pb based on the standard SHF model; Second row: Same as the first row but based on the KIDS model using $K_{sym}=-156$ MeV; Third row: Same as the second row but using $K_{sym}=0$ MeV; Bottom row: Same as the second and the third rows but letting $K_{sym}$ change as an independent variable.} \label{fig8}
\end{figure}

\begin{figure}[ht]
\includegraphics[scale=0.1]{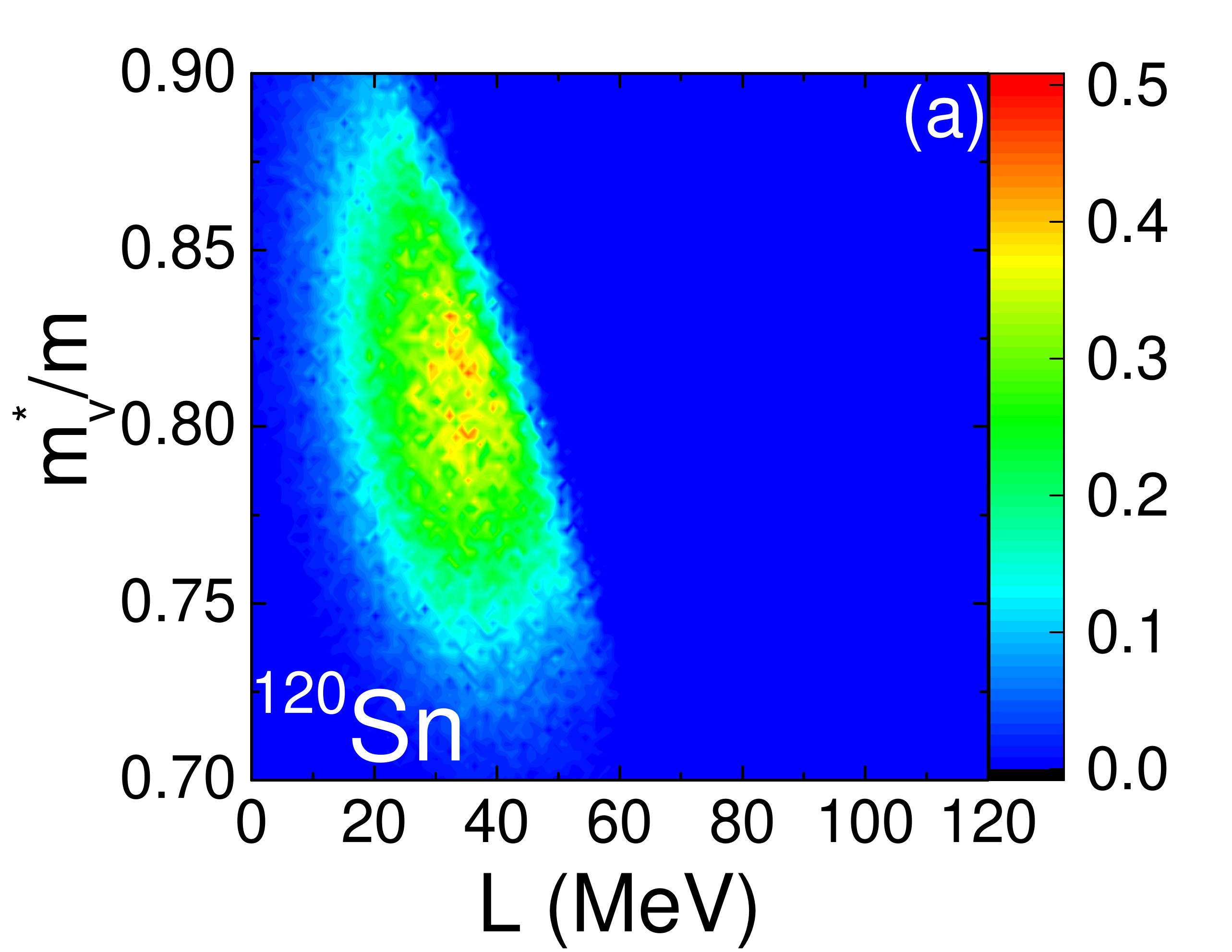}\includegraphics[scale=0.1]{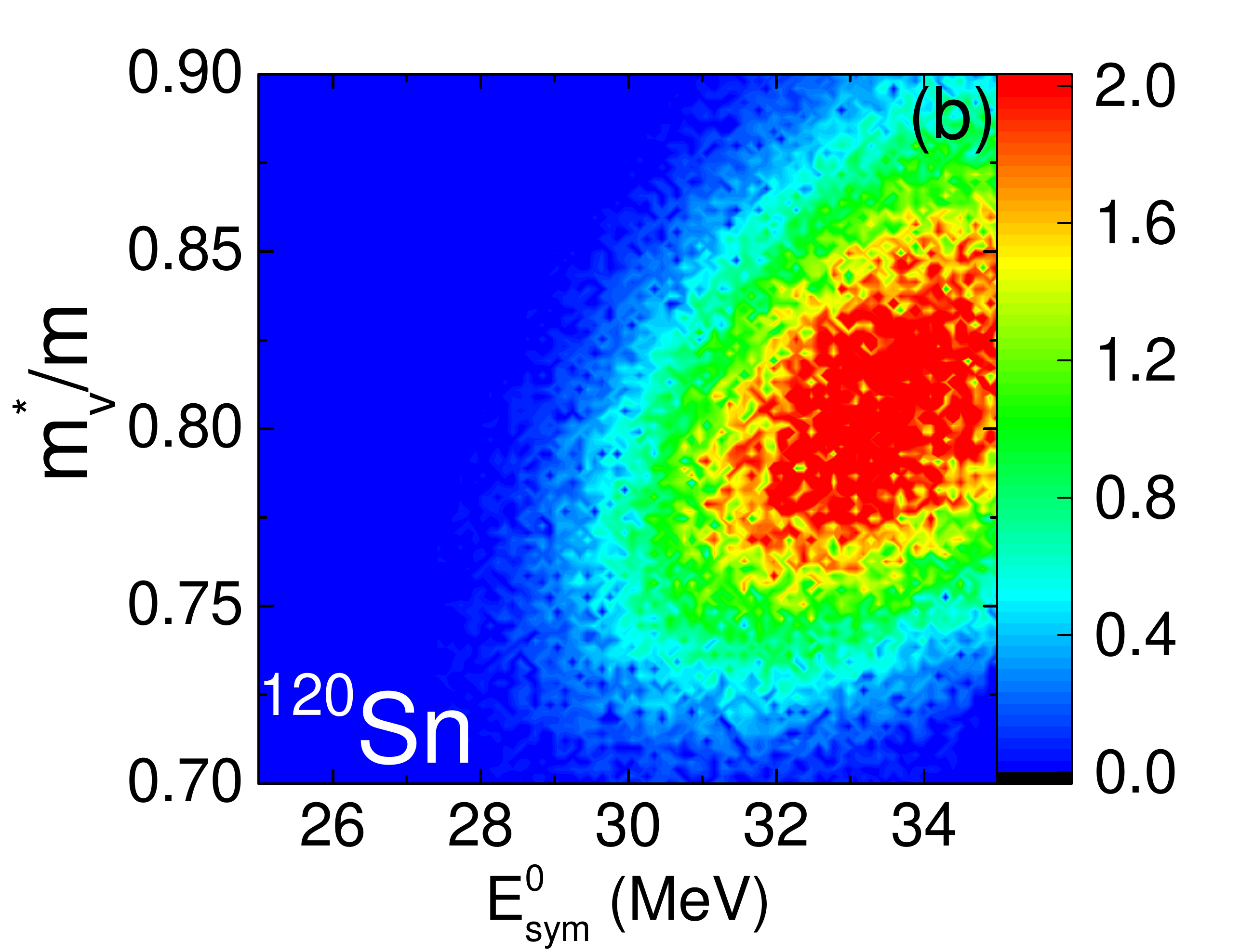}\includegraphics[scale=0.1]{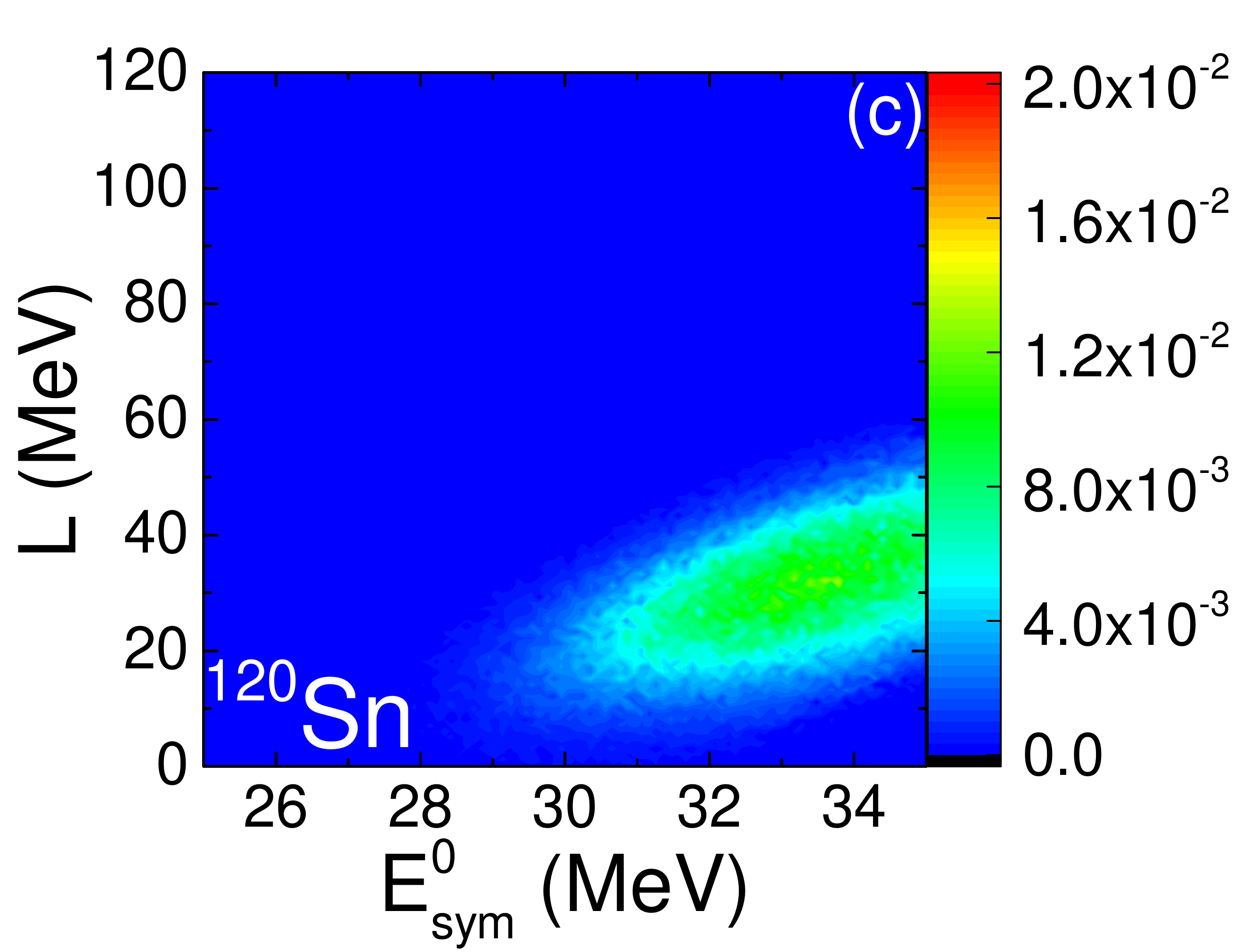}\\
\includegraphics[scale=0.1]{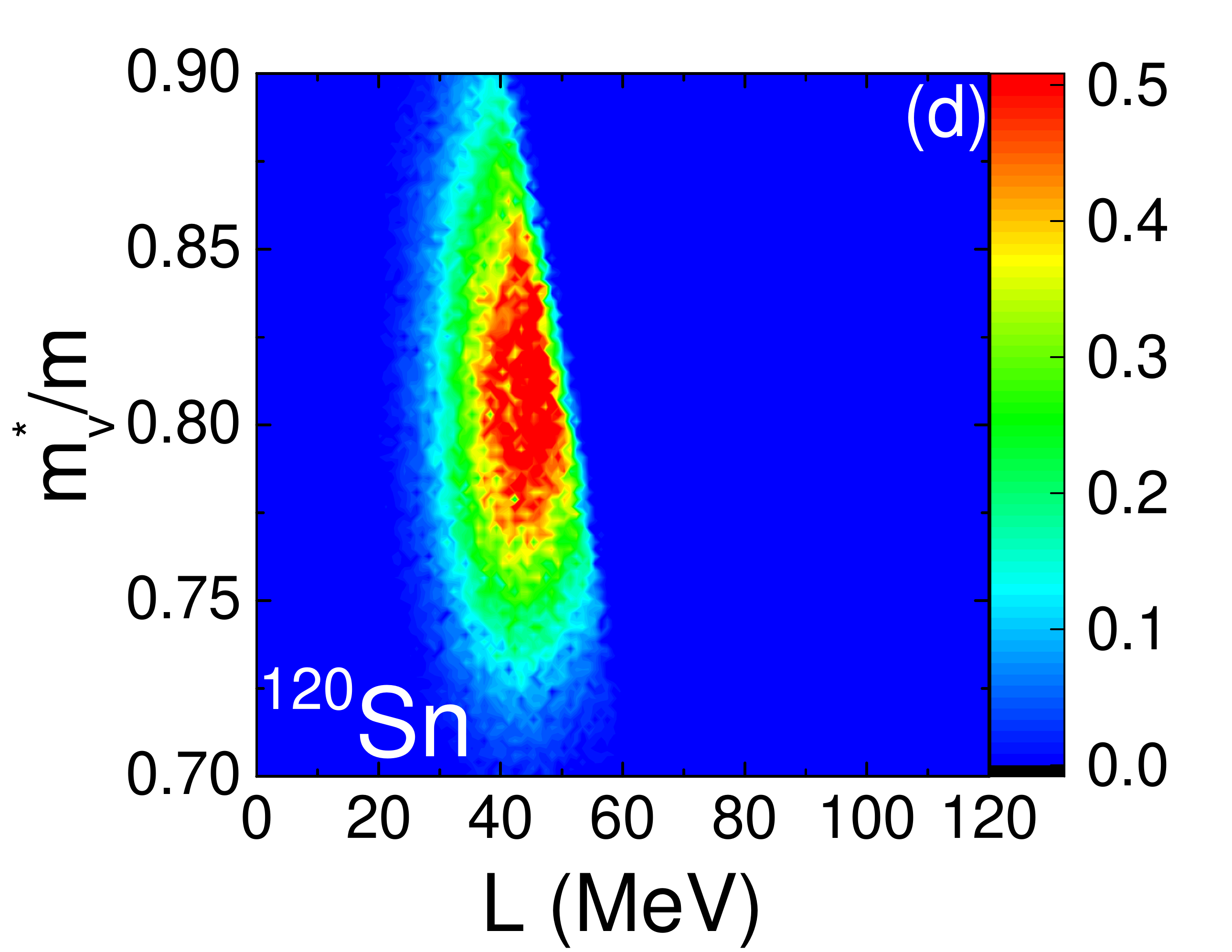}\includegraphics[scale=0.1]{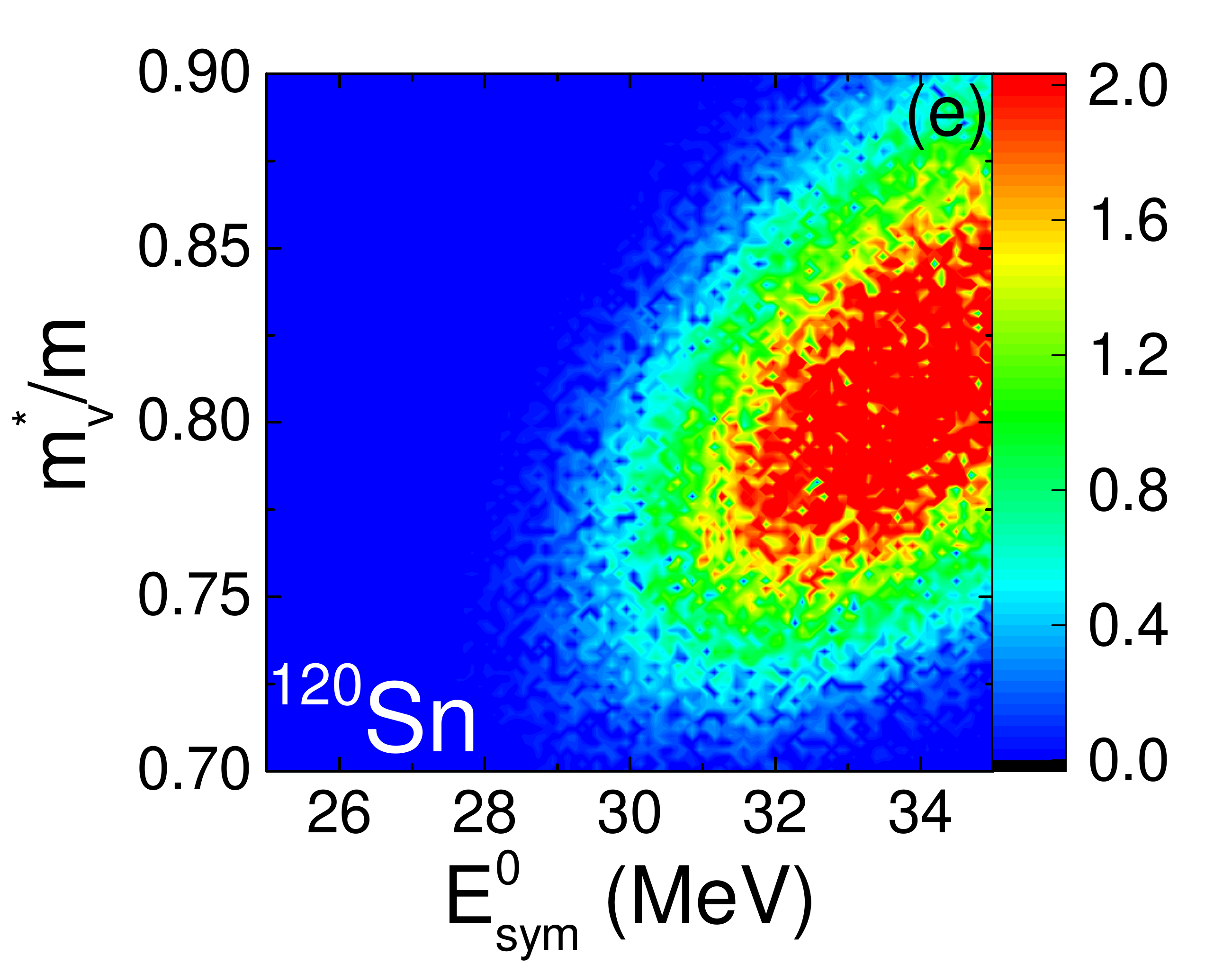}\includegraphics[scale=0.1]{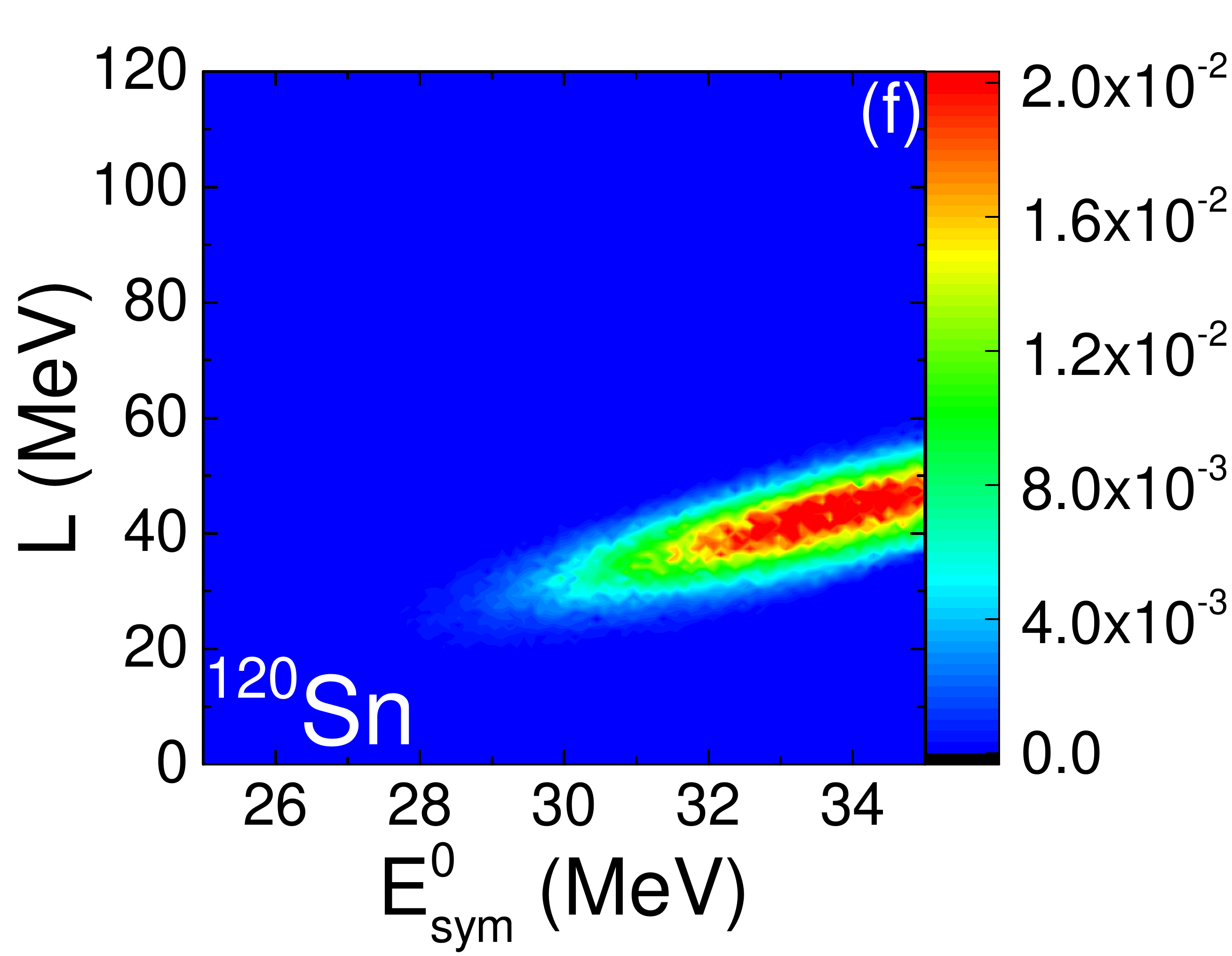}\\
\includegraphics[scale=0.1]{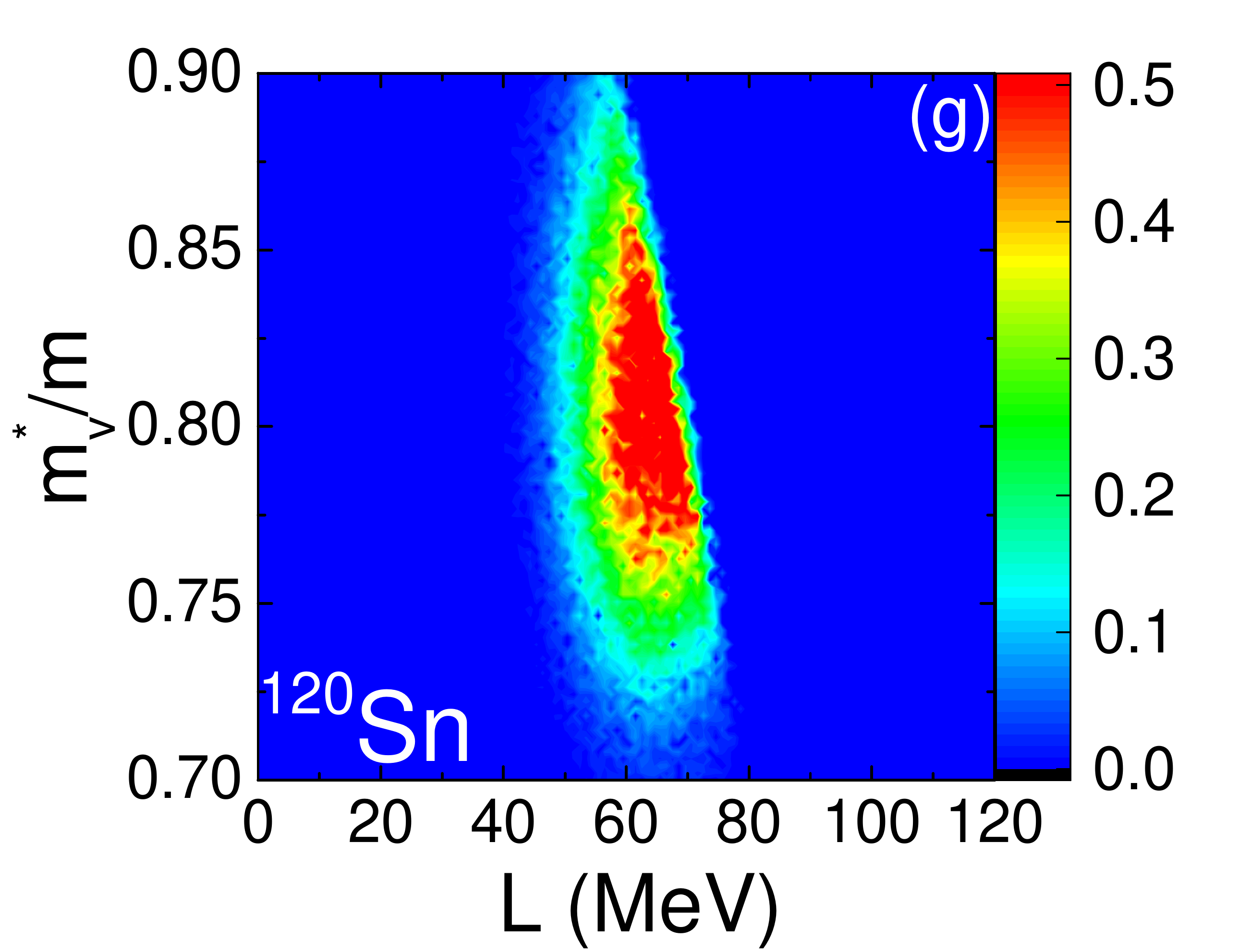}\includegraphics[scale=0.1]{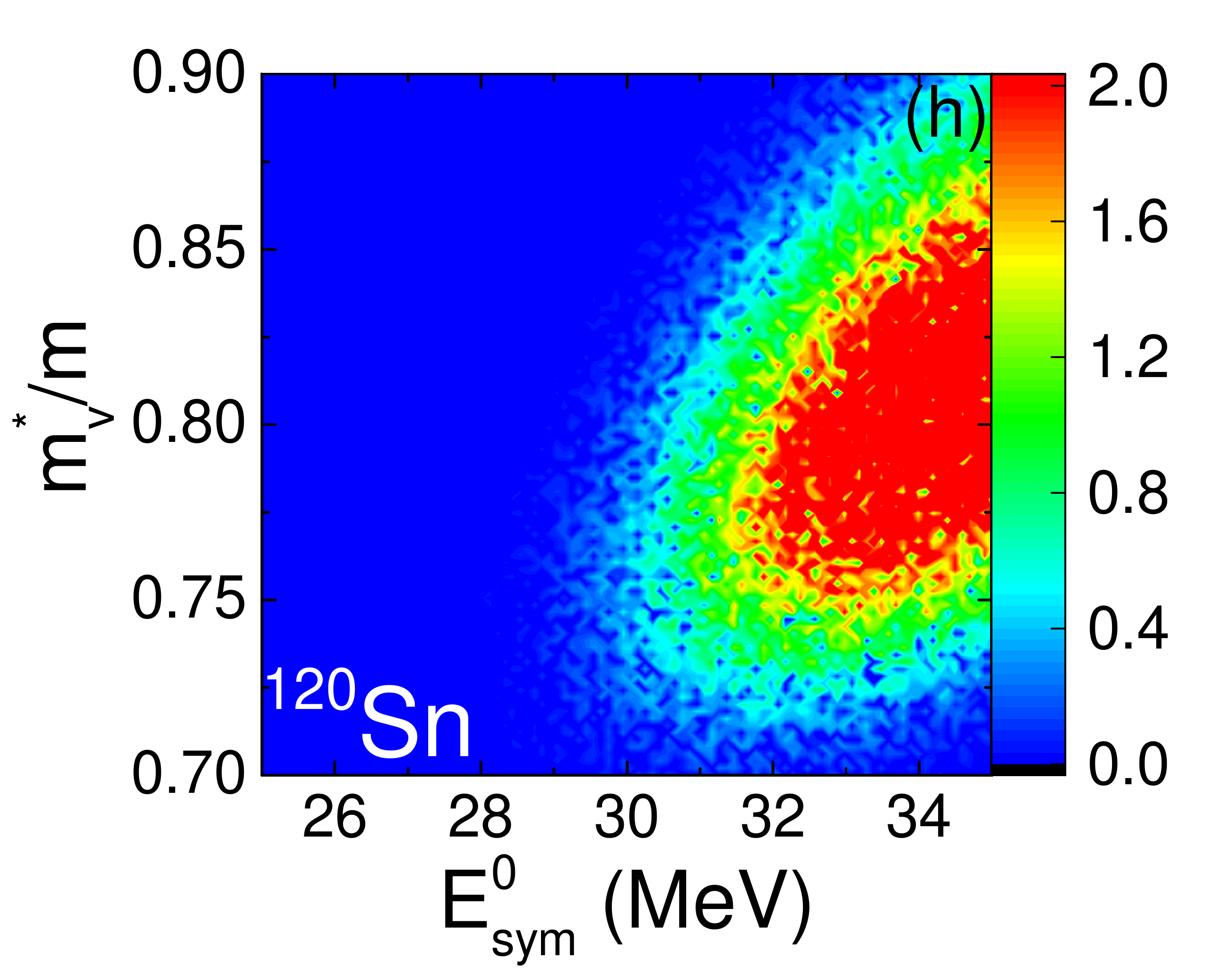}\includegraphics[scale=0.1]{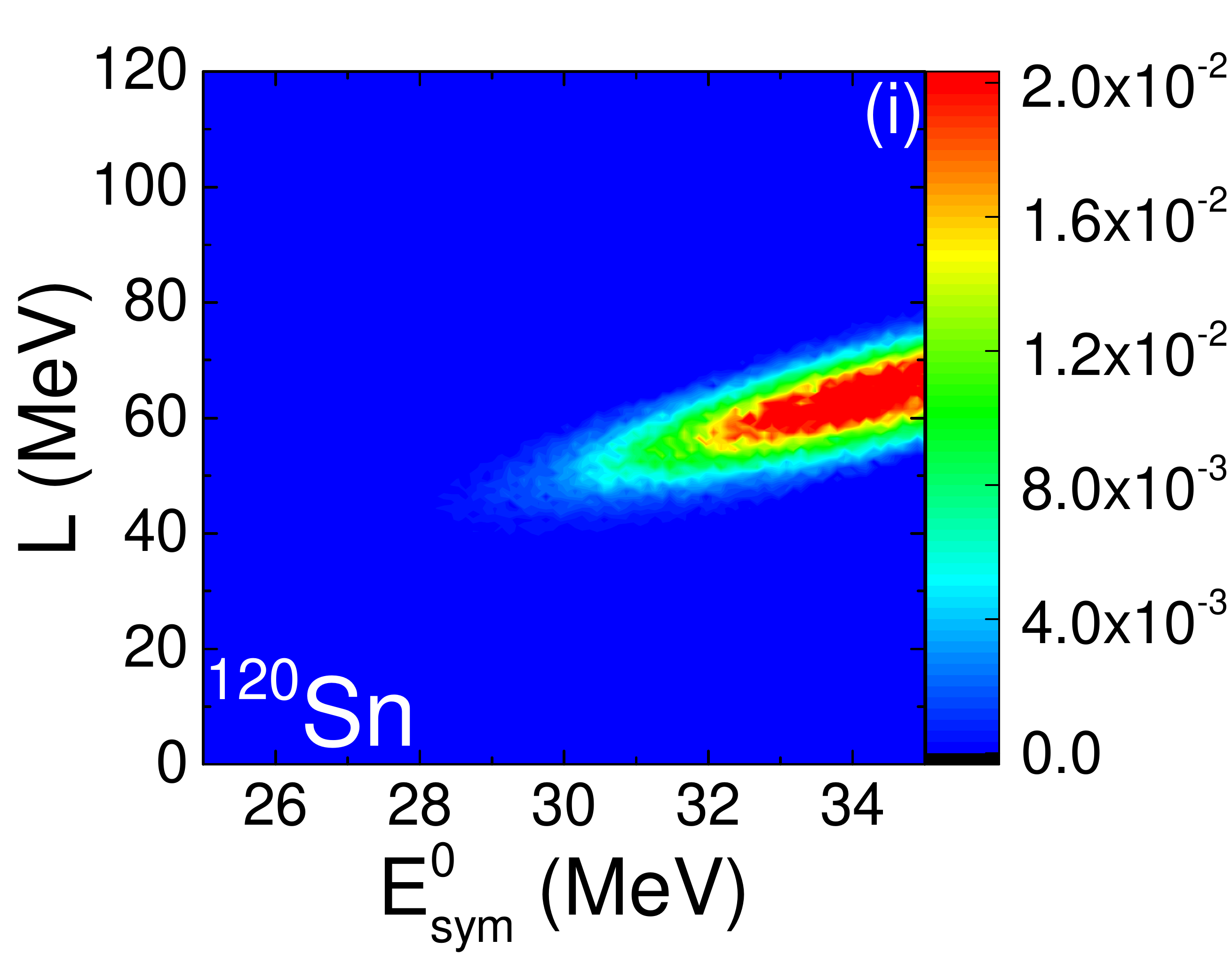}\\
\includegraphics[scale=0.1]{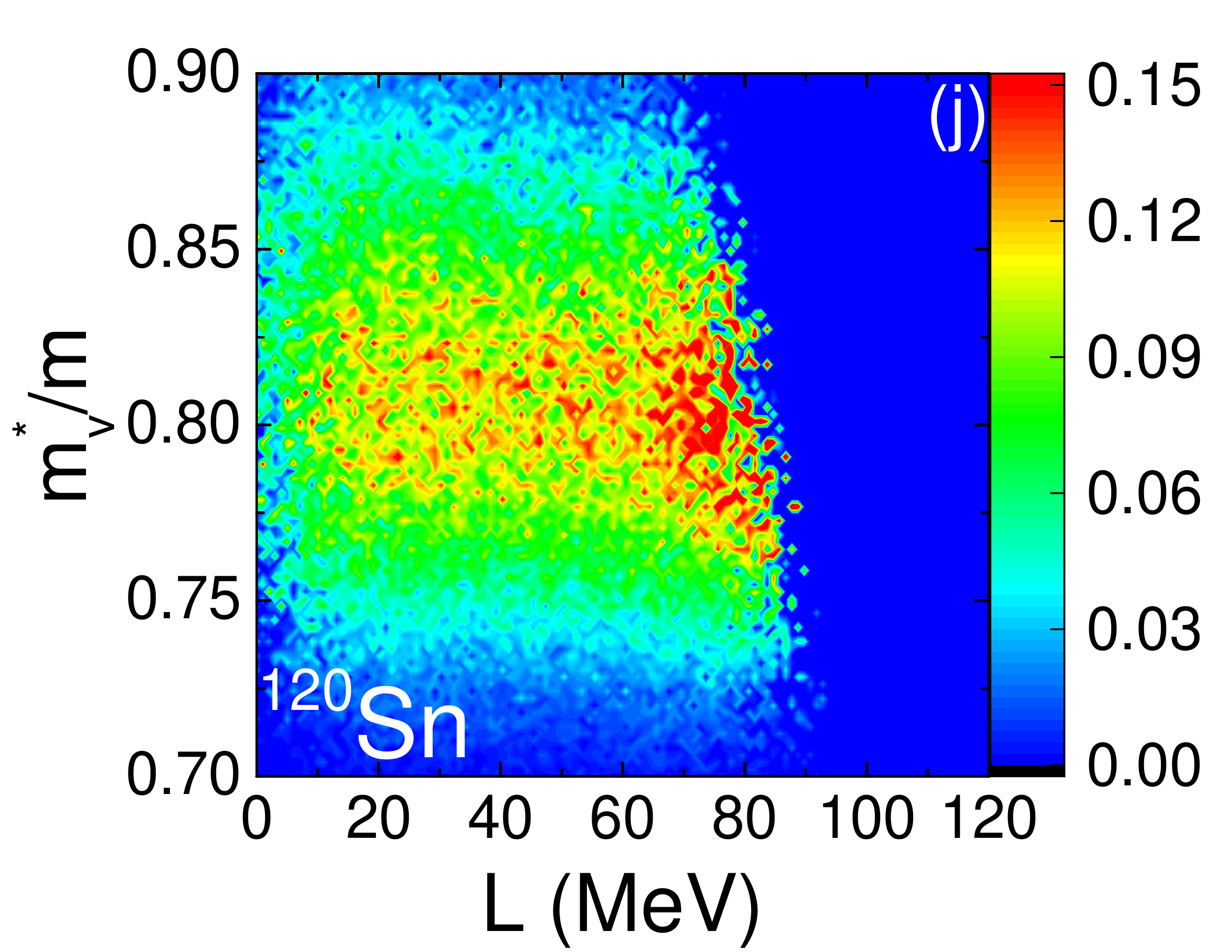}\includegraphics[scale=0.1]{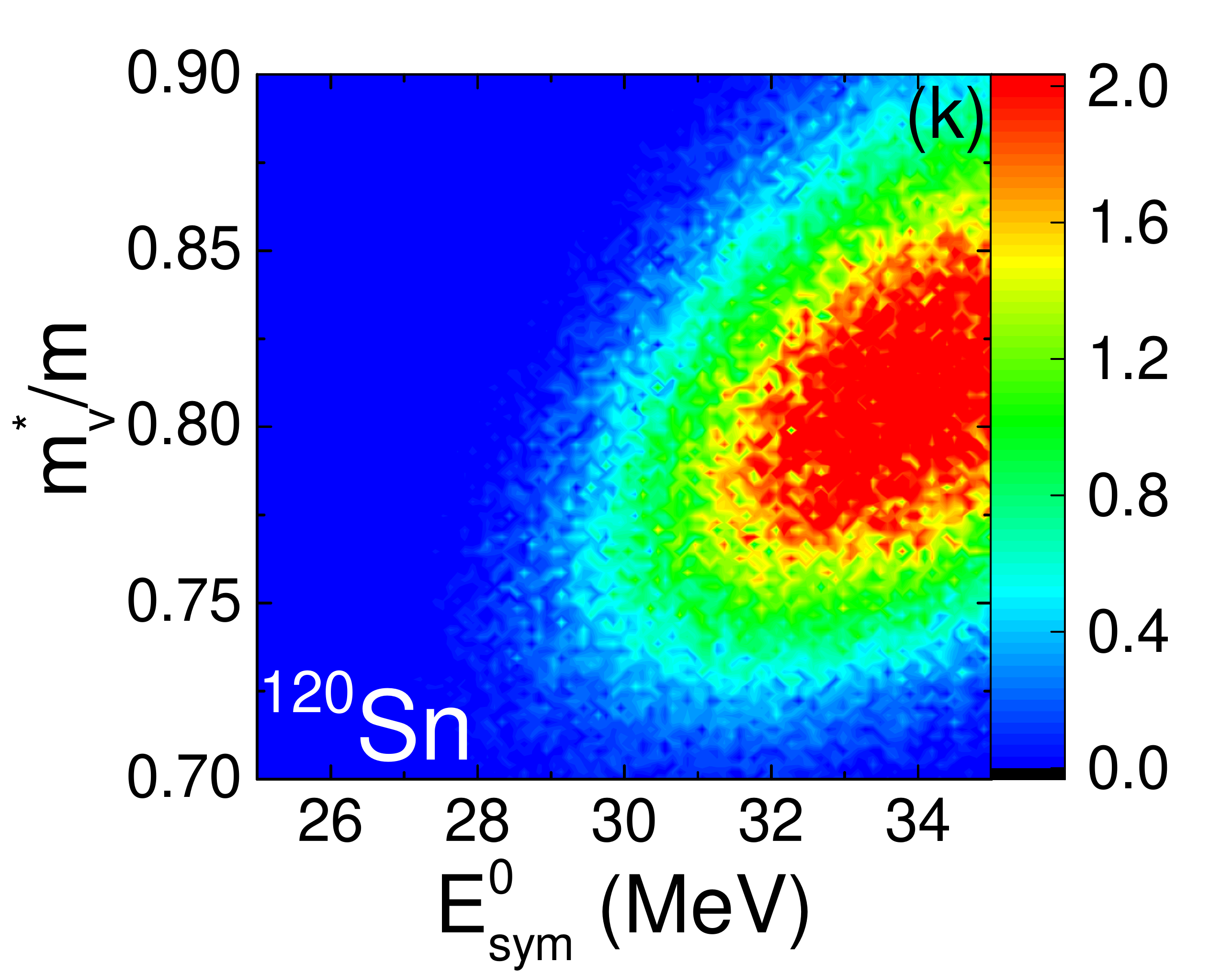}\includegraphics[scale=0.1]{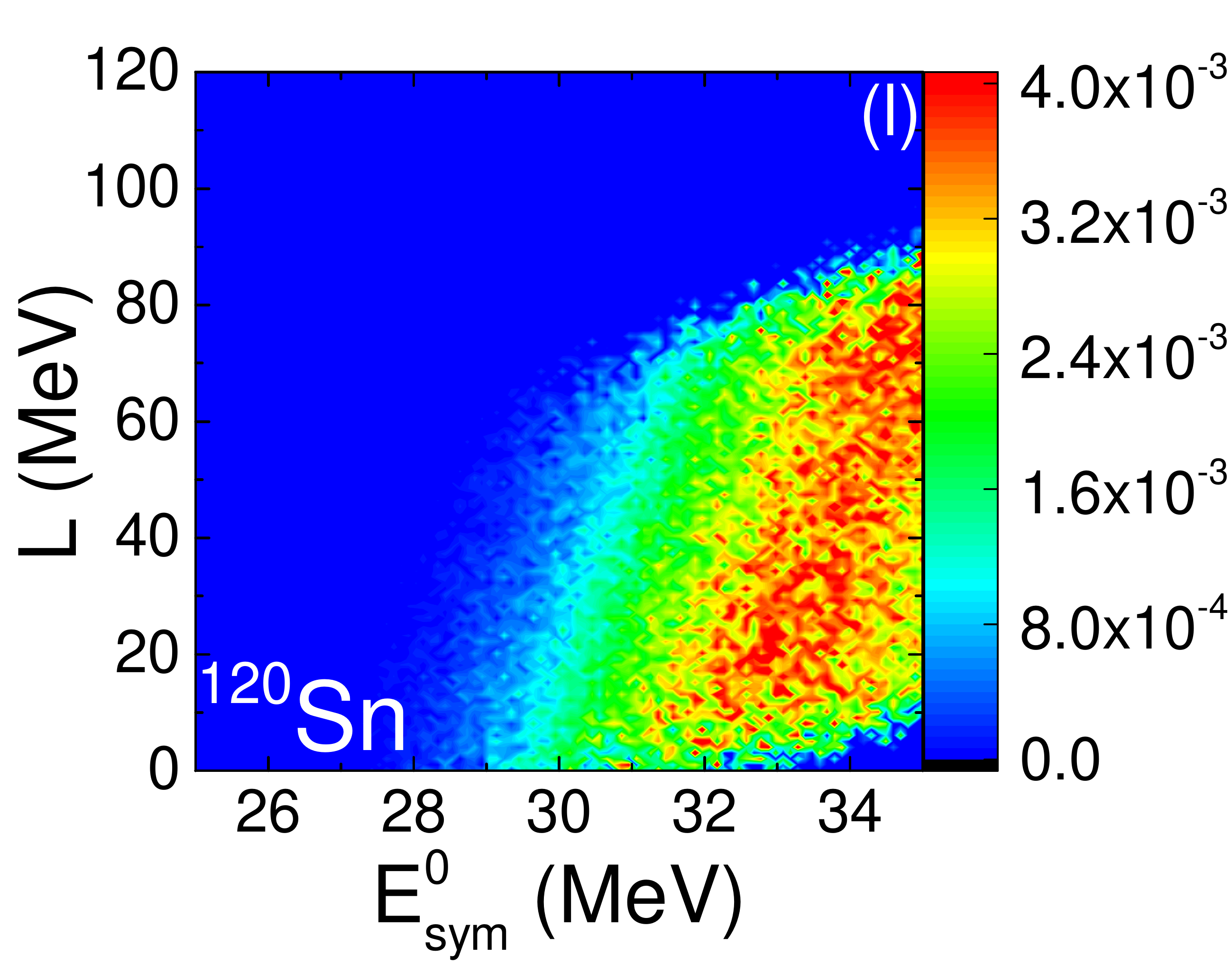}
	\caption{Same as Fig.~\ref{fig7} but under the constraints of the nuclear structure data of $^{120}$Sn.} \label{fig9}
\end{figure}

\begin{figure}[ht]
\includegraphics[scale=0.1]{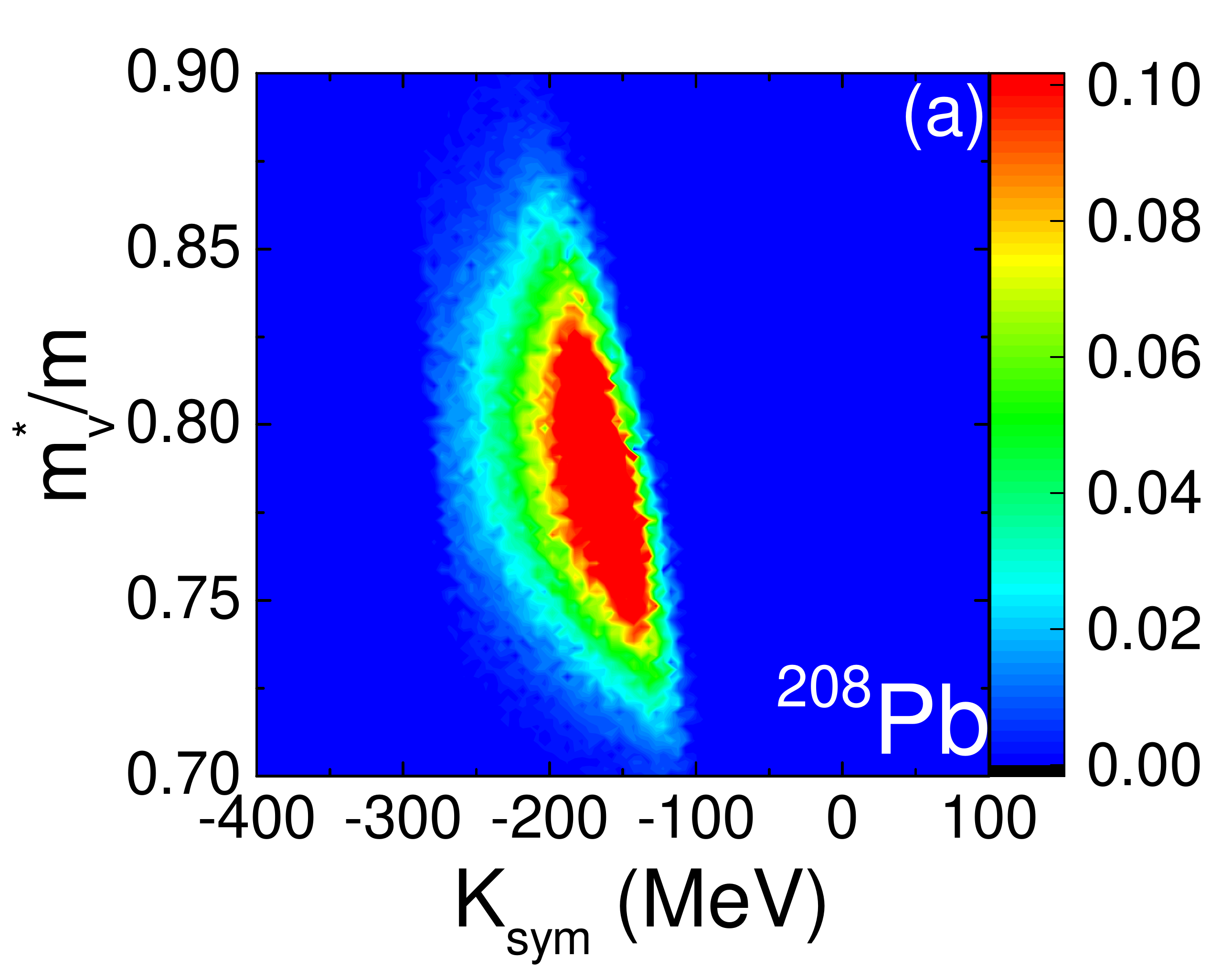}\includegraphics[scale=0.1]{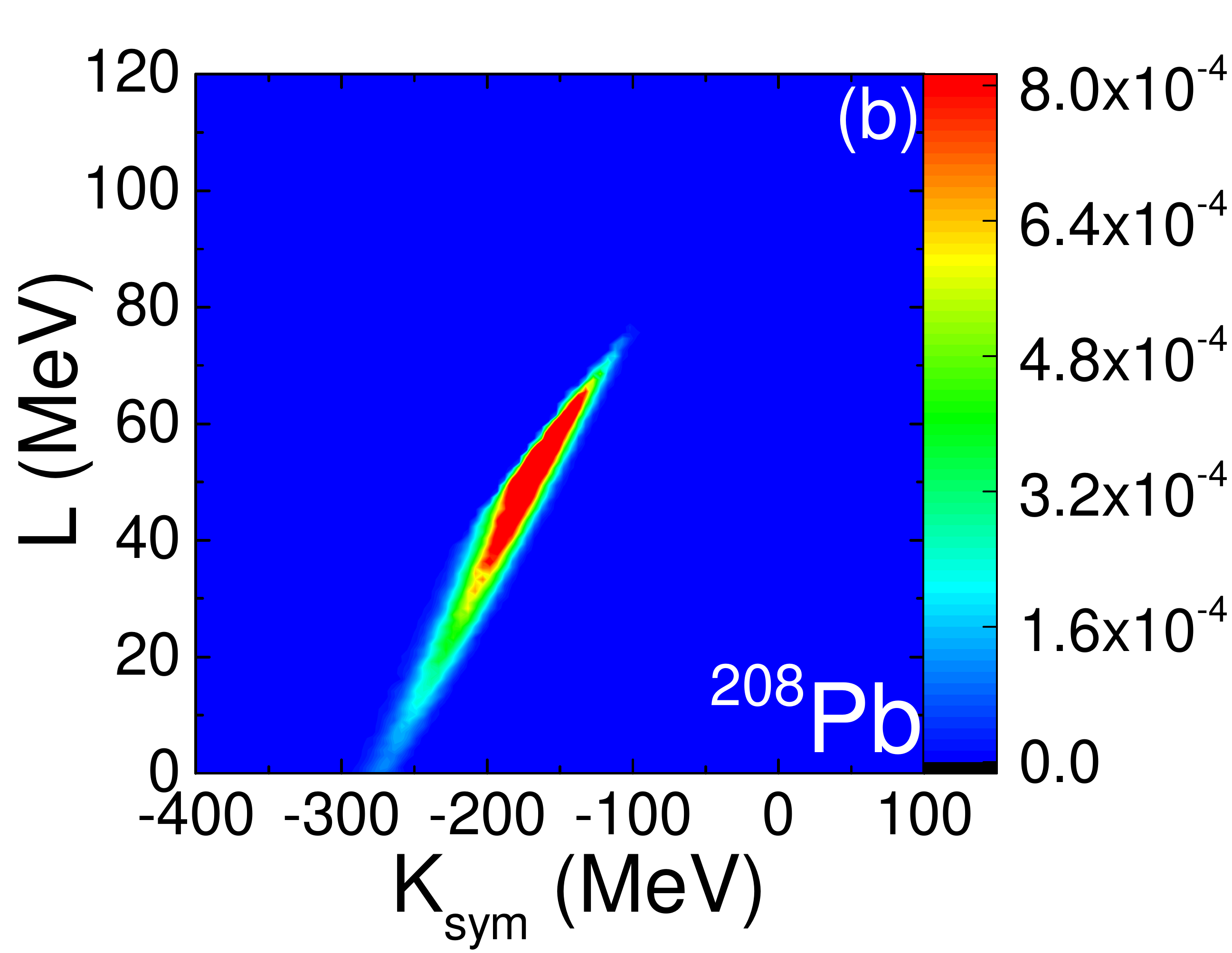}\includegraphics[scale=0.1]{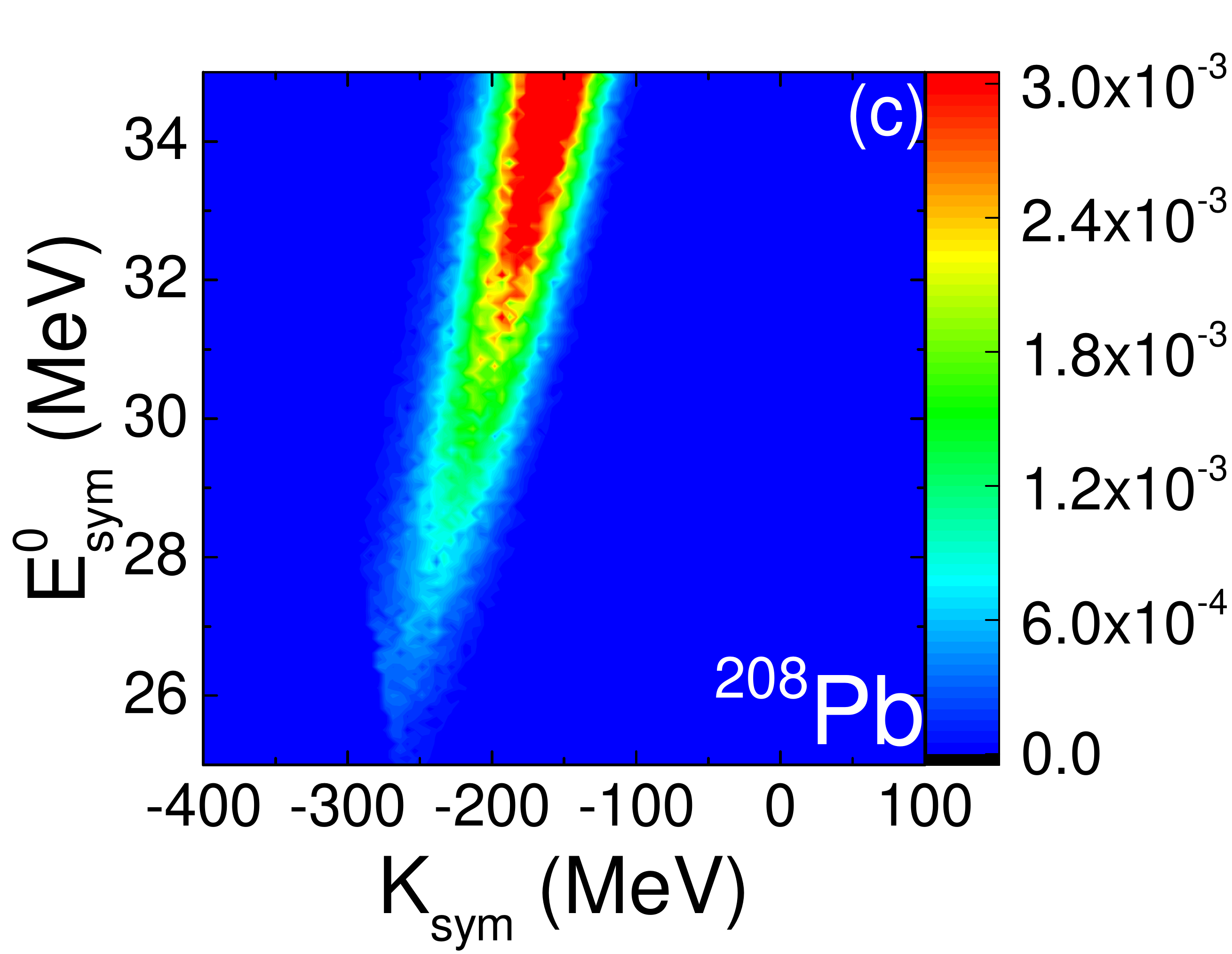}\\
\includegraphics[scale=0.1]{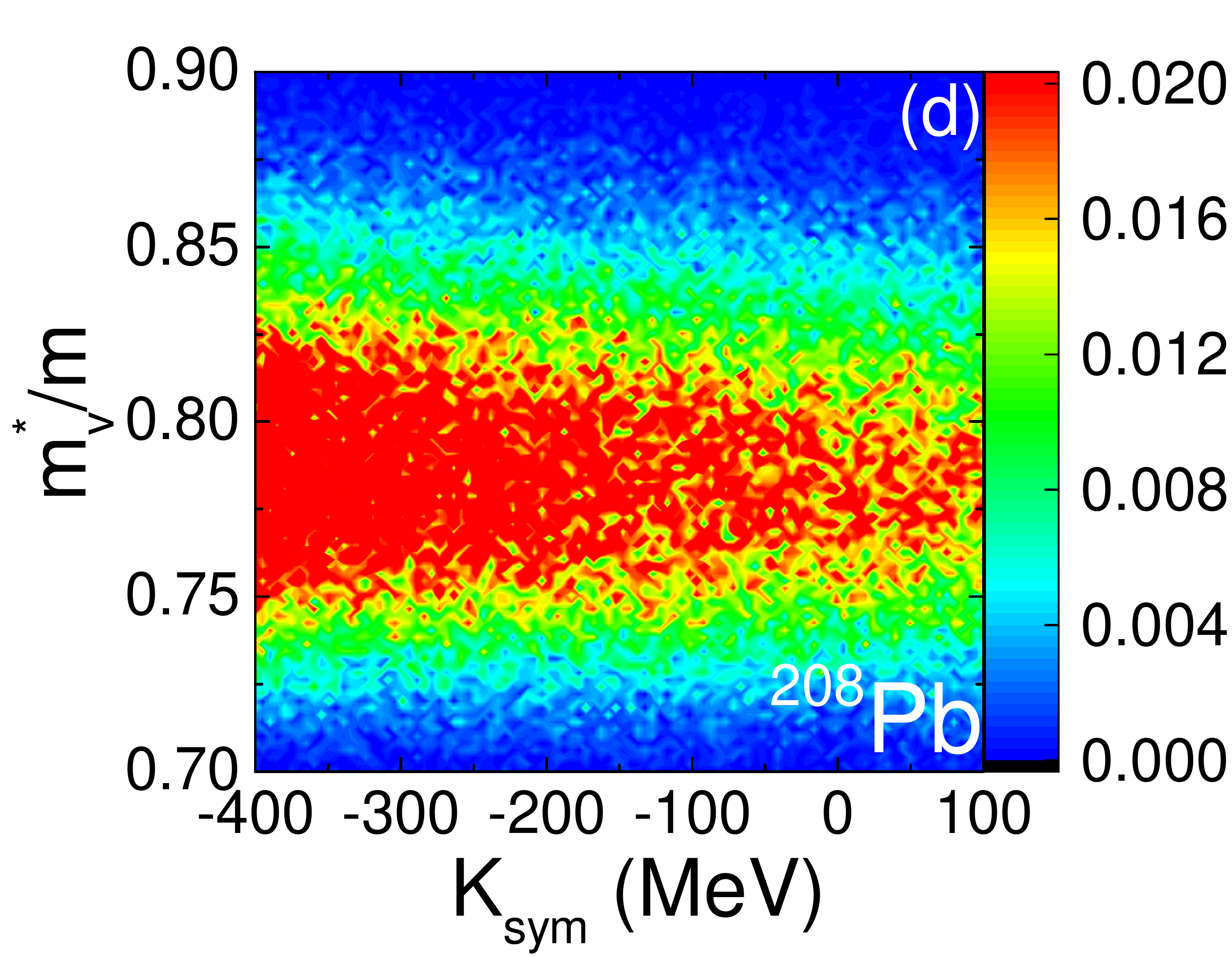}\includegraphics[scale=0.1]{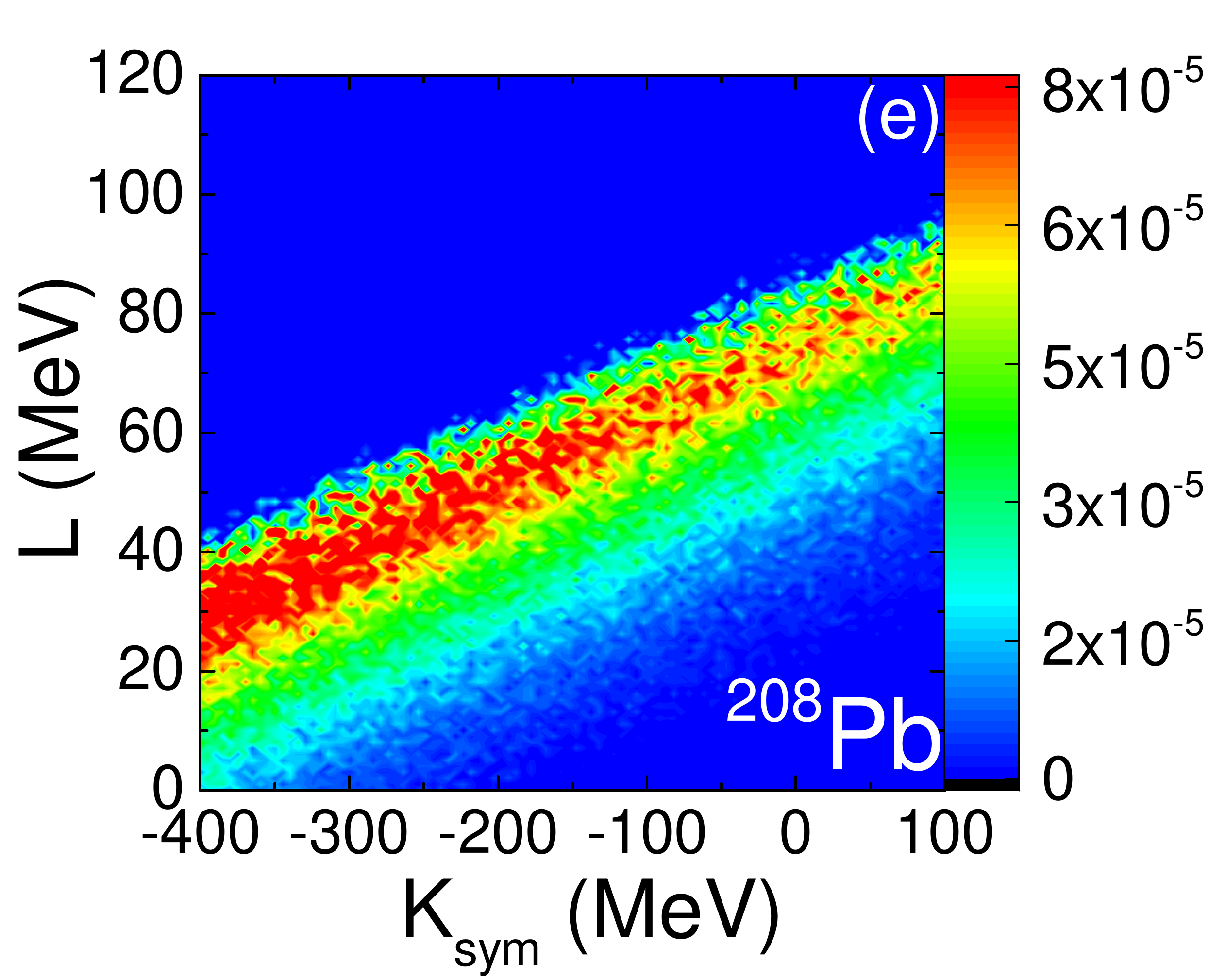}\includegraphics[scale=0.1]{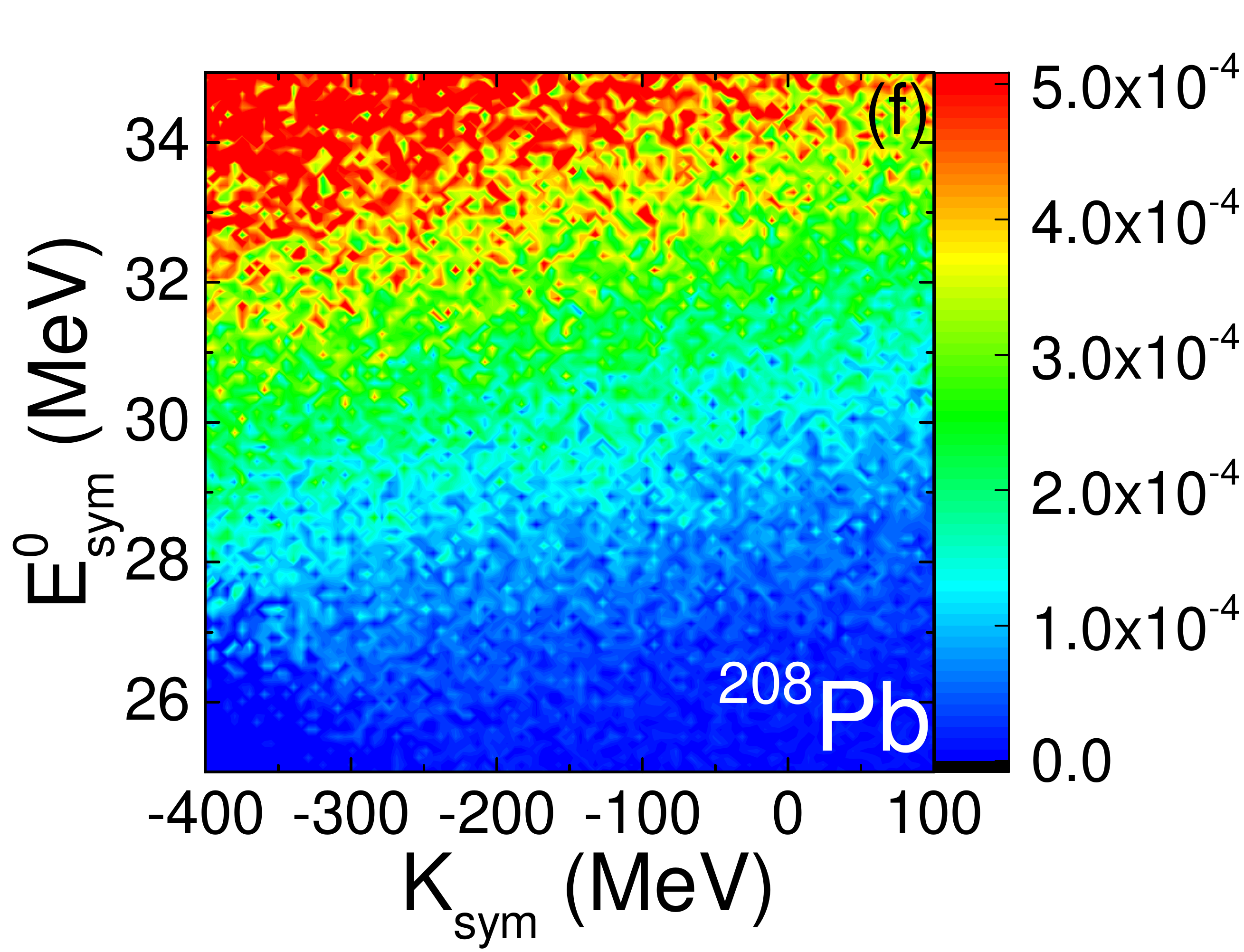}\\
\includegraphics[scale=0.1]{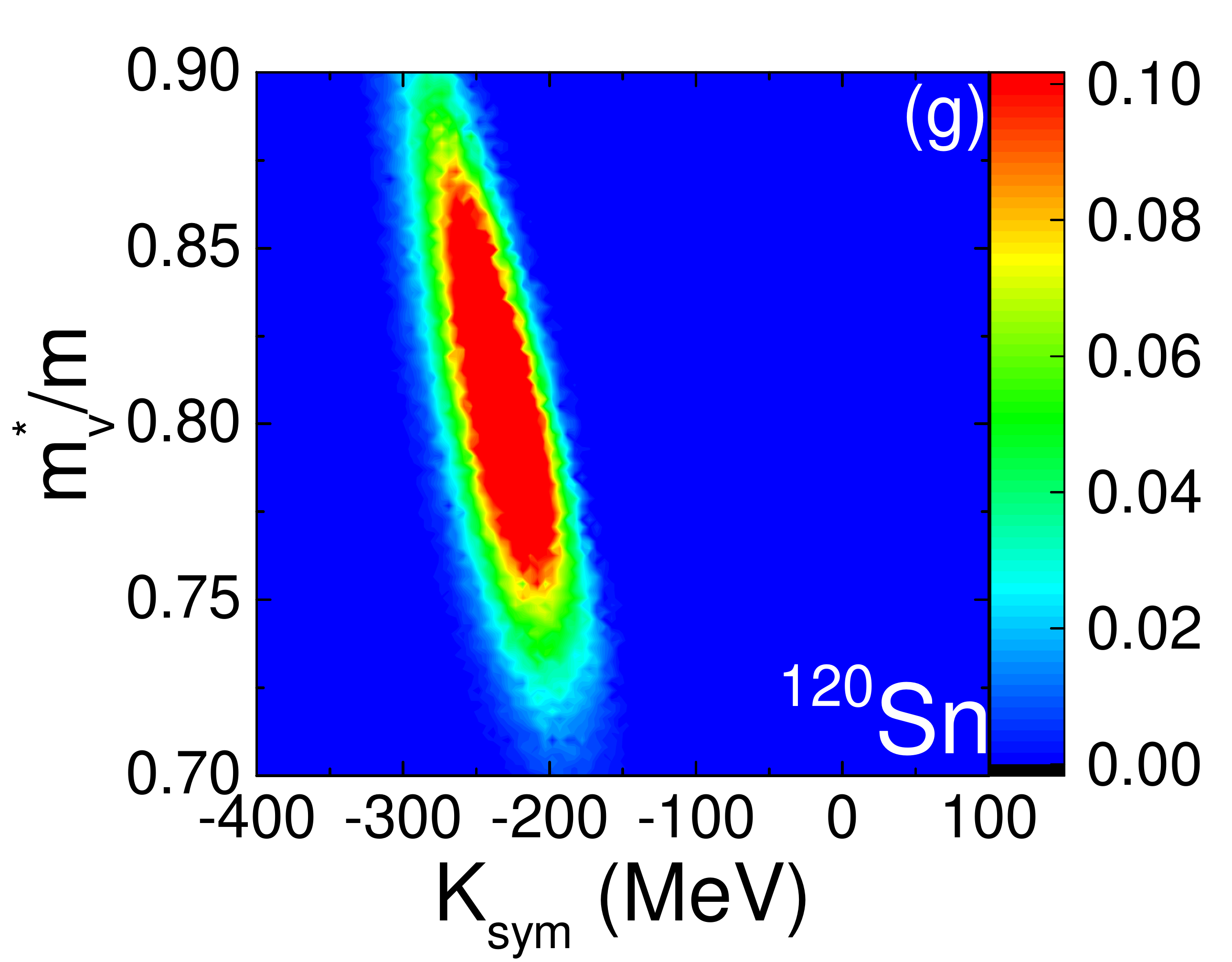}\includegraphics[scale=0.1]{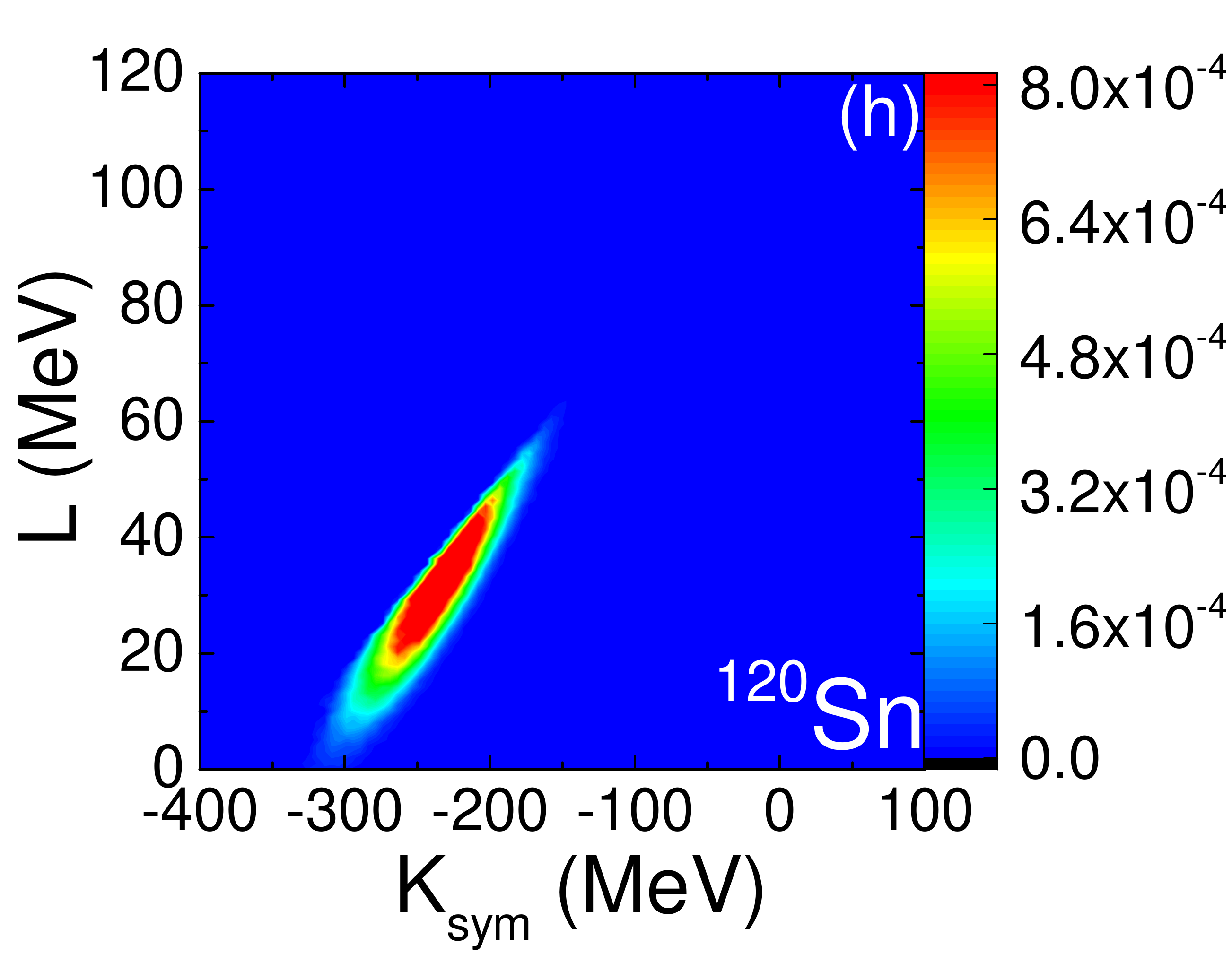}\includegraphics[scale=0.1]{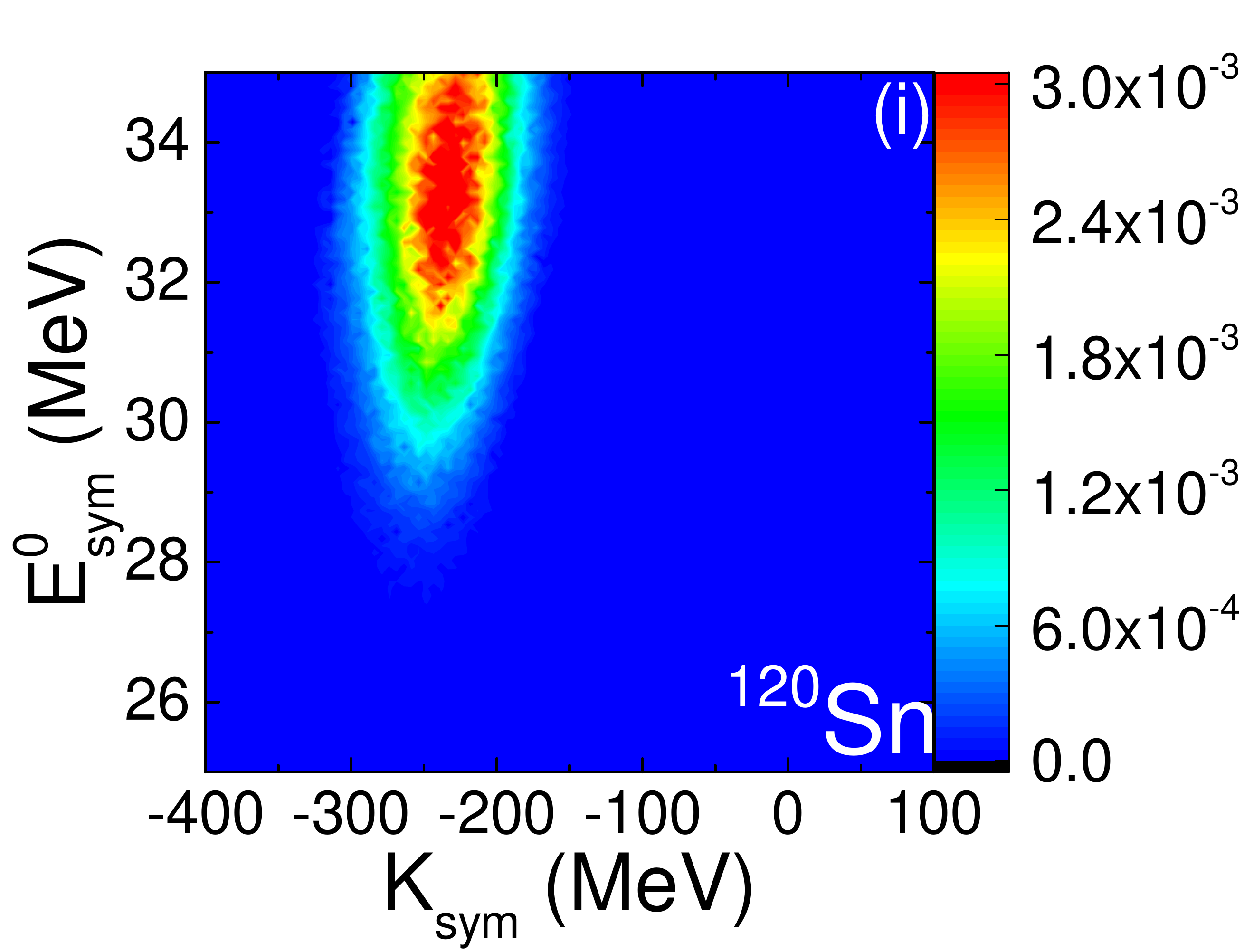}\\
\includegraphics[scale=0.1]{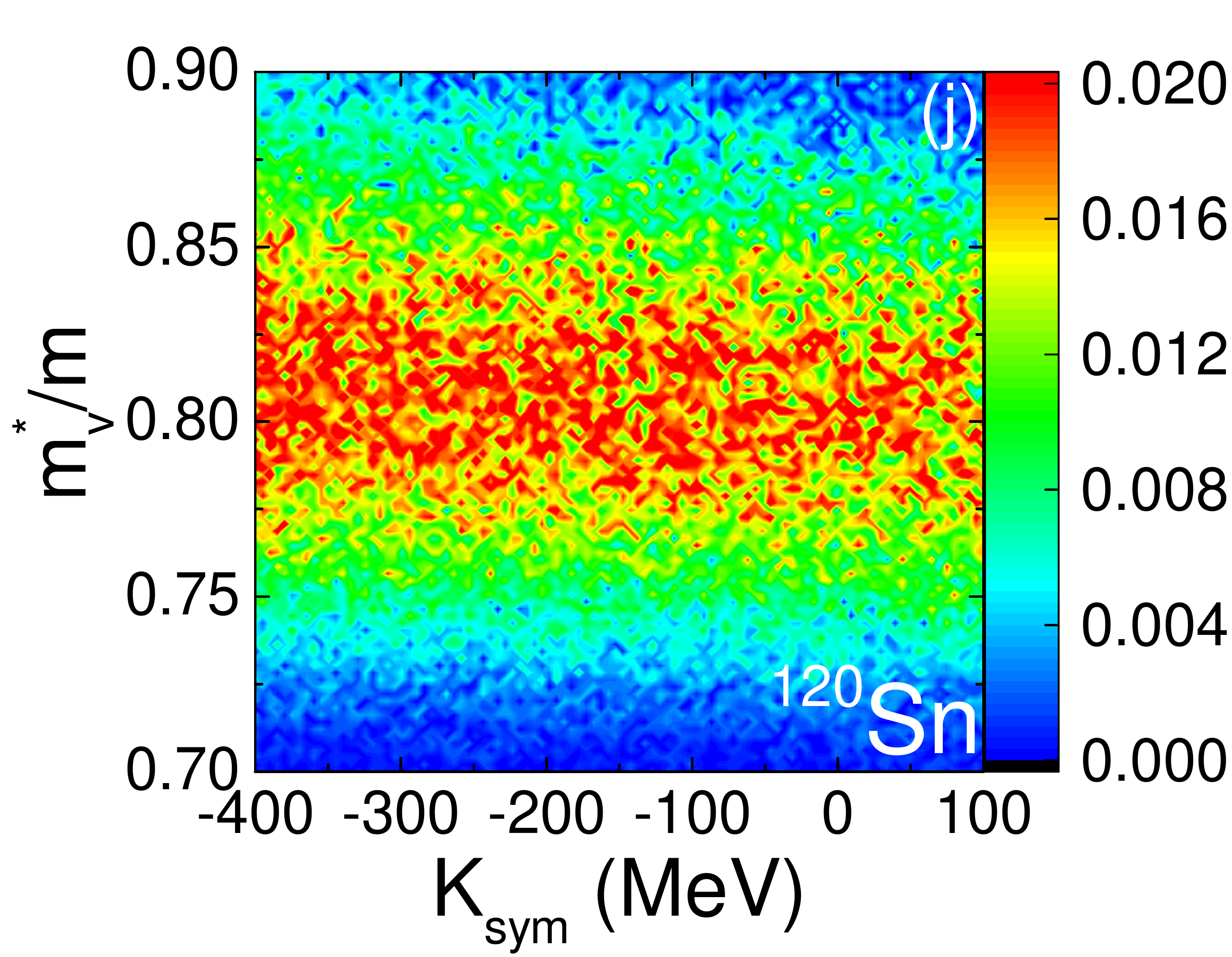}\includegraphics[scale=0.1]{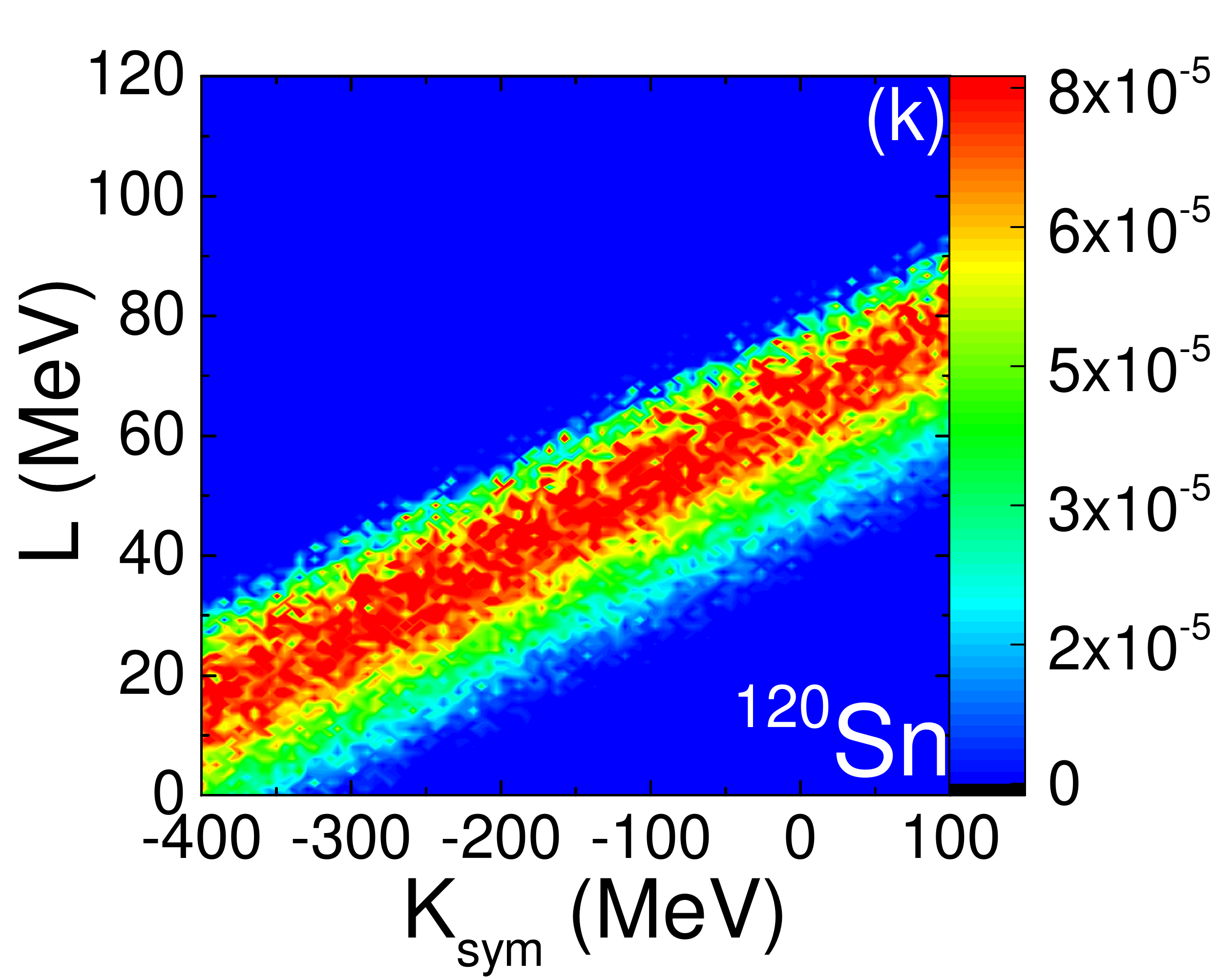}\includegraphics[scale=0.1]{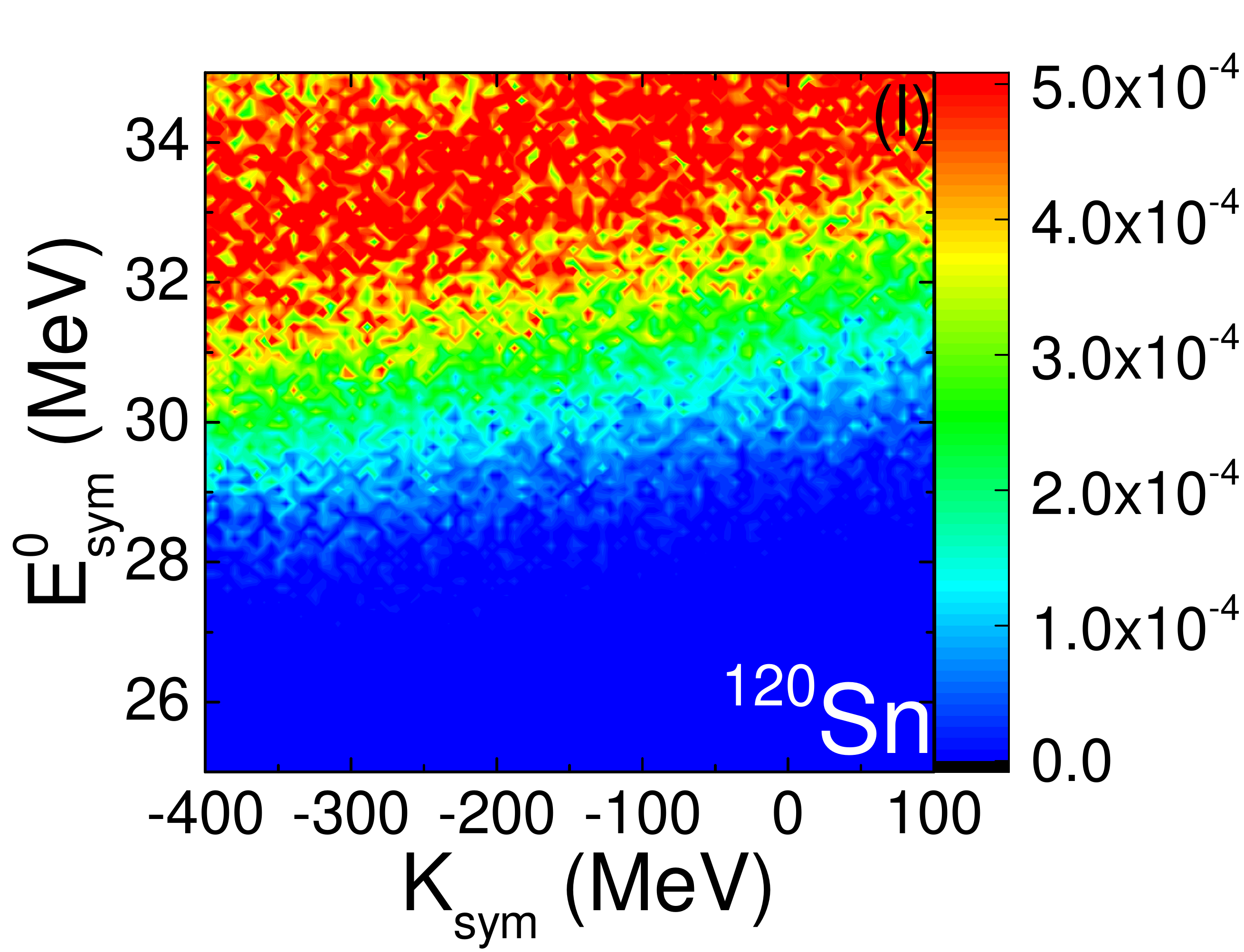}
	\caption{First row: Posterior correlated PDFs between $K_{sym}$ and $m_v^\star$, $L$, and $E_{sym}^0$ under the constraints of $\Delta r_{np}$, $E_{-1}$, and $\alpha_D$ in $^{208}$Pb based on the standard SHF model; Second row: Same as the first row but based on the KIDS model; Third row: Same as the first row but under the constraints of the nuclear structure data of $^{120}$Sn; Fourth row: Same as the second row but under the constraints of the nuclear structure data of $^{120}$Sn. } \label{fig10}
\end{figure}

\begin{figure}[h]
\includegraphics[scale=0.21]{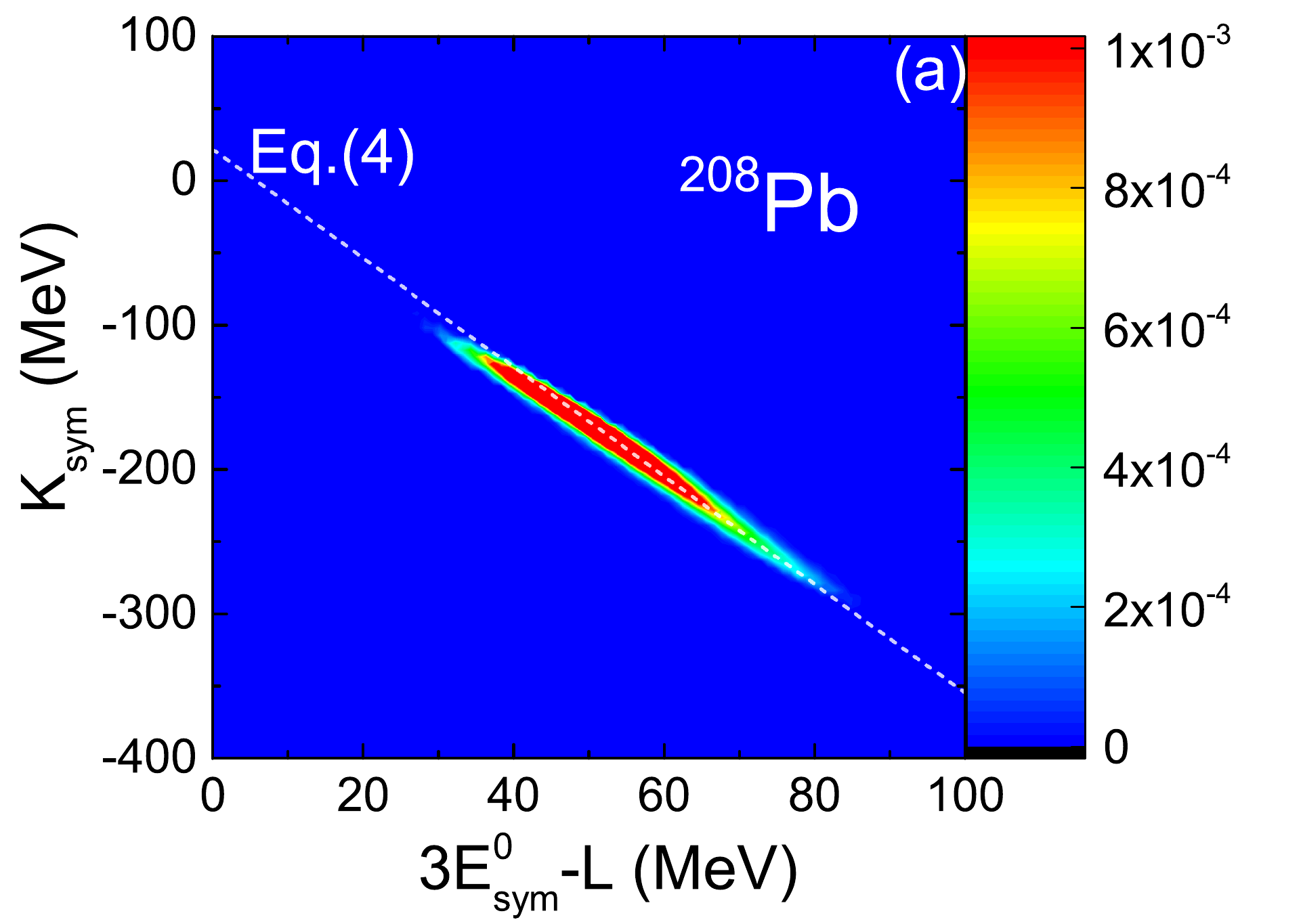}
\includegraphics[scale=0.21]{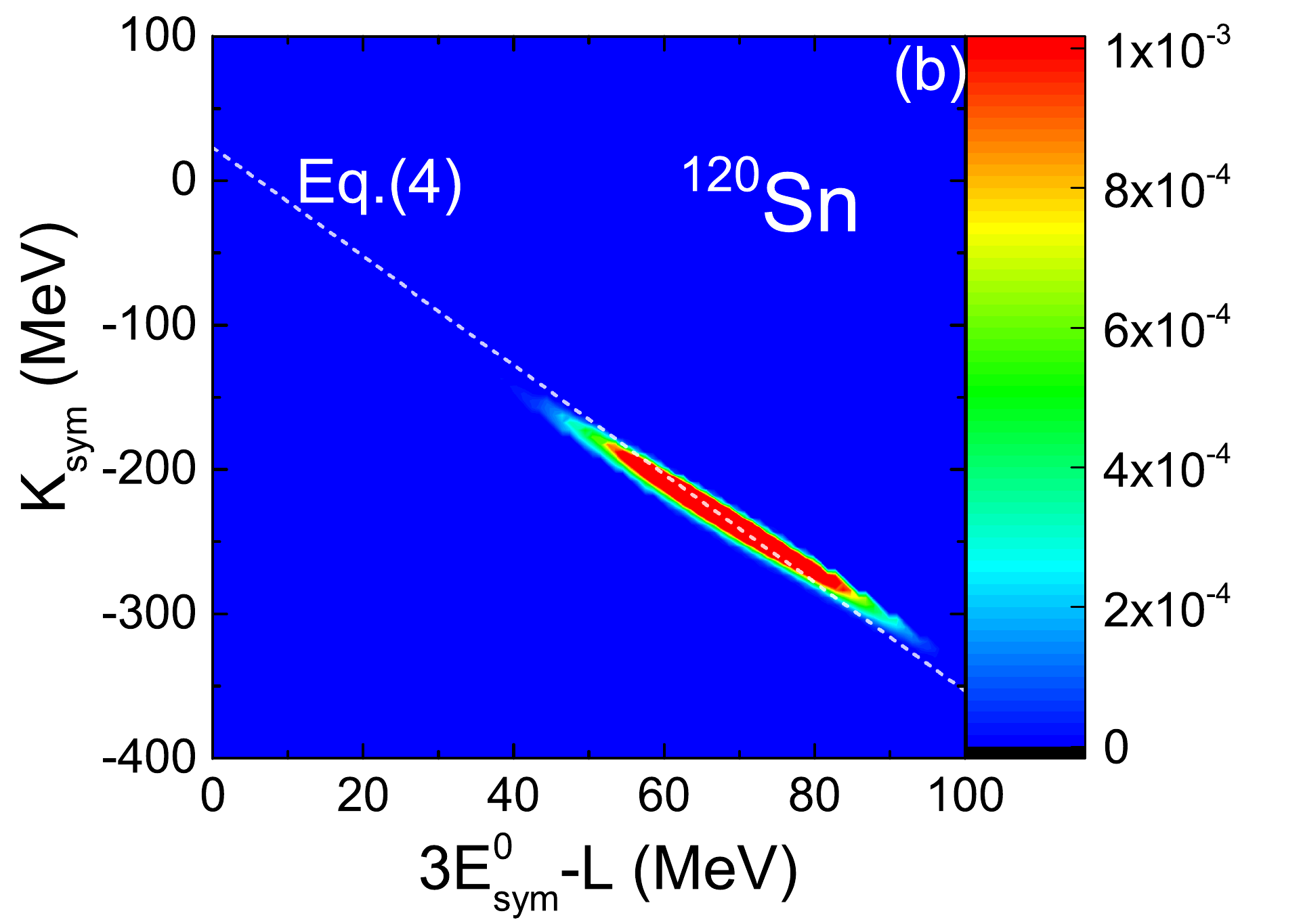}\\
\includegraphics[scale=0.21]{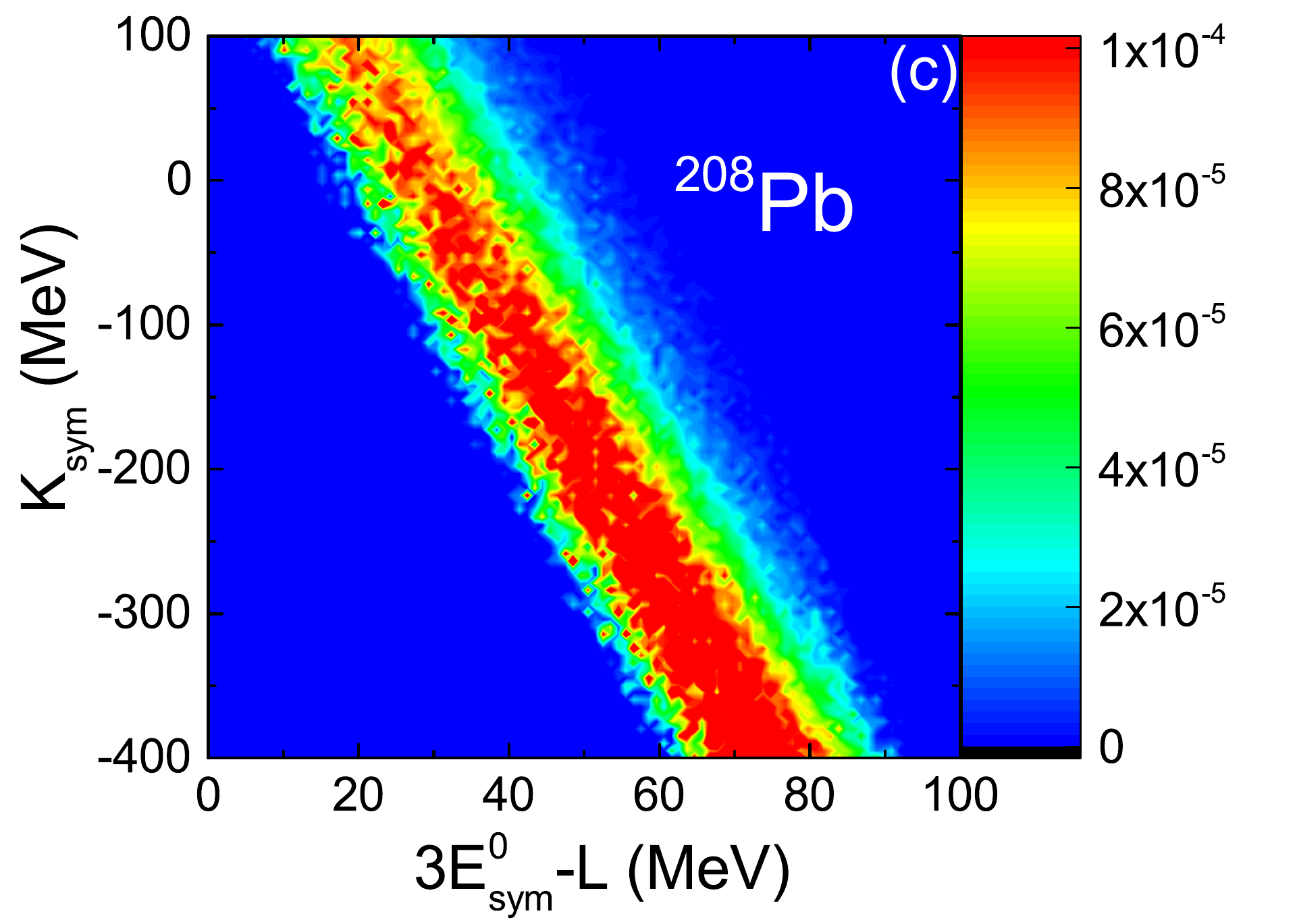}
\includegraphics[scale=0.21]{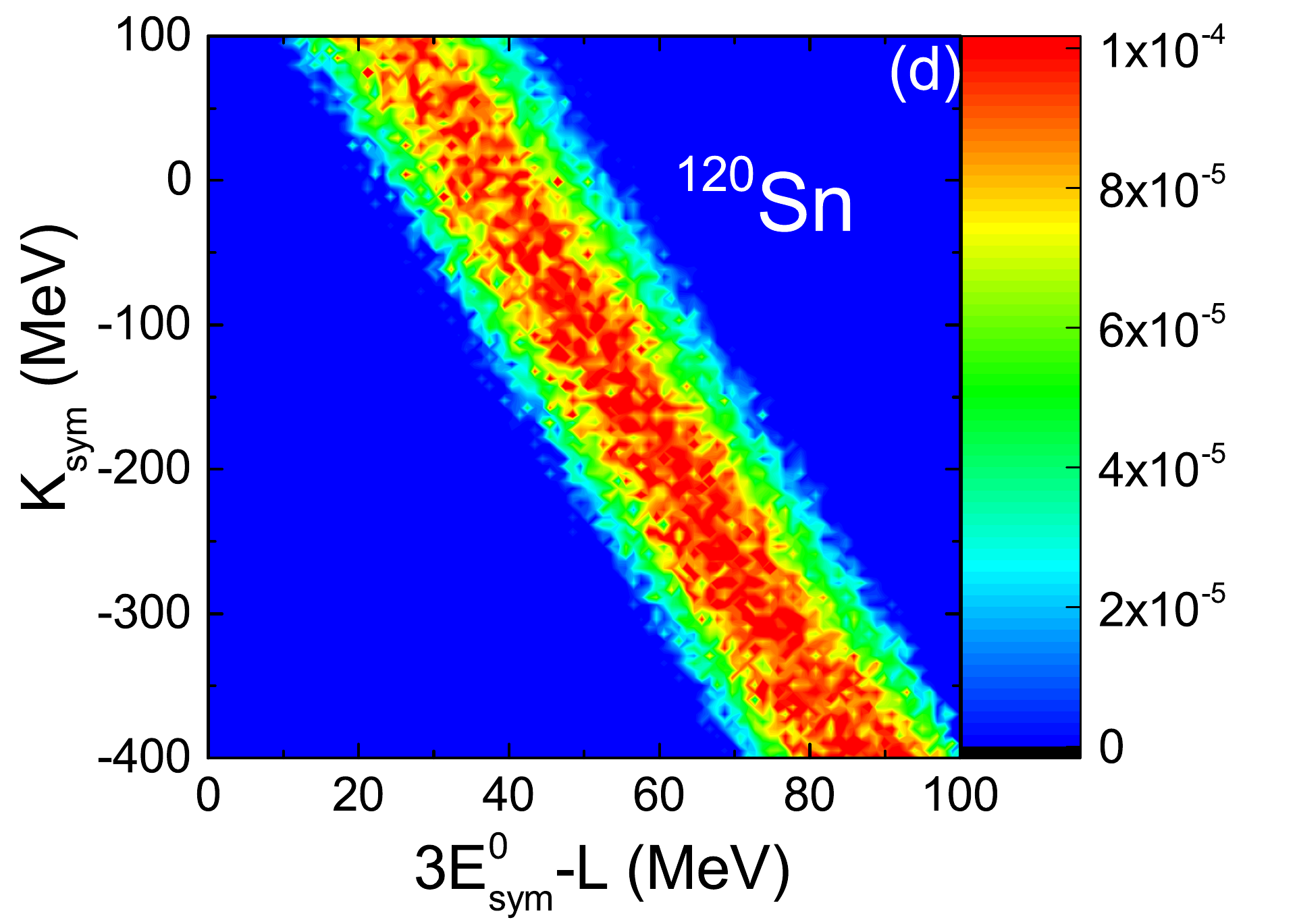}
	\caption{Posterior correlated PDFs between $K_{sym}$ and $3E_{sym}^0-L$ under the constraints of $\Delta r_{np}$, $E_{-1}$, and $\alpha_D$ in $^{208}$Pb (left) and $^{120}$Sn (right) based on the standard SHF (upper) and KIDS (lower) model.} \label{fig11}
\end{figure}

\begin{figure*}[ht]
\includegraphics[scale=0.8]{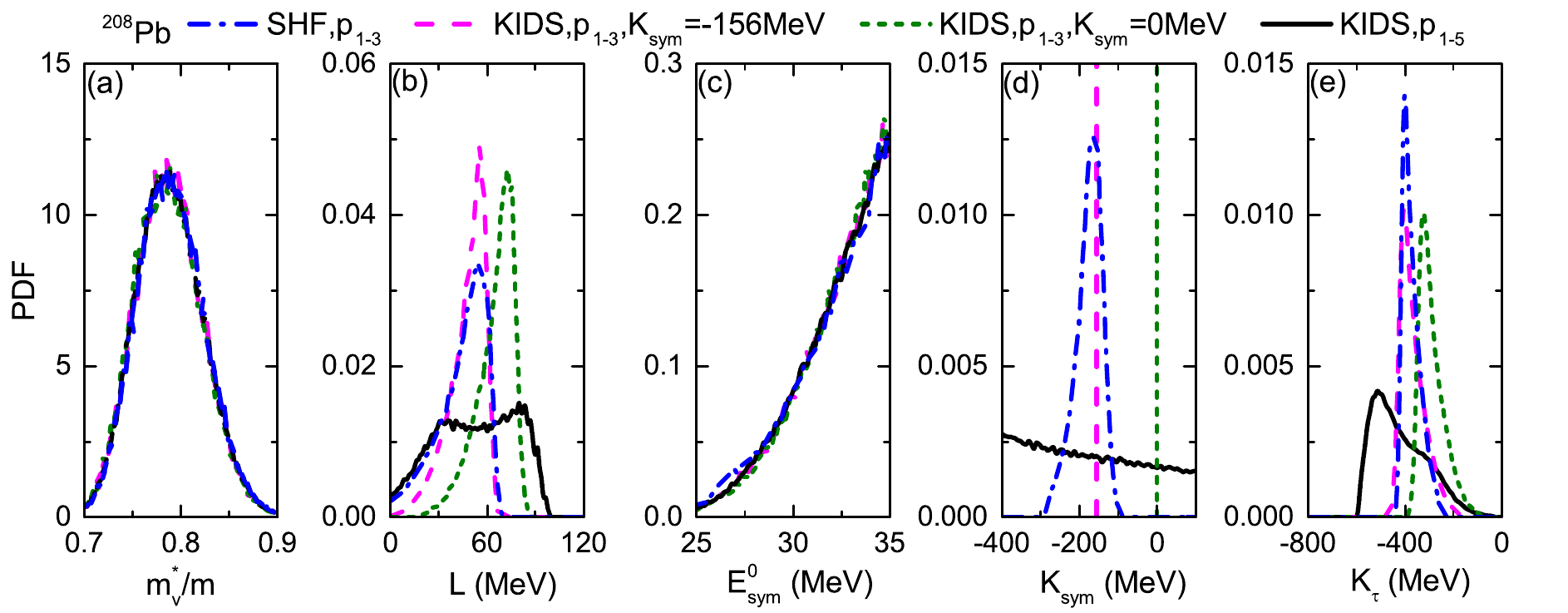}\\
\includegraphics[scale=0.8]{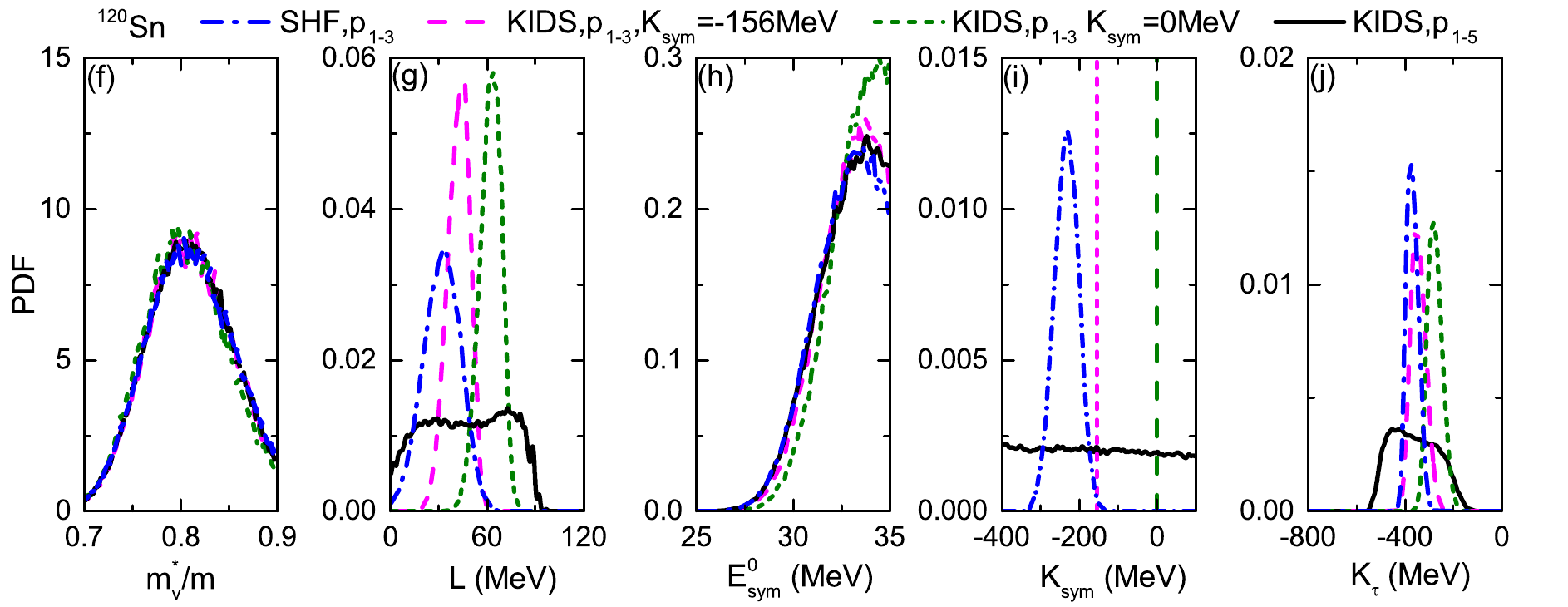}\\
	\caption{Posterior PDFs of $m_v^\star/m$, $L$, $E_{sym}^0$, $K_{sym}$, and $K_\tau$ under the constraints of $\Delta r_{np}$, $E_{-1}$, and $\alpha_D$ in $^{208}$Pb (upper) and $^{120}$Sn (lower) from the four scenarios in Figs.~\ref{fig8} and \ref{fig9}. } \label{fig12}
\end{figure*}

In addition to the neutron-skin thickness data, we now incorporate the constraint of the IVGDR data, i.e., the centroid energy $E_{-1}$ and the electric polarizability $\alpha_D$ from the IVGDR in $^{208}$Pb and $^{120}$Sn. By varying $m_v^\star$, $L$, and $E_{sym}^0$, their posterior correlated PDFs under the constraints of $\Delta r_{np}$, $E_{-1}$, and $\alpha_D$ are compared in Fig.~\ref{fig8} for $^{208}$Pb and in Fig.~\ref{fig9} for $^{120}$Sn from the Bayesian analysis, based on the standard SHF model and the KIDS model with fixed $K_{sym}=-156$ and 0 MeV, respectively. Comparing especially the $E_{sym}^0 - L$ correlations in Figs.~\ref{fig8} and \ref{fig9} to those in Fig.~\ref{fig4}, one sees that the more precise IVGDR data dominate the results, compared with the $\Delta r_{np}$ data with a larger error bar. This can be seen from the positive correlation between $L$ and $E_{sym}$ under the constraint of the IVGDR data~\cite{Xu20,Xu21} in the standard SHF model, and it can again be understood from the positive (negative) correlation between $E_{-1}$ and $E_{sym}^0$ ($E_{-1}$ and $L$) as well as the negative (positive) correlation between $\alpha_D$ and $E_{sym}^0$ ($\alpha_D$ and $L$) in Figs.~\ref{fig2} and \ref{fig3}. The large $L$ values in the $L-m_v^\star$ plane are ruled out, and this is also seen in the $E_{sym}^0-L$ plane as a result of the limited prior range for $E_{sym}^0$. Due to the sensitivity of $E_{-1}$ to $m_v^\star$, $m_v^\star$ is also constrained. Again, the correlated PDFs in the $L-m_v^\star$ planes and in the $E_{sym}^0-L$ plane are also affected by the fixed value of $K_{sym}$, while those in the $E_{sym}^0-m_v^\star$ plane are not affected by much. The correlated PDFs from the standard SHF model, with $K_{sym}$ dependent on $L$, $E_{sym}^0$, and other quantities, have similar shapes but are more diffusive, compared with those from the KIDS model for a fixed $K_{sym}$.

Under the constraints of both $\Delta r_{np}$ and IVGDR data, we have further incorporated the independent variables $K_{sym}$ and $Q_{sym}$ based on the KIDS model, and the resulting correlated PDFs of interest for both $^{208}$Pb and $^{120}$Sn are shown in the bottom rows of Figs.~\ref{fig8} and \ref{fig9}, respectively. Again, the correlated PDFs in the $L-m_v^\star$ plane and in the $E_{sym}^0-L$ plane are basically the superpositions of those at different fixed $K_{sym}$, while those in the $E_{sym}^0-m_v^\star$ plane are approximately independent of $K_{sym}$, compared with those in the second and third rows of Figs.~\ref{fig8} and \ref{fig9}. The correlated PDFs between $K_{sym}$ and $m_v^\star$, $L$, and $E_{sym}^0$ from such analyses are displayed in Fig.~\ref{fig10} for both the standard SHF and KIDS models. For the standard SHF model, $K_{sym}$ is not an independent variable but can be obtained from other parameters through Eq.~(\ref{Ksymshf}), so it is understandable that the correlated PDFs are narrow and sharp. For the KIDS model, we observe similar positive correlations between $L$ and $K_{sym}$ as in Fig.~\ref{fig5}. This can again be understood from the sensitivity study shown in Figs.~\ref{fig2} and \ref{fig3}, where $\Delta r_{np}$ and $\alpha_D$ increase with increasing $L$ but decrease with increasing $K_{sym}$, and $E_{-1}$ decreases with increasing $L$ but increases with increasing $K_{sym}$. With the same constraints from the $\Delta r_{np}$ and IVGDR data, the correlated PDFs are much more constrained in the standard SHF than in the KIDS model, due to more independent variables and flexibility in the latter case.

The linear anticorrelation between $K_{sym}$ and $3E_{sym}^0-L$ has been found to be a general one in various EDFs~\cite{Mon17}. In the standard SHF model, it can be attributed to Eq.~(\ref{Ksymshf}) with given $\alpha$ (through Eq.~(\ref{alpha})), $\rho_0$, and the nucleon effective masses (see also Ref.~\cite{Gil21}). Under the constraints of $\Delta r_{np}$ and IVGDR data for $^{208}$Pb and $^{120}$Sn, the posterior correlated PDFs between $K_{sym}$ and $3E_{sym}^0-L$ are displayed in Fig.~\ref{fig11} based on the standard SHF and KIDS model. The white line for the standard SHF model is from Eq.~(\ref{Ksymshf}) obtained using default values of $\rho_0$, $E_0$, $K_0$, $m_s^\star$, and the MAP value of $m_v^\star$. One sees that the correlated PDFs are consistent with Eq.~(\ref{Ksymshf}) within the restricted range of $3E_{sym}^0-L$, while the small deviations are likely due to the diffusive PDF of $m_v^\star$. Interestingly, without the intrinsic relation as Eq.~(\ref{Ksymshf}), the KIDS model also gives a linear anticorrelation between $K_{sym}$ and $3E_{sym}^0-L$ under the constraint of the nuclear structure data, but with the correlated PDFs more diffusive and with a different slope. Some differences in the correlated PDFs are also observed from the $^{208}$Pb and $^{120}$Sn data in both standard SHF and KIDS models. Additional constraints from astrophysical observables would further reduce the range of $K_{sym}$ and $3E_{sym}^0-L$ as shown in Ref.~\cite{Gil21}.

Integrating the other physics variable in the correlated PDFs leads to the one-dimensional PDFs of $m_v^\star/m$, $L$, $E_{sym}^0$, and $K_{sym}$ shown in Fig.~\ref{fig12} for both $^{208}$Pb and $^{120}$Sn for the four scenarios discussed in Figs.~\ref{fig9} and \ref{fig10}, where the PDF of
\begin{equation}\label{ktau}
K_\tau = K_{sym}-6L-\frac{Q_0}{K_0}L
\end{equation}
characterizing the isospin-dependence of the incompressibility of nuclear matter~\cite{Che09} is also displayed. One sees that the PDFs of $m_v^\star$ and $E_{sym}^0$ are not much affected by the EDF or the value of $K_{sym}$, and the corresponding PDFs are similar to those obtained in Ref.~\cite{Xu21}, while those of $L$, $K_{sym}$, and $K_\tau$ can be different in different scenarios. The data favors a large value of $E_{sym}^0$ but limited by its prior range $25 \sim 35$ MeV deduced from various earlier analyses~\cite{Li13,Oer17}. The $L$ is mostly constrained within moderate values from the IVGDR data, while the $^{208}$Pb data still leads to a slightly larger $L$ value than $^{120}$Sn attributed to the $\Delta r_{np}$ data by PREXII. In the KIDS model, although the combined data of $\Delta r_{np}$, $E_{-1}$, and $\alpha_D$ are unable to constrain $K_{sym}$, they help to constrain $K_\tau$, whose PDF is affected by both $L$ and $K_{sym}$. In the standard SHF model, $K_{sym}$ depends on $L$, $E_{sym}^0$, etc, and can be constrained from the combined data of $\Delta r_{np}$ and IVGDR, with the $^{208}$Pb ($^{120}$Sn) data favoring a larger (smaller) $K_{sym}$ value. The standard SHF model also gives much sharper PDFs of $L$ and $K_\tau$ as a result of less independent parameters.

\subsection{Bayesian inference on $\Delta r_{np}$, $E_{-1}$, $\alpha_D$, and $E_{ISGMR}$}

\begin{figure*}[ht]
\includegraphics[scale=0.2]{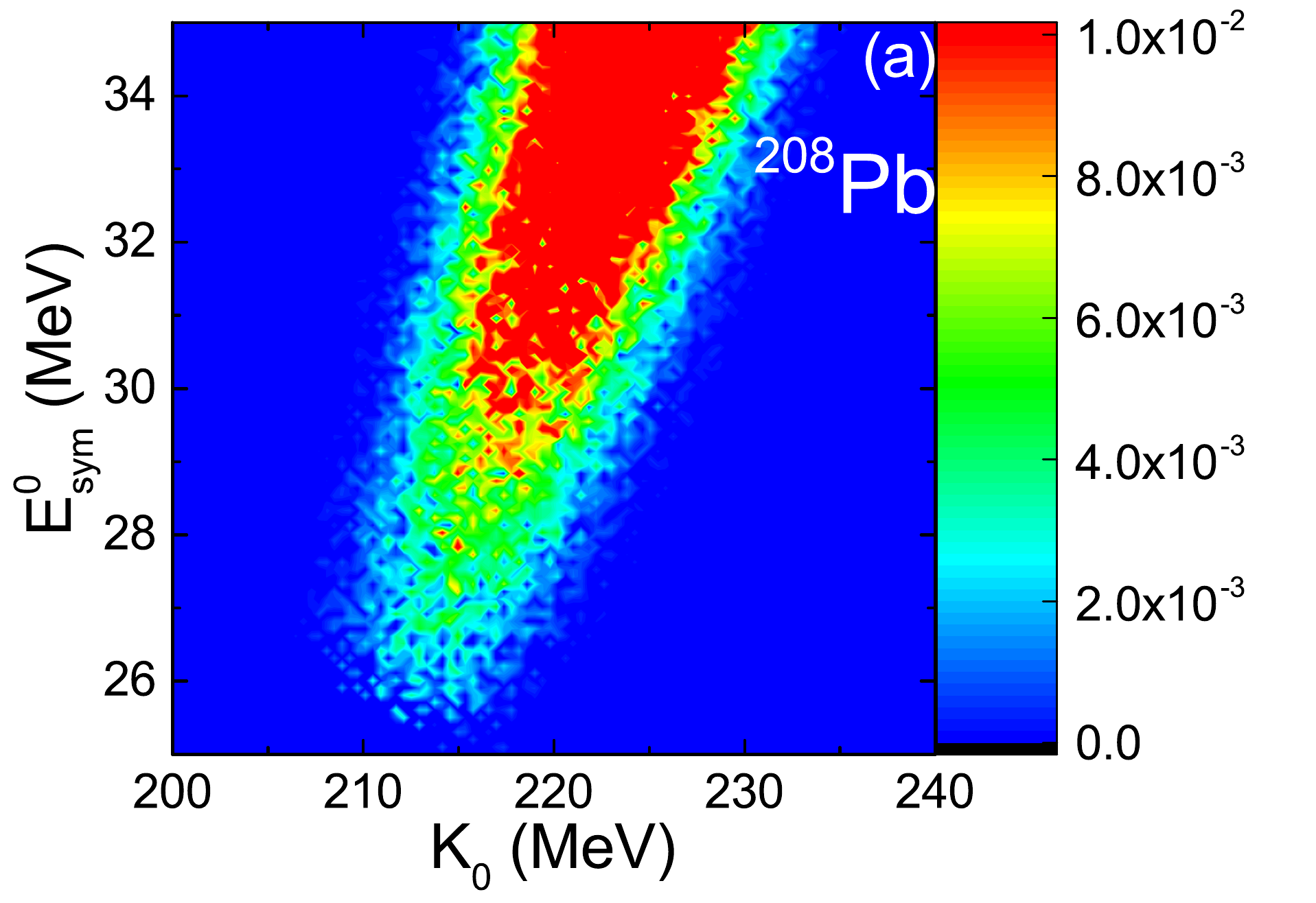}\includegraphics[scale=0.2]{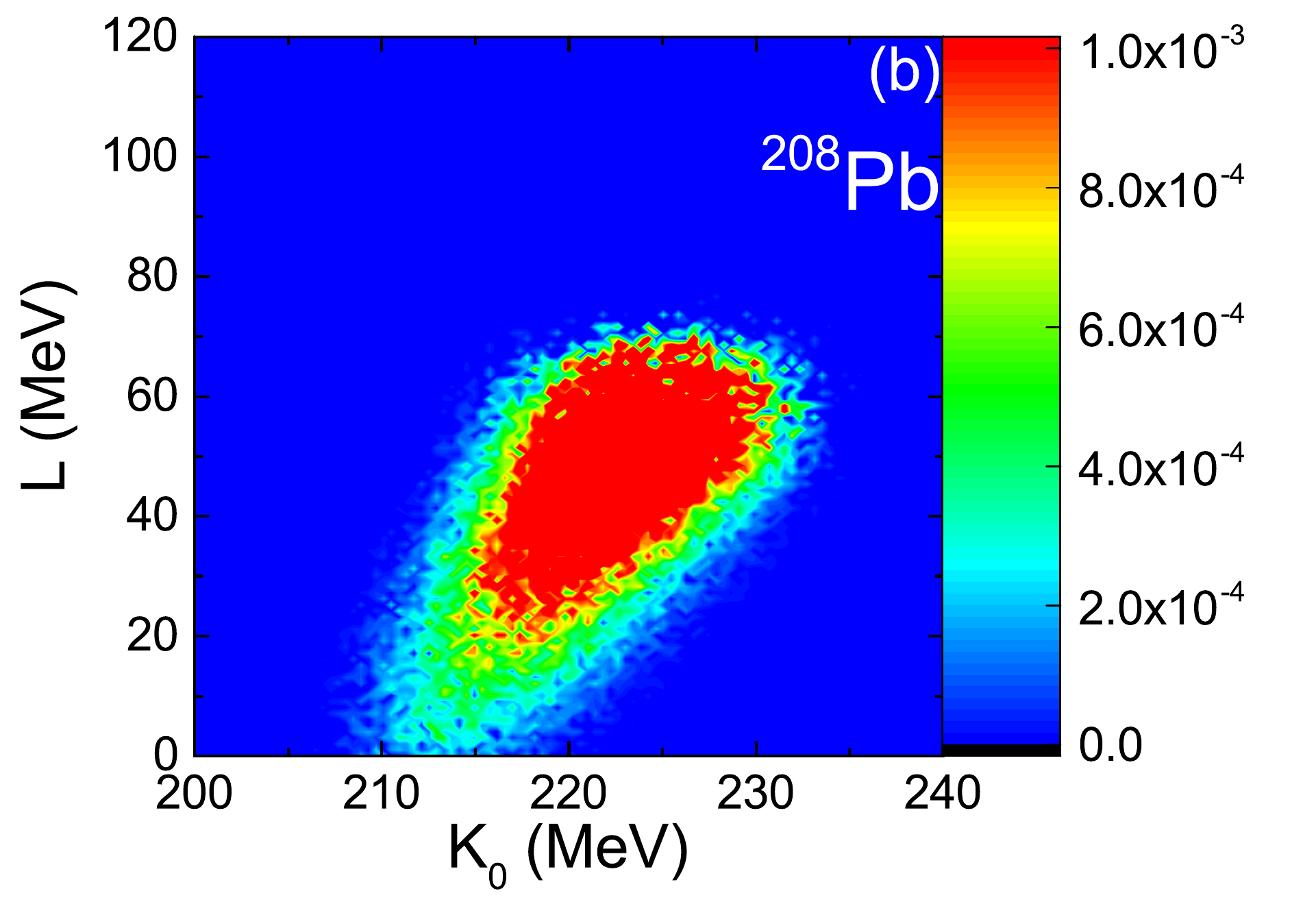}\includegraphics[scale=0.2]{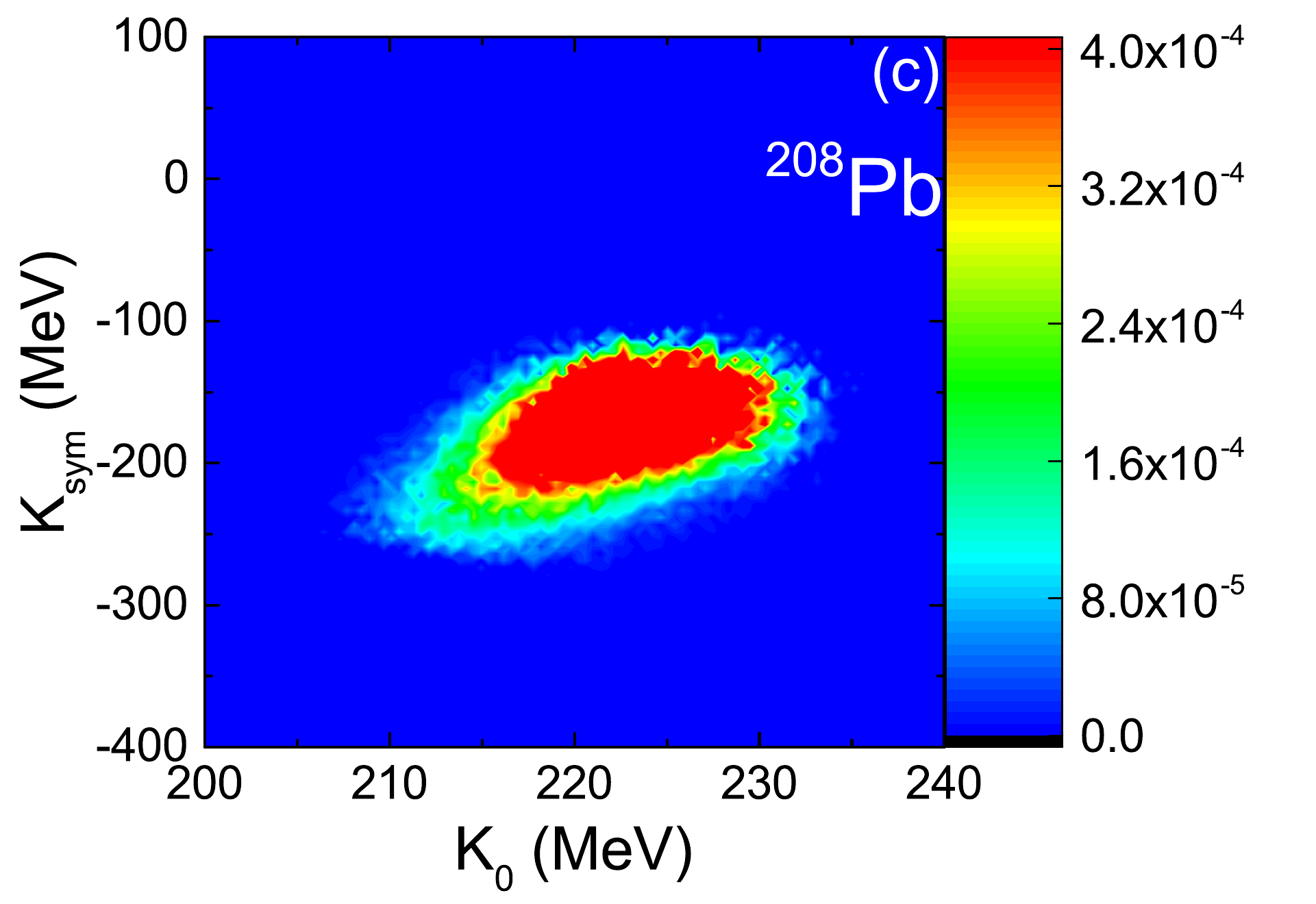}\includegraphics[scale=0.2]{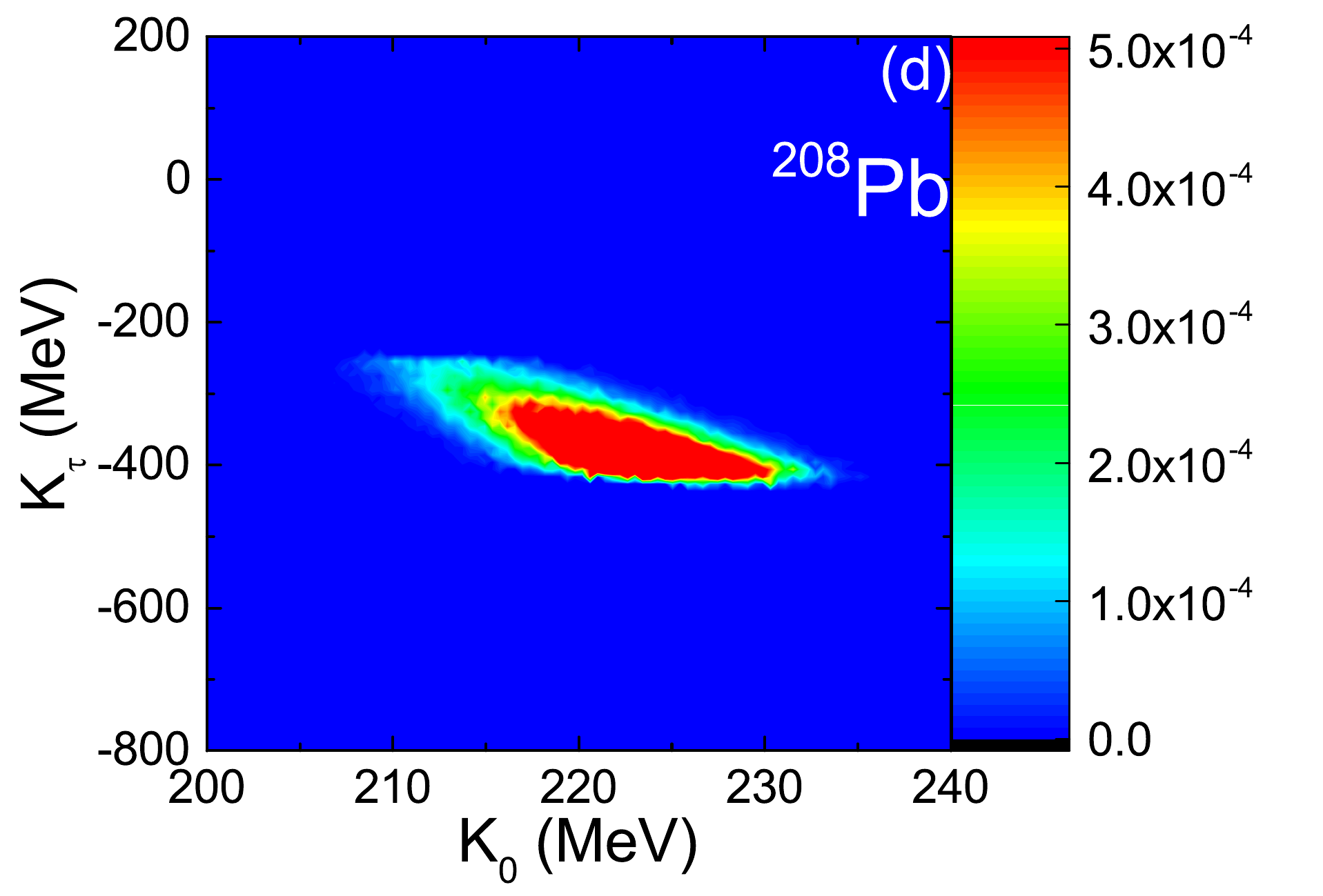}\\
\includegraphics[scale=0.2]{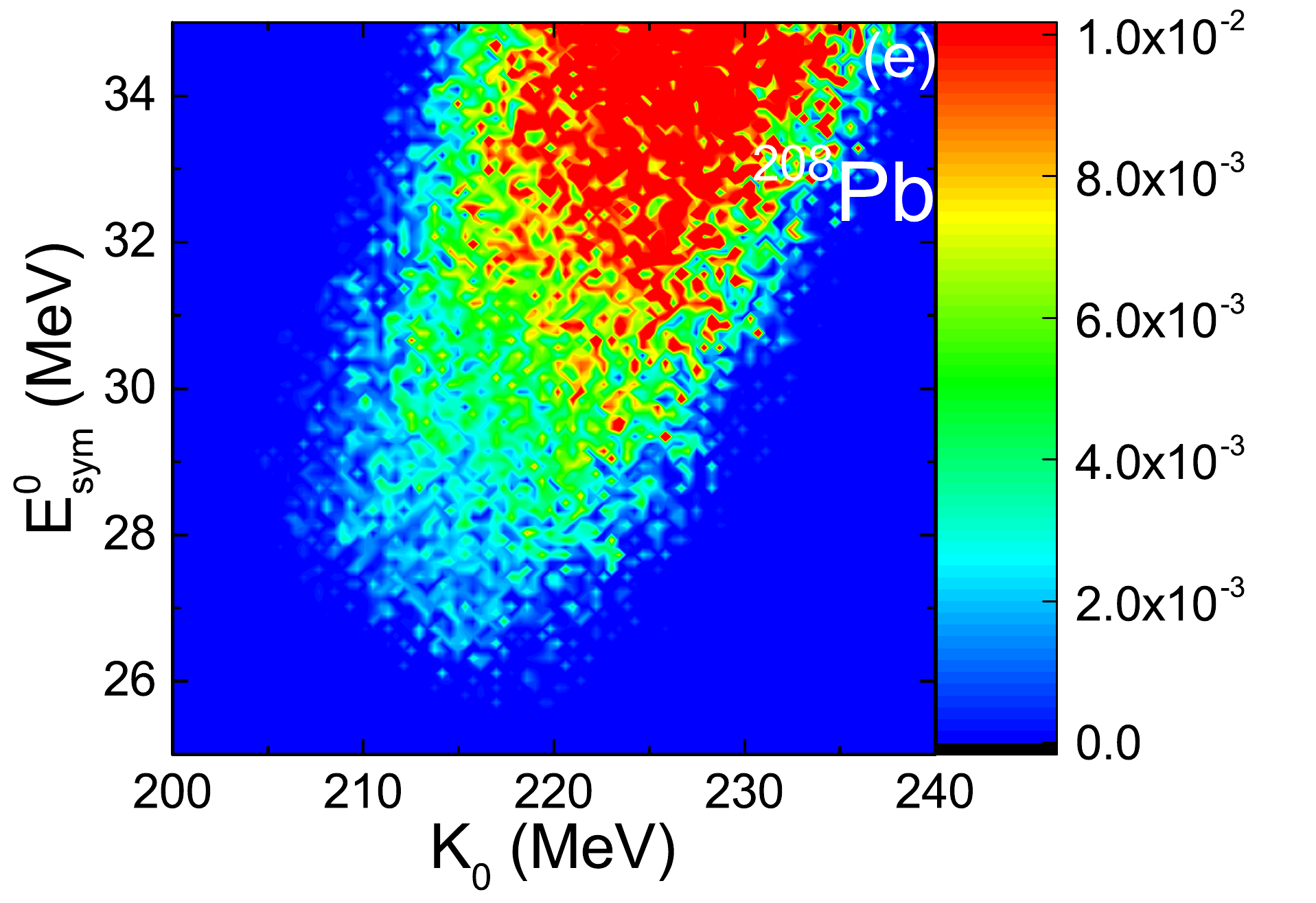}\includegraphics[scale=0.2]{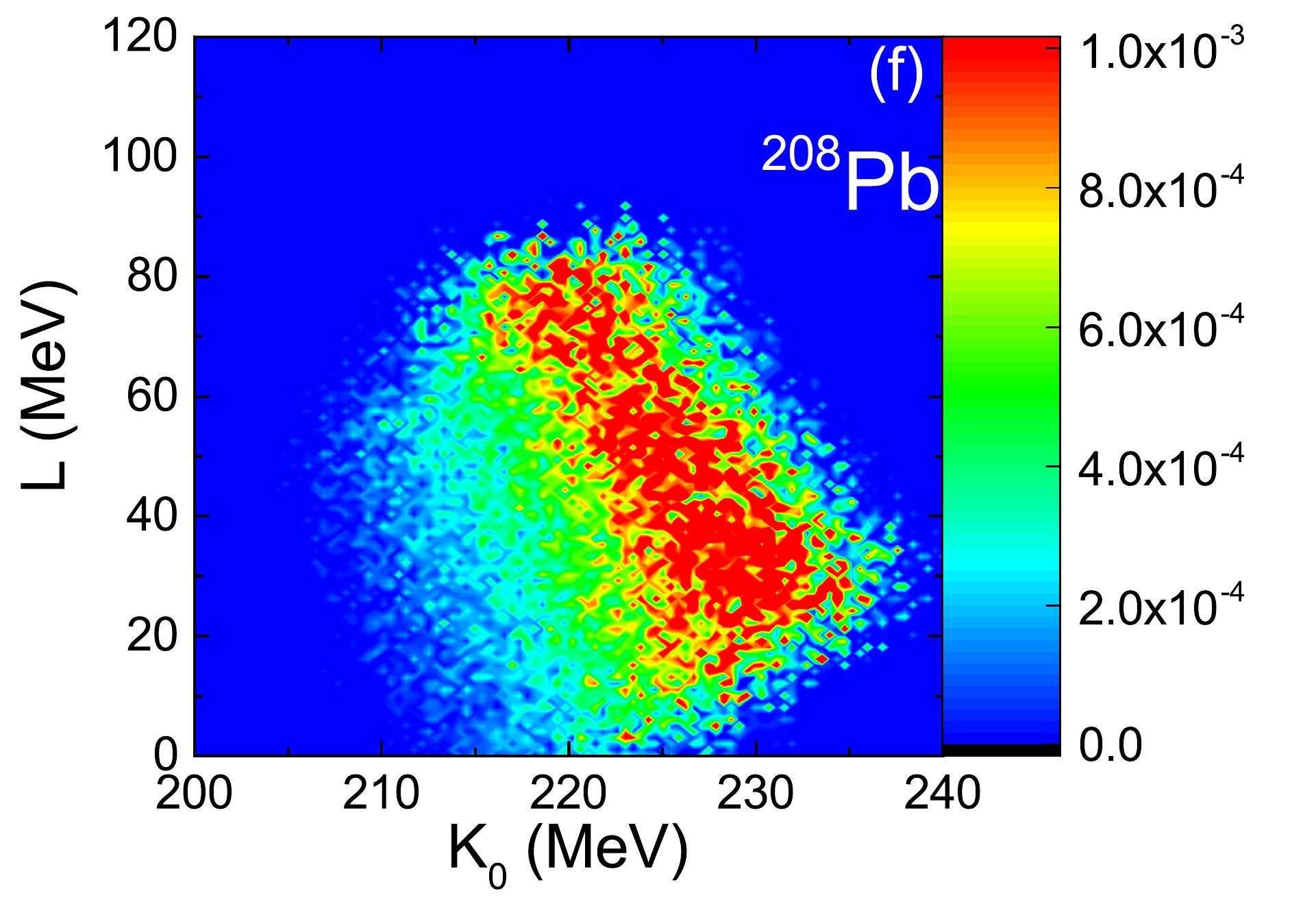}\includegraphics[scale=0.2]{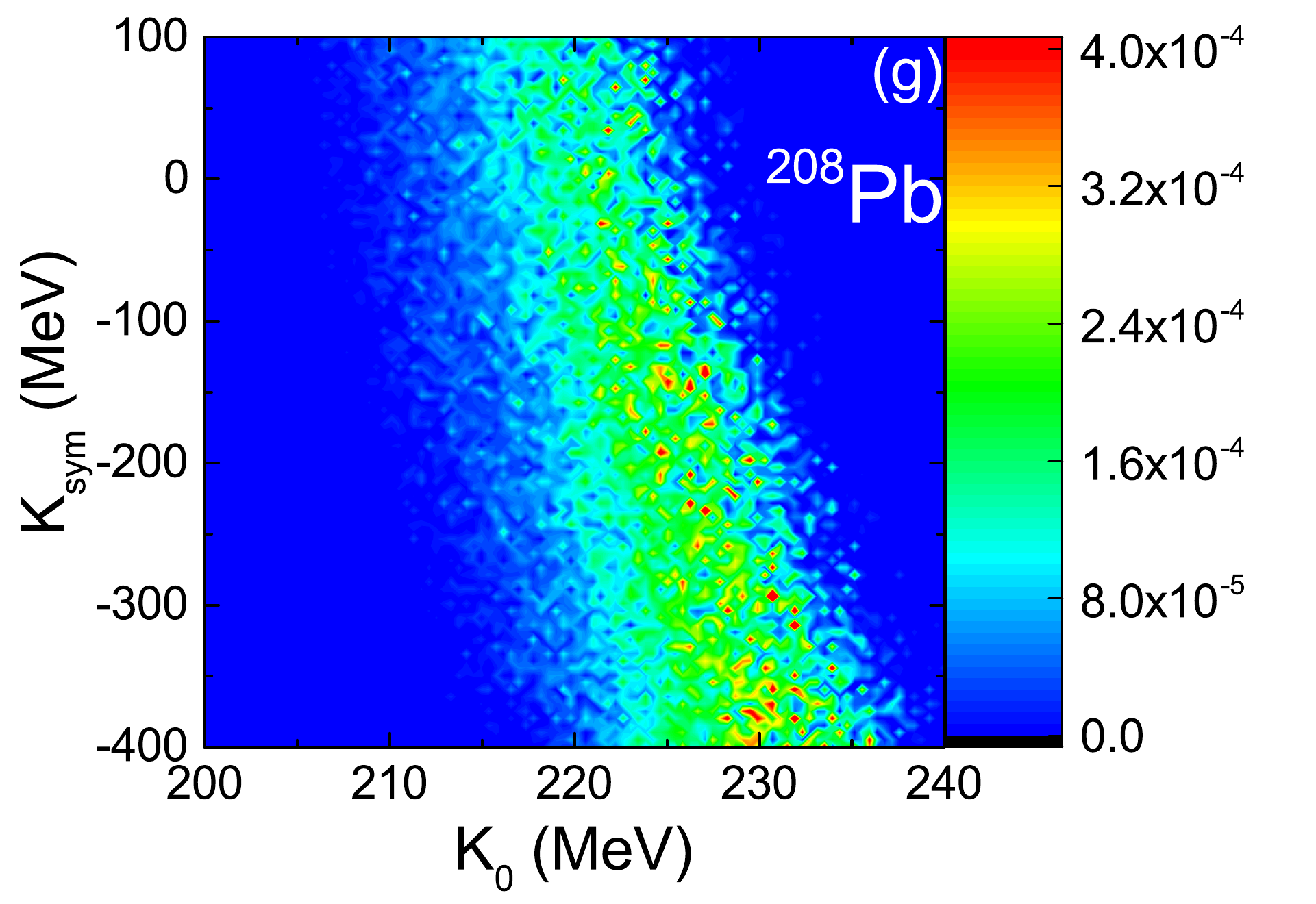}\includegraphics[scale=0.2]{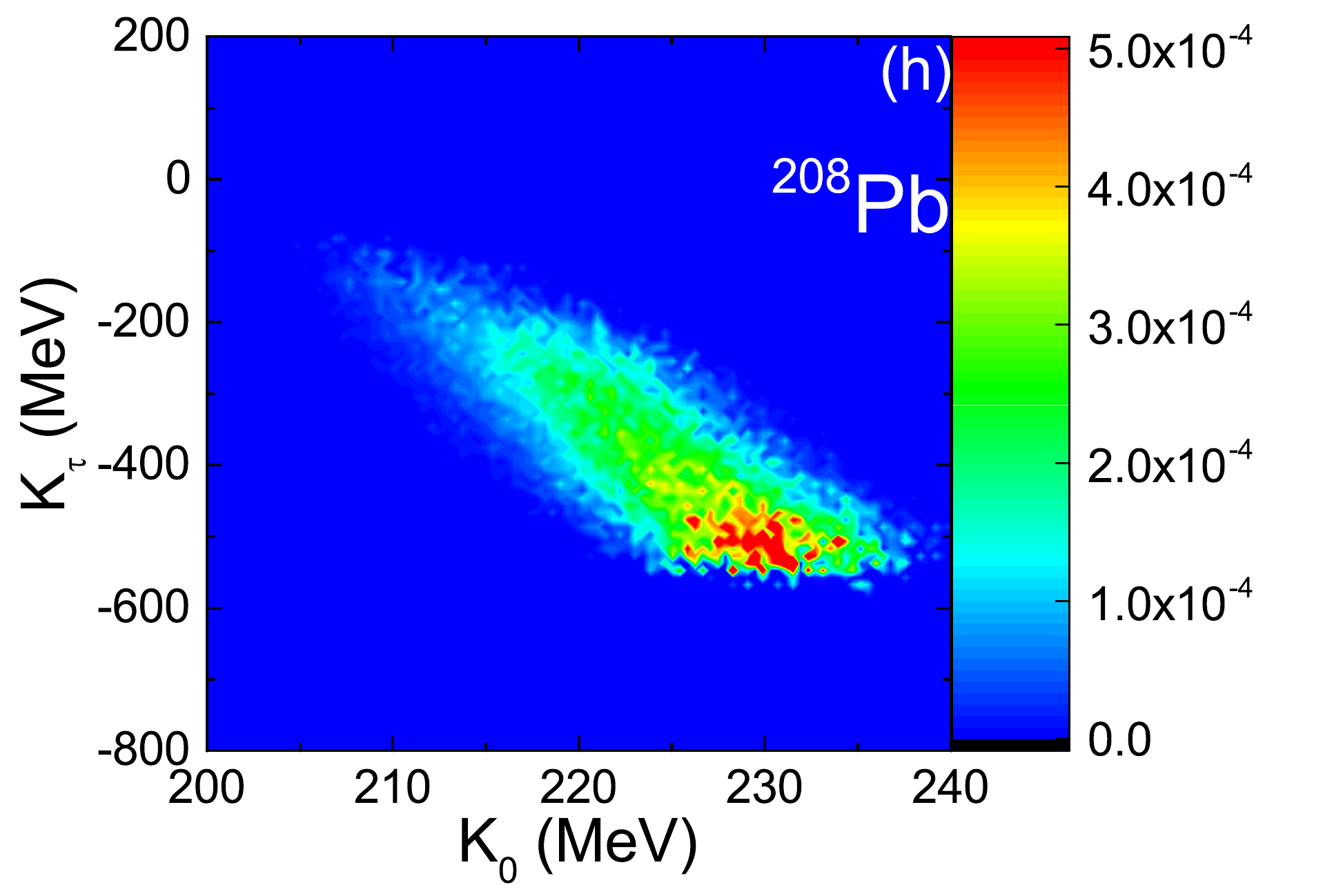}
	\caption{First row: Posterior correlated PDFs between $K_0$ and $L$, $E_{sym}^0$, $K_{sym}$, and $K_\tau$ under the constraints of $\Delta r_{np}$, $E_{-1}$, $\alpha_D$, and $E_{ISGMR}$ in $^{208}$Pb based on the standard SHF model; Second row: Same as the first row but based on the KIDS model.} \label{fig13}
\end{figure*}

\begin{figure*}[ht]
\includegraphics[scale=0.2]{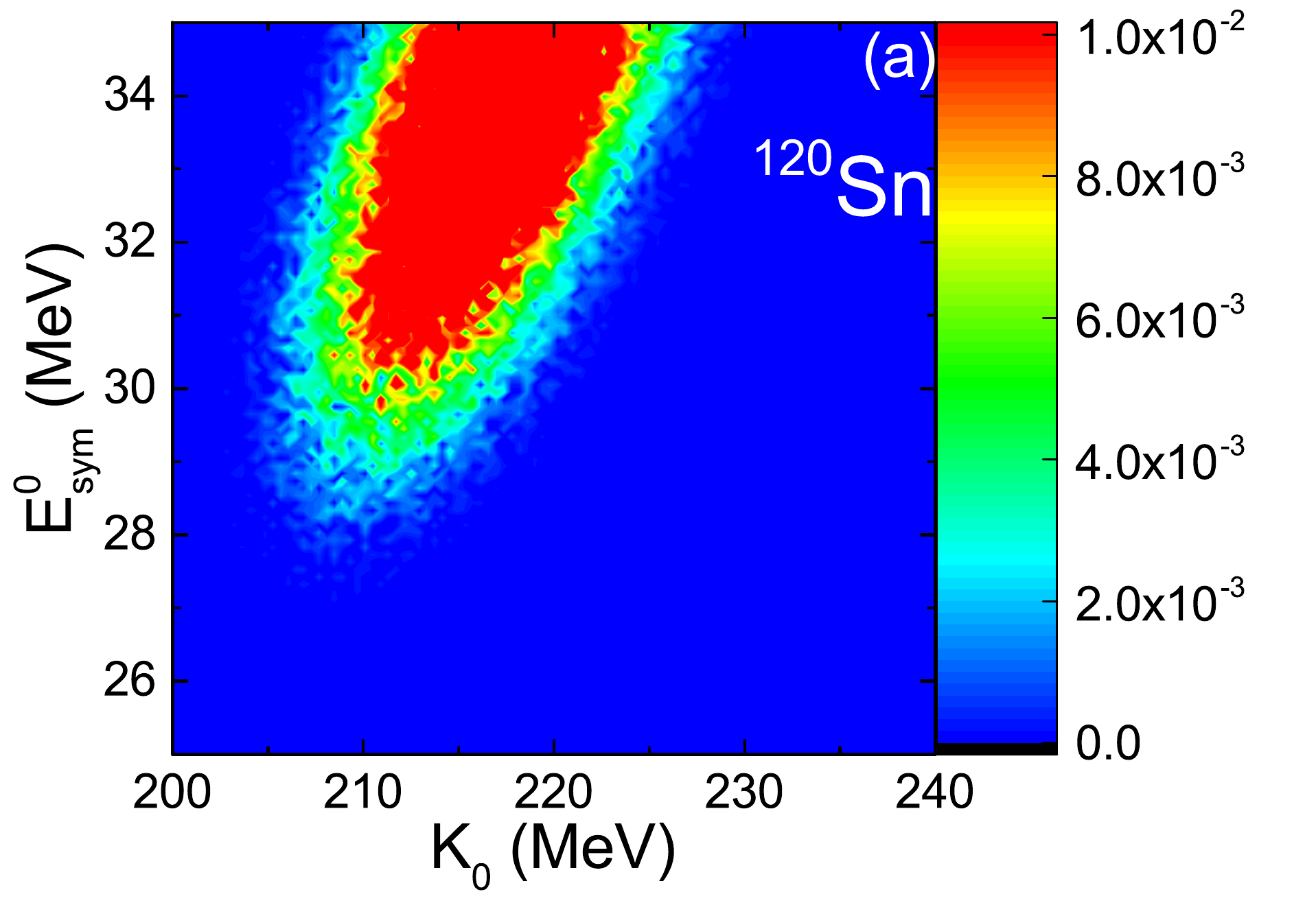}\includegraphics[scale=0.2]{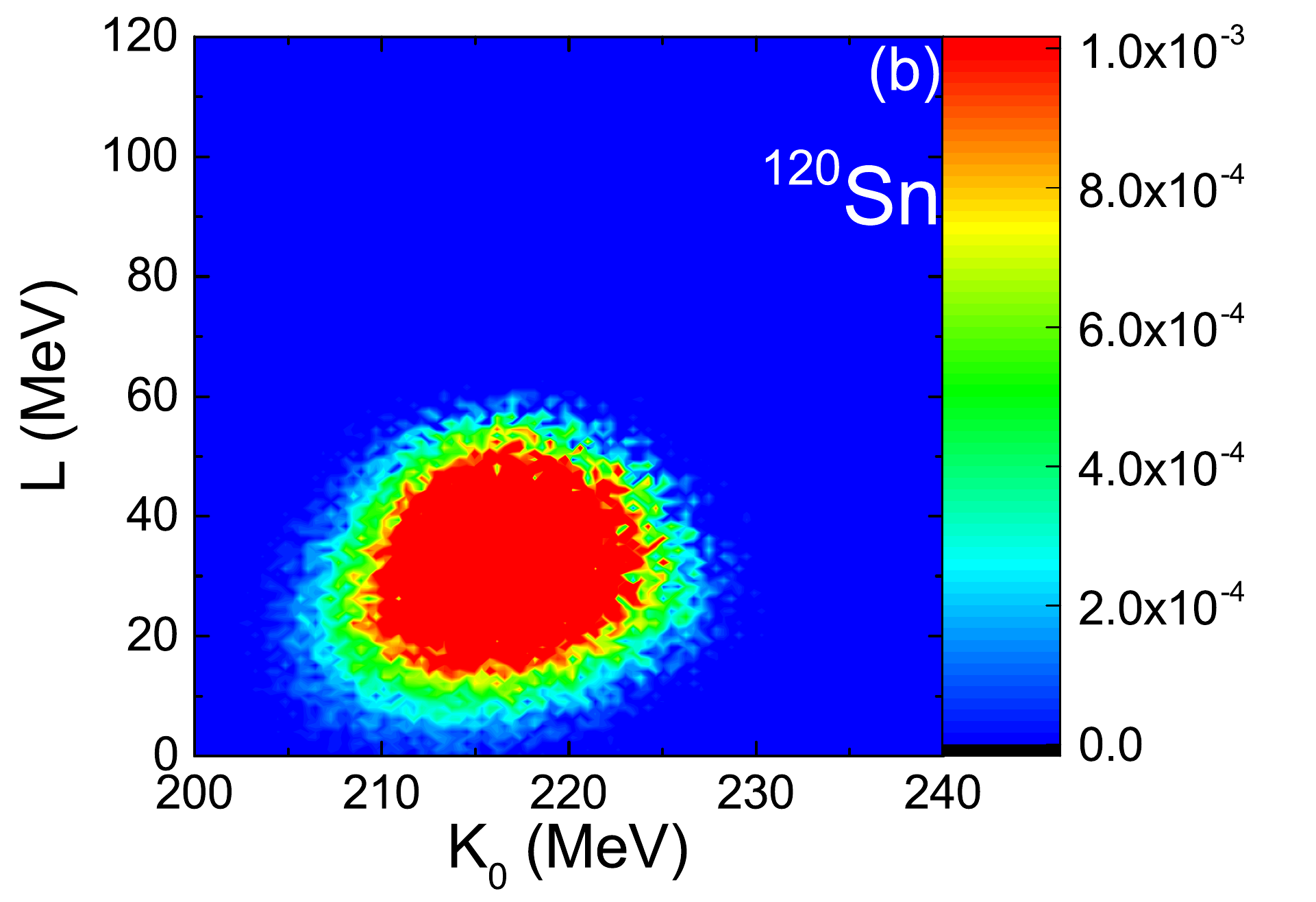}\includegraphics[scale=0.2]{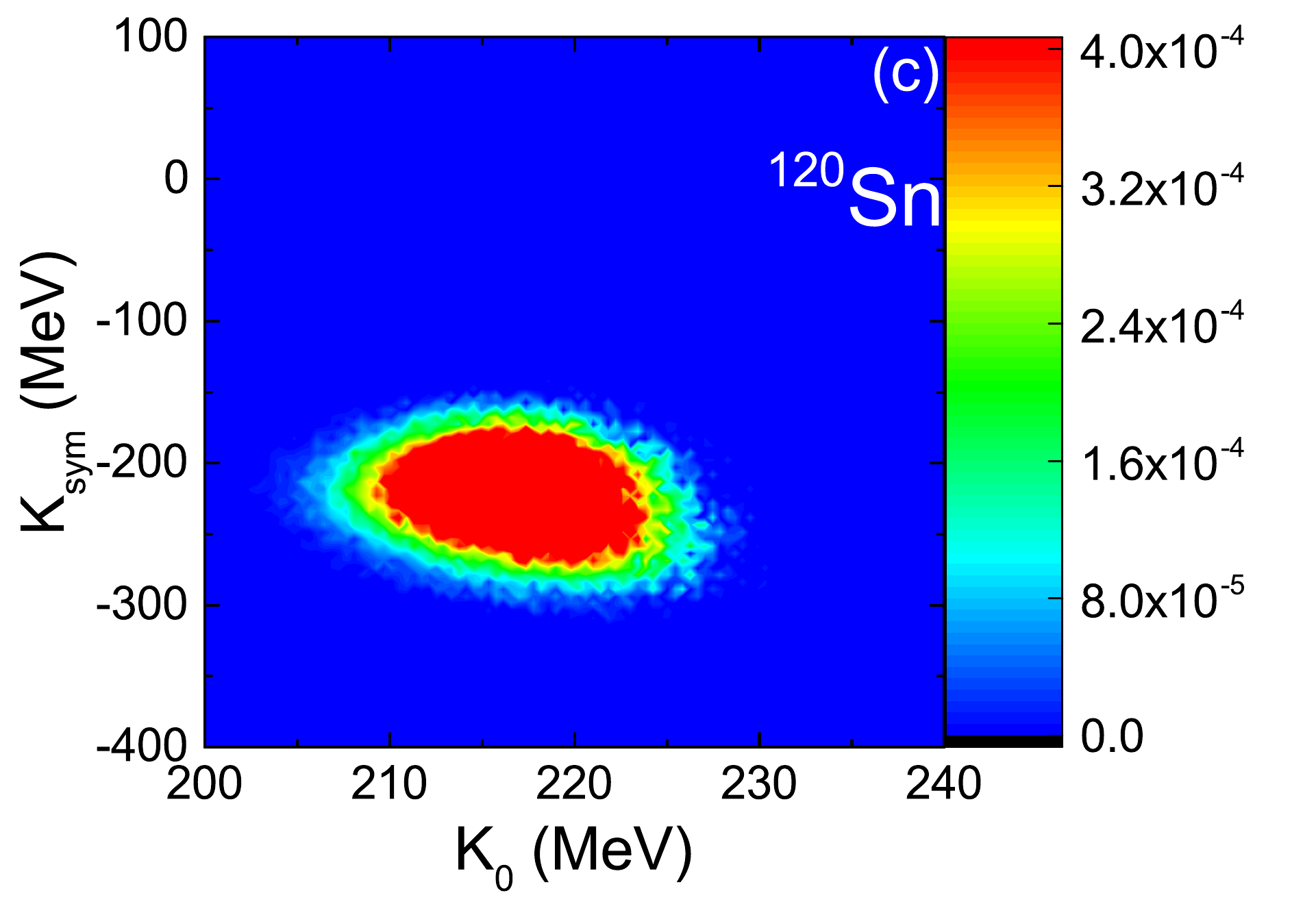}\includegraphics[scale=0.2]{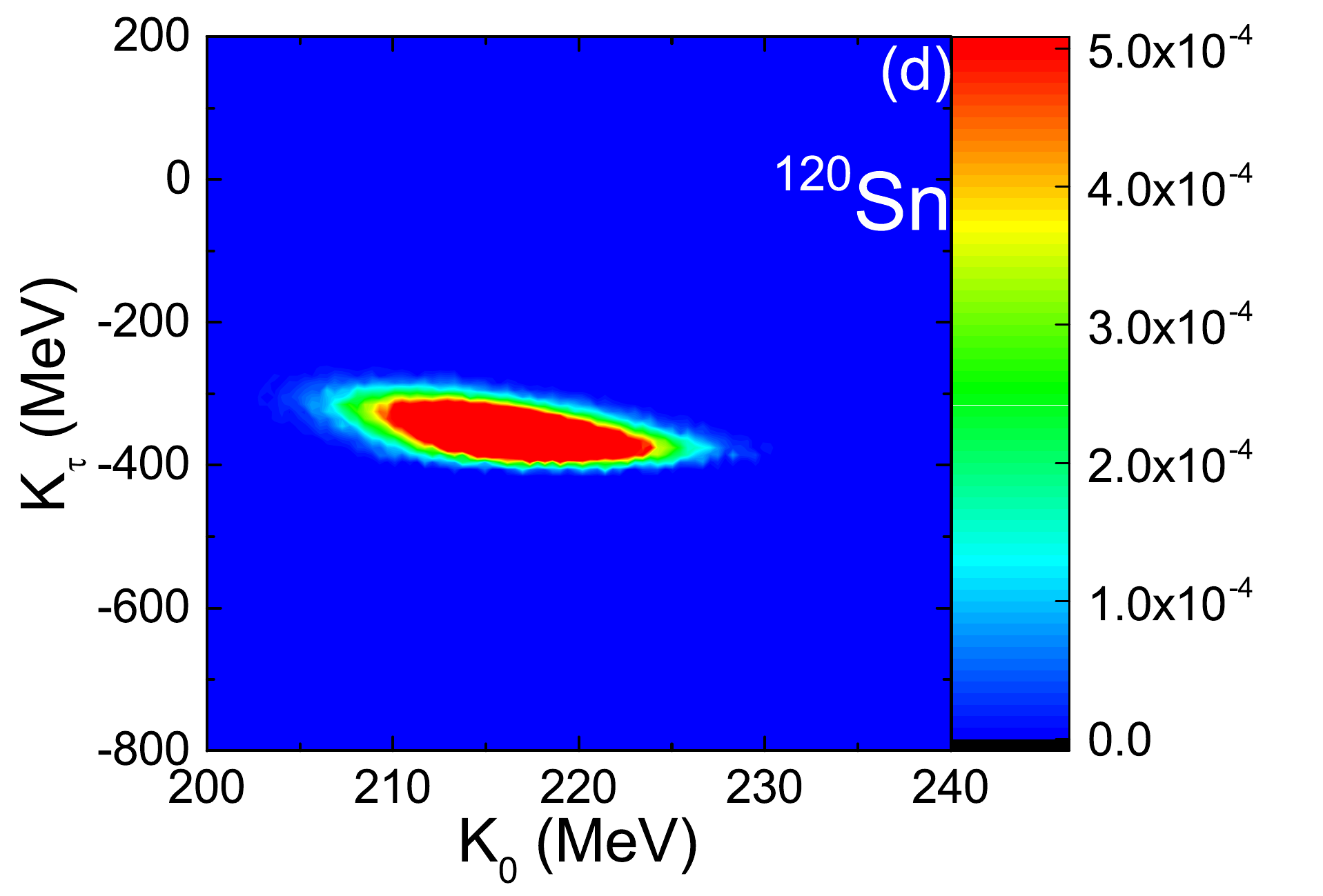}\\
\includegraphics[scale=0.2]{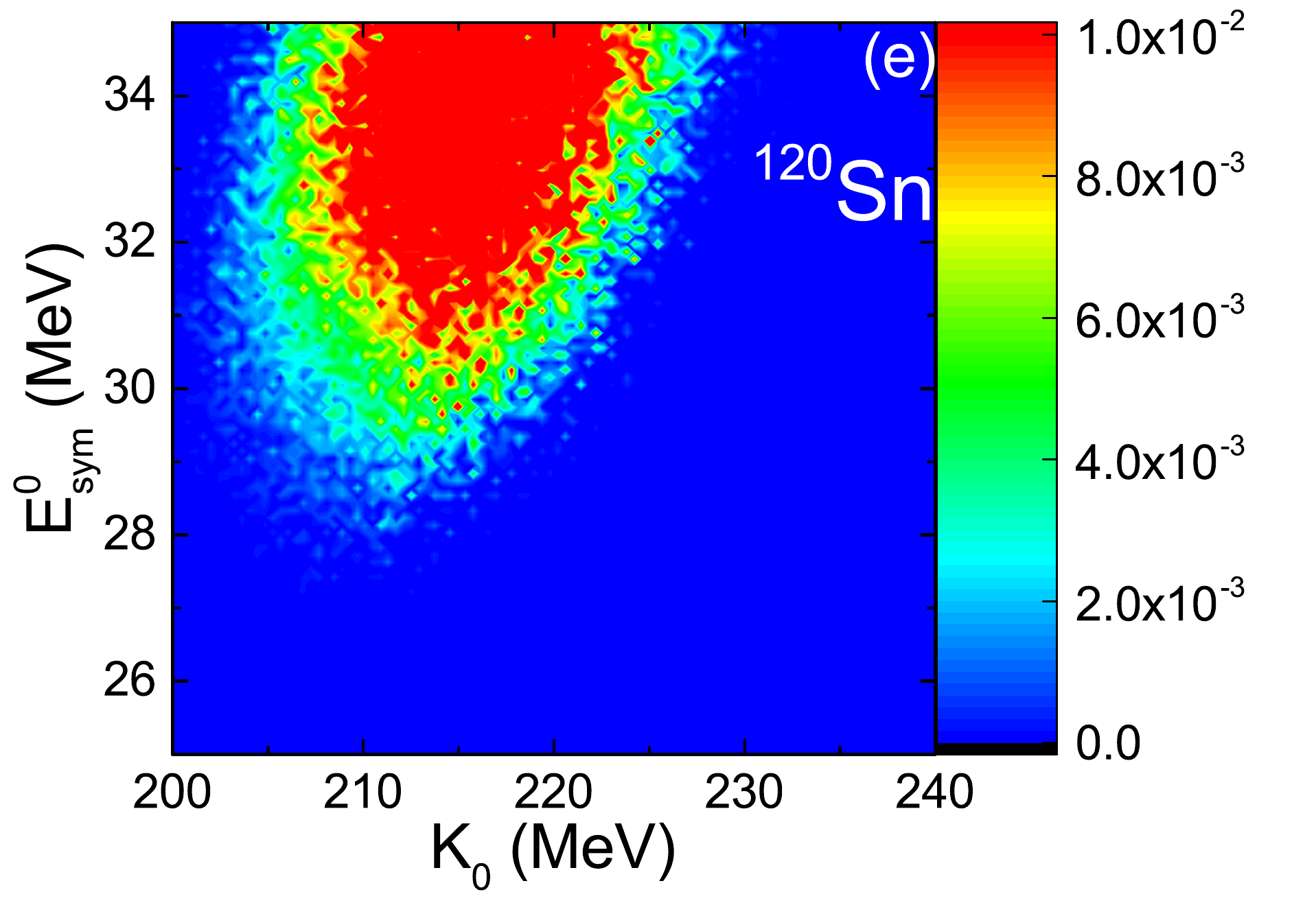}\includegraphics[scale=0.2]{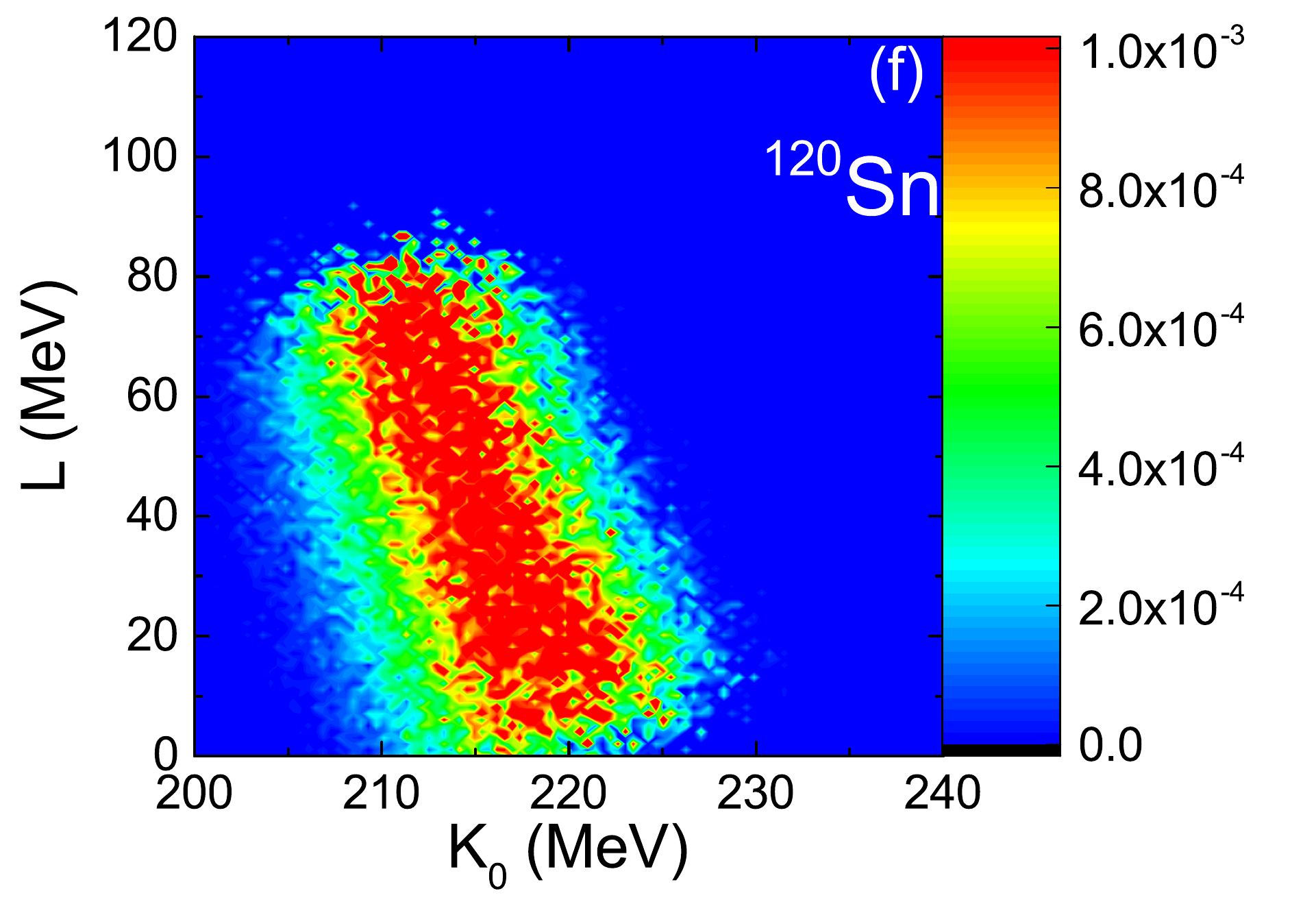}\includegraphics[scale=0.2]{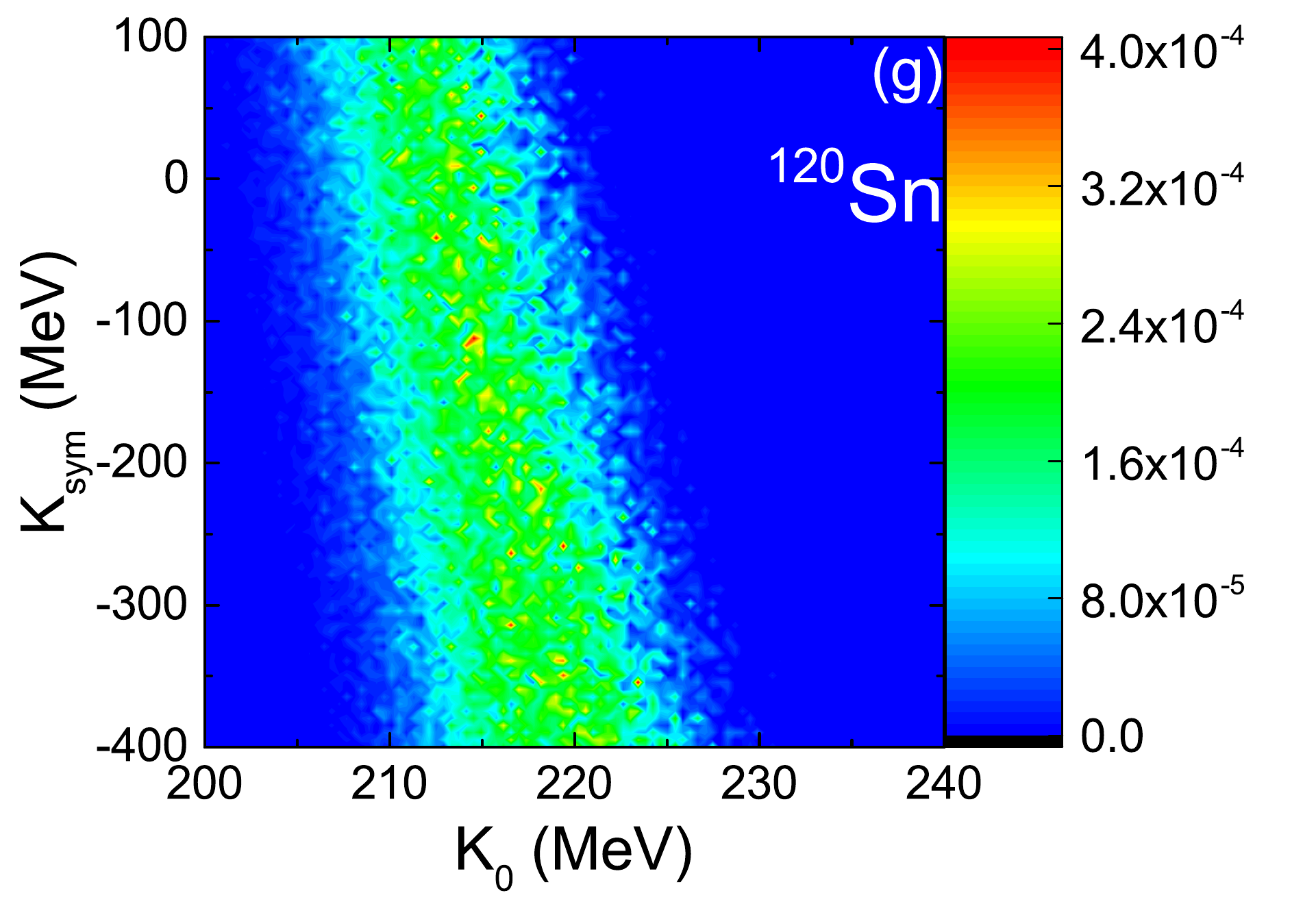}\includegraphics[scale=0.2]{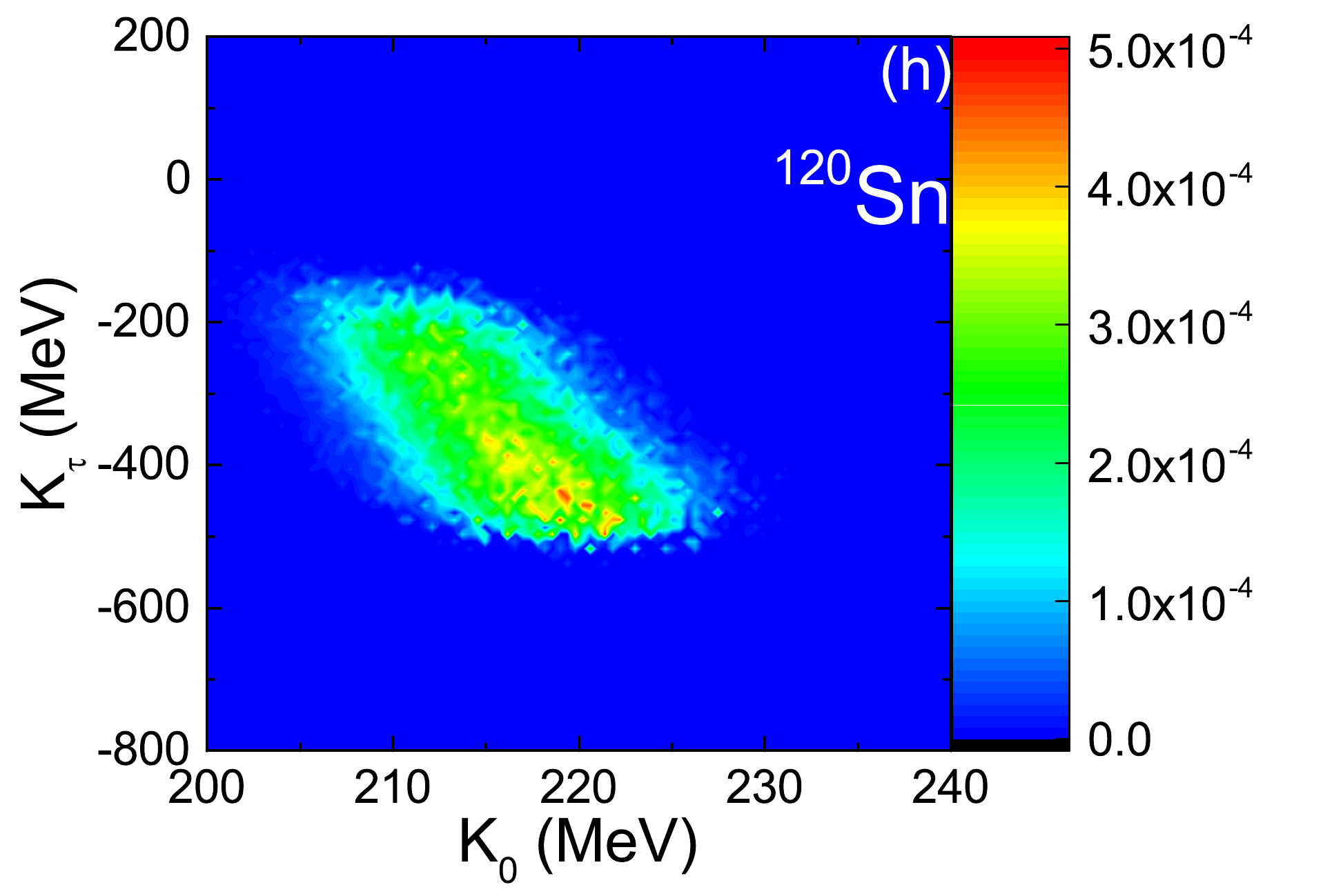}
	\caption{Same as Fig.~\ref{fig13} but under the constraints of the nuclear structure data of $^{120}$Sn.} \label{fig14}
\end{figure*}

We now further incorporate the ISGMR data and add the incompressibility $K_0$ as an independent variable in the Bayesian analysis. Since the excitation energy $E_{ISGMR}$ of ISGMR is rather insensitive to isovector parameters as shown in Figs.~\ref{fig2} and \ref{fig3}, the posterior PDFs of isovector parameters as well as their correlations are not much affected. On the other hand, since both $^{208}$Pb and $^{120}$Sn are neutron-rich nuclei, where $E_{ISGMR}$ is affected by both $K_0$ and $K_\tau$, one expects that there are correlations between $K_0$ and isovector EOS parameters~\cite{Xu21b}, and these correlations hamper us from constraining accurately $K_0$ from the ISGMR in neutron-rich nuclei~\cite{Col14}. Figures~\ref{fig13} and \ref{fig14} display the posterior correlated PDFs between $K_0$ and $L$, $E_{sym}^0$, $K_{sym}$, and $K_\tau$ under the constraints of $\Delta r_{np}$, $E_{-1}$, $\alpha_D$, and $E_{ISGMR}$ in $^{208}$Pb and $^{120}$Sn, respectively, based on the standard SHF and KIDS model. In the standard SHF model we vary $m_v^\star$, $E_{sym}^0$, $L$, and $K_0$ in the Bayesian analysis, while in the KIDS model we vary the additional $K_{sym}$ and $Q_{sym}$ as independent variables besides $m_v^\star$, $E_{sym}^0$, $L$, and $K_0$. With more independent variables, the PDFs are generally more diffusive. In addition, although the weak positive $K_0-E_{sym}^0$ correlations are observed in all scenarios, there are significant differences in the correlated PDFs in the $K_0-L$ plane and in the $K_0-K_{sym}$ plane in different scenarios. In the standard SHF model, weak positive $K_0-L$ and $K_0-K_{sym}$ correlations are observed for $^{208}$Pb, but there are almost no correlations between $K_0$ and $L$ for $^{120}$Sn. The stronger correlation in the $^{208}$Pb case is likely due to its larger isospin asymmetry compared with $^{120}$Sn. After incorporating $K_{sym}$ as an independent variable, negative $K_0-L$ and $K_0-K_{sym}$ correlations are observed for both $^{208}$Pb and $^{120}$Sn based on the KIDS model compared with those based on the standard SHF model, where the value of $K_{sym}$ depends on $L$, $E_{sym}^0$, etc. Interestingly, despite the different $K_0-L$ and $K_0-K_{sym}$ correlations in the standard SHF and KIDS model, both models give the similar $K_0-K_\tau$ correlations, though the correlated PDF based on the KIDS model is more diffusive compared with that based on the standard SHF model. The weak negative $K_0-K_\tau$ correlation under the constraint of $E_{ISGMR}$ is completely understandable, since both $K_0$ and $K_\tau$ contribute positively to $E_{ISGMR}$. We haven't observed nontrivial correlation between $K_0$ and $m_v^\star$ or $Q_{sym}$.

\begin{figure*}[ht]
\includegraphics[scale=0.4]{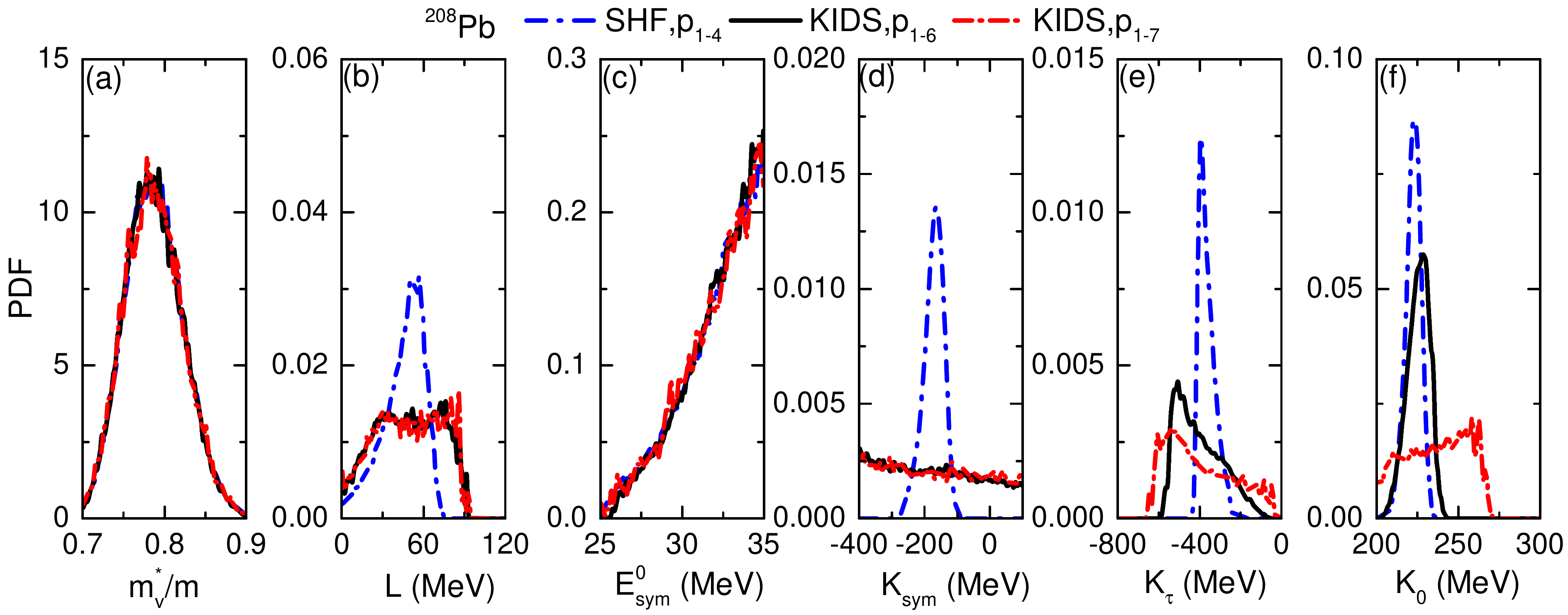}\\
\includegraphics[scale=0.4]{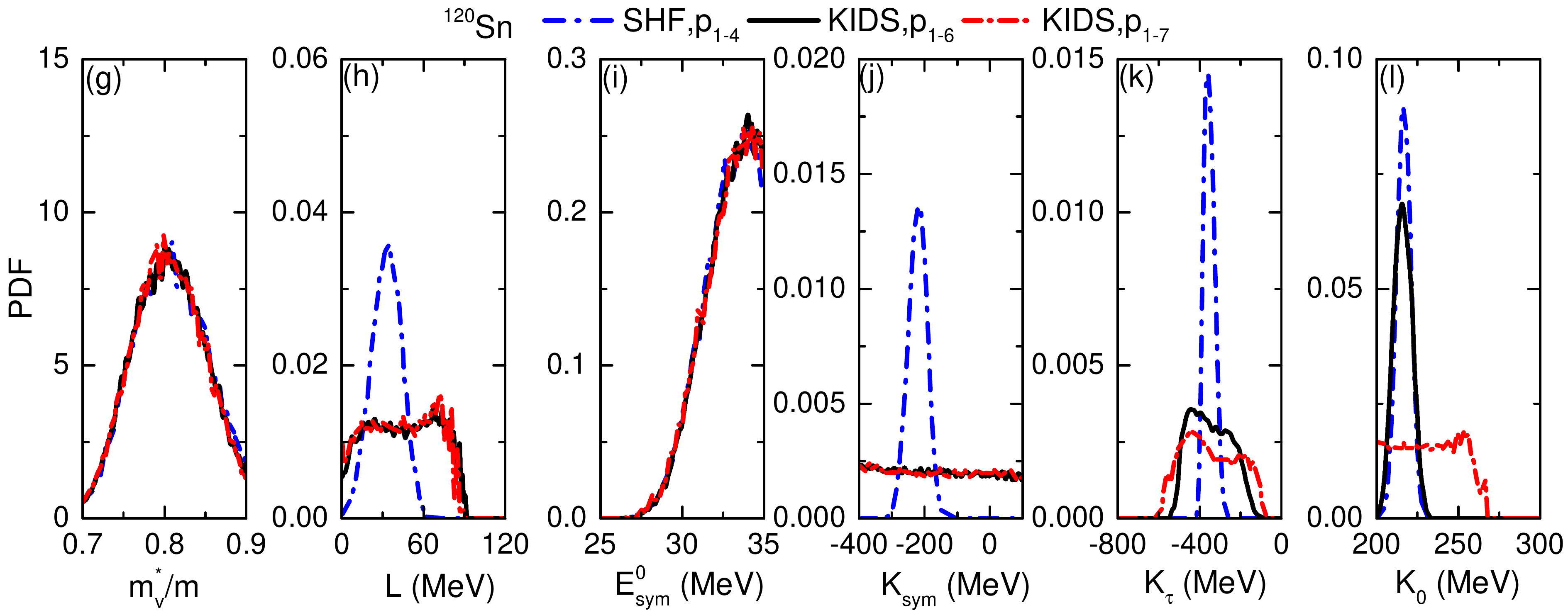}
	\caption{Upper: Posterior PDFs of $m_v^\star/m$, $L$, $E_{sym}^0$, $K_{sym}$, $K_\tau$, and $K_0$ under the constraints of $\Delta r_{np}$, $E_{-1}$, $\alpha_D$, and $E_{ISGMR}$ in $^{208}$Pb; Lower: Same as the upper panels but under the constraints of the nuclear structure data of $^{120}$Sn. Results from adjusting different numbers of independent variables based on the standard SHF model and the KIDS model are compared.} \label{fig15}
\end{figure*}

\begin{figure}[h]
\includegraphics[scale=0.21]{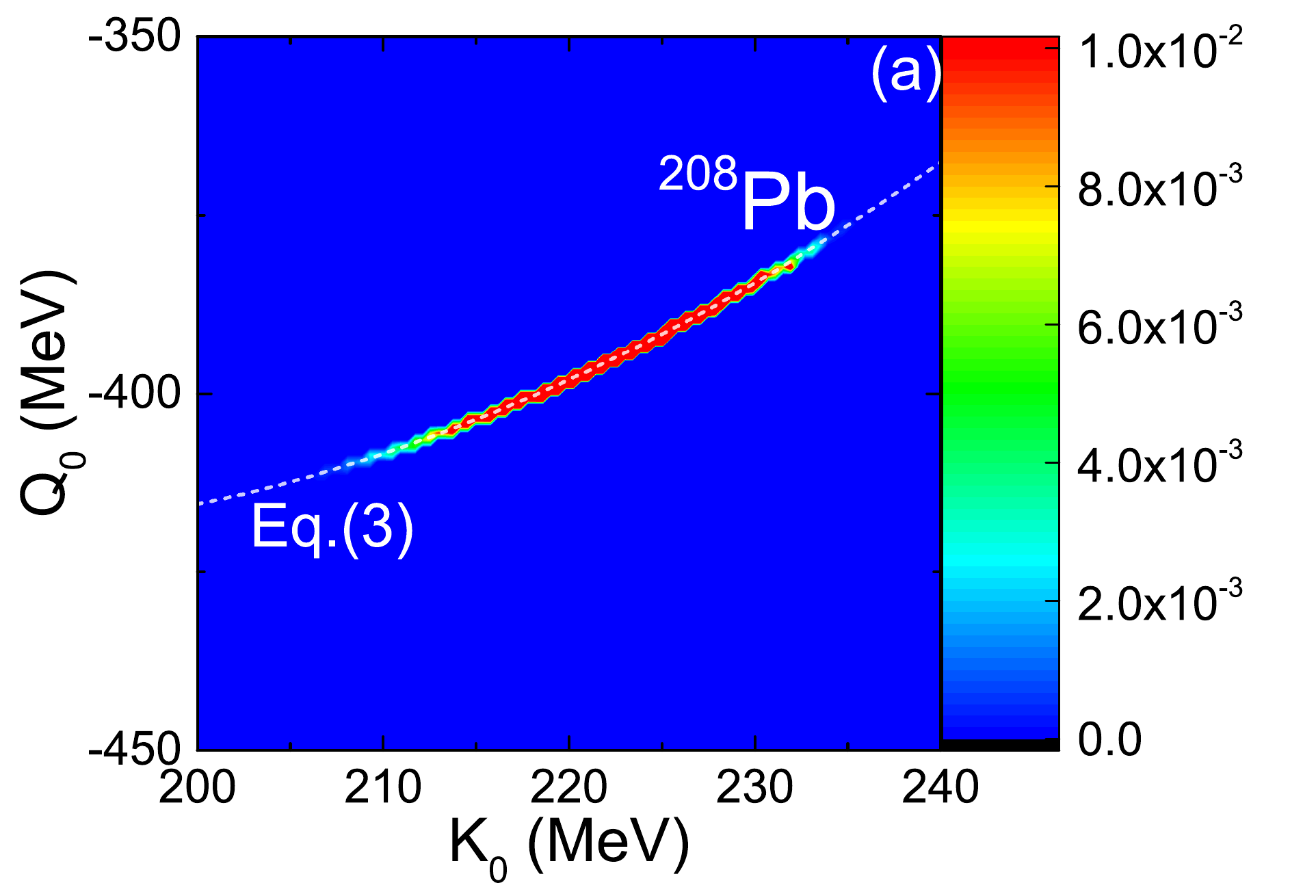}
\includegraphics[scale=0.21]{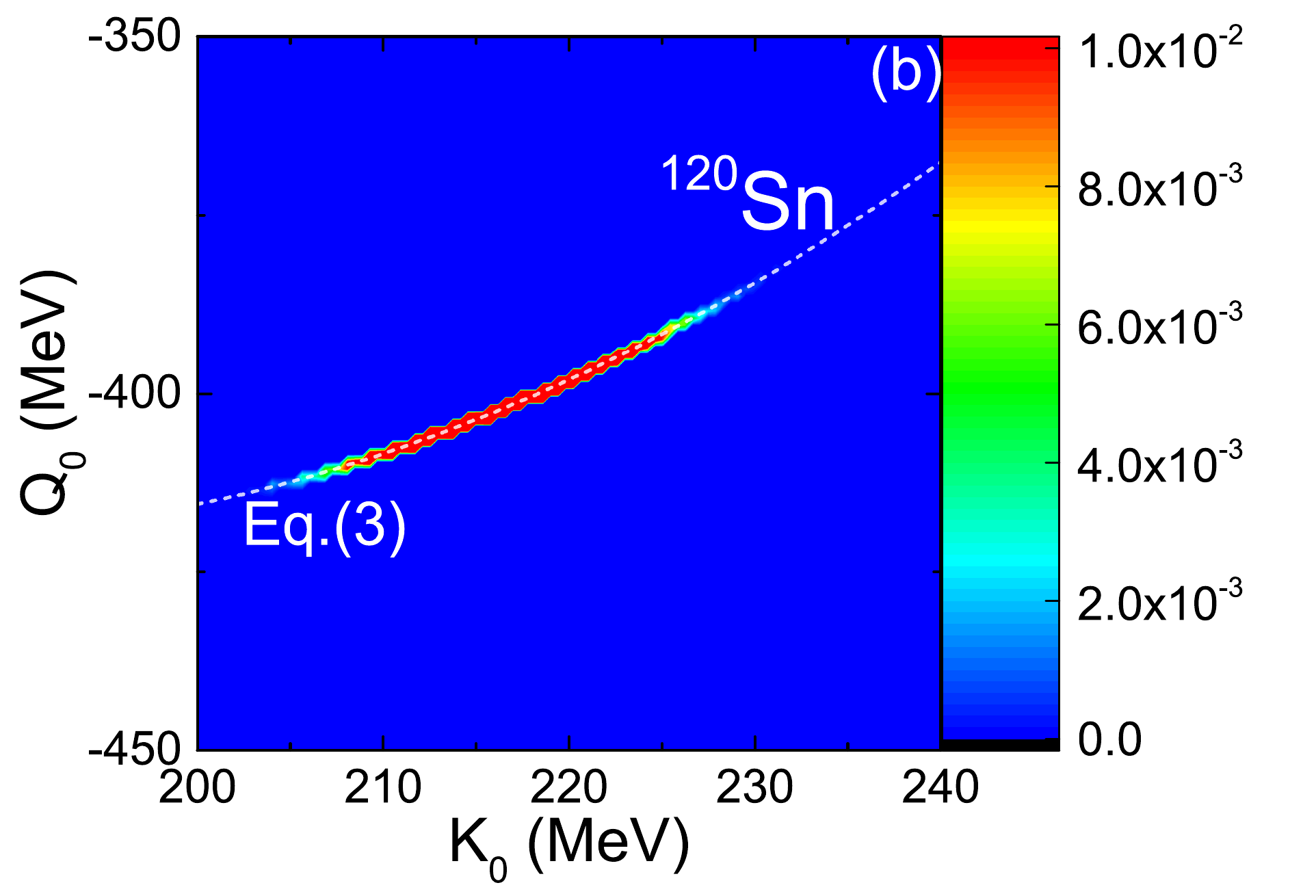}\\
\includegraphics[scale=0.21]{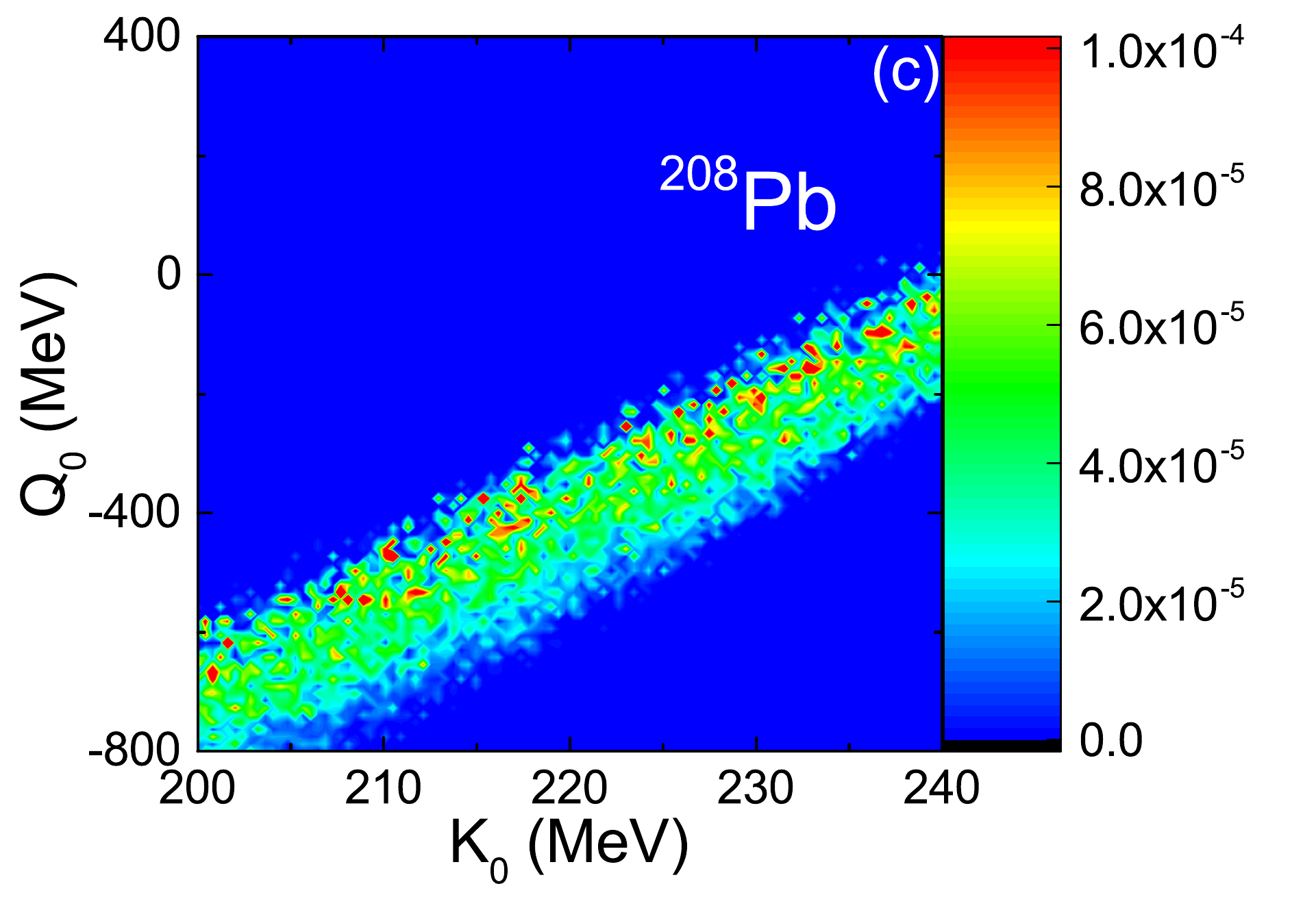}
\includegraphics[scale=0.21]{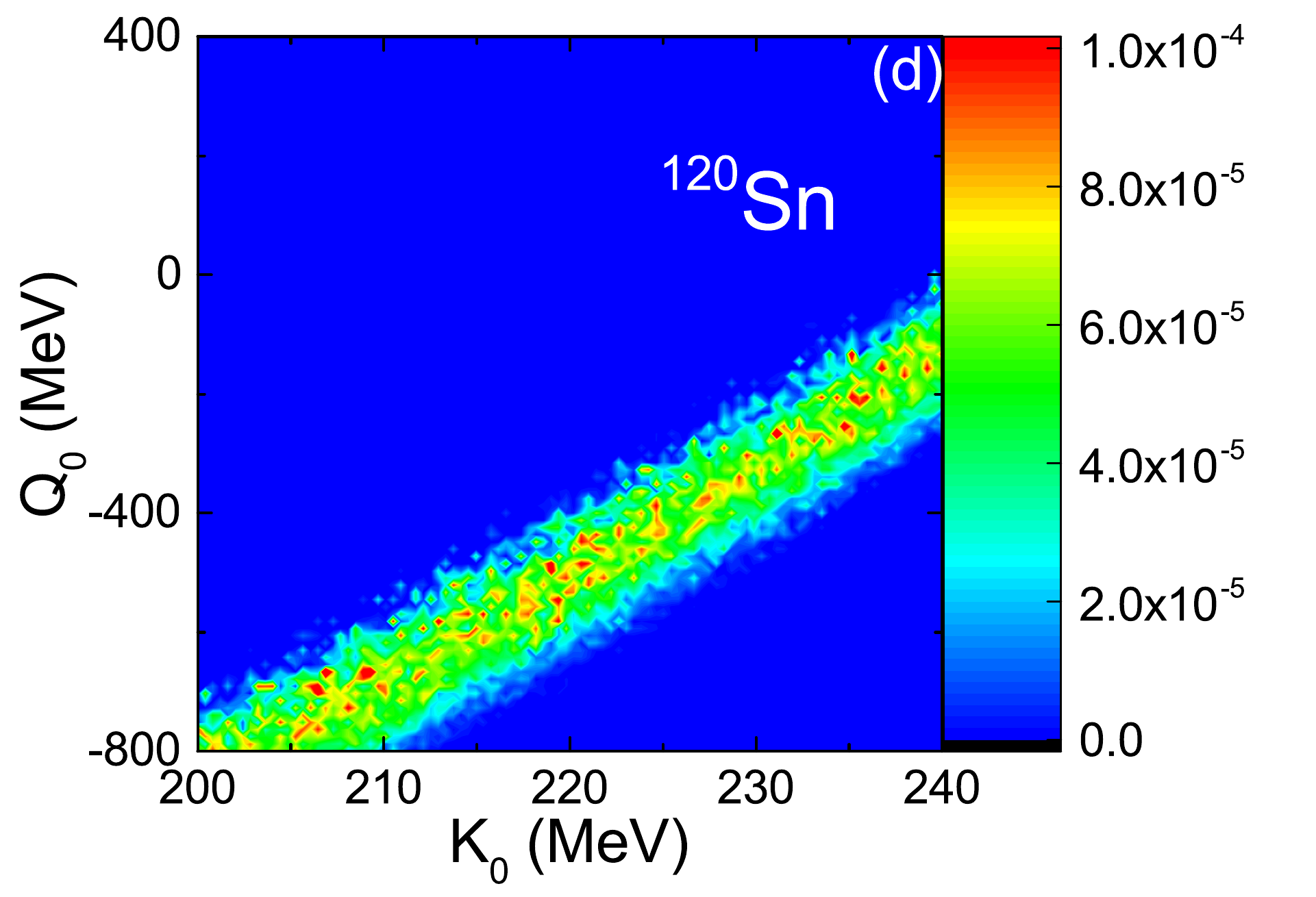}
	\caption{Posterior correlated PDFs between $K_0$ and $Q_0$ under the constraints of $\Delta r_{np}$, $E_{-1}$, $\alpha_D$, and $E_{ISGMR}$ in $^{208}$Pb (left) and $^{120}$Sn (right) based on the standard SHF (upper) and KIDS (lower) model.} \label{fig16}
\end{figure}

\begin{figure*}[ht]
\includegraphics[scale=0.6]{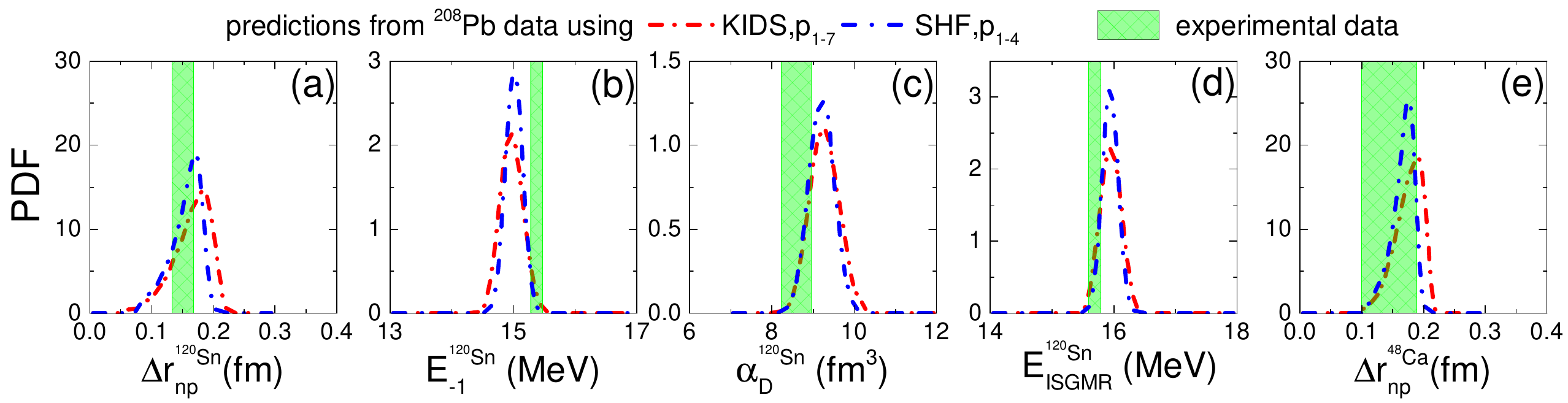}\\
\includegraphics[scale=0.6]{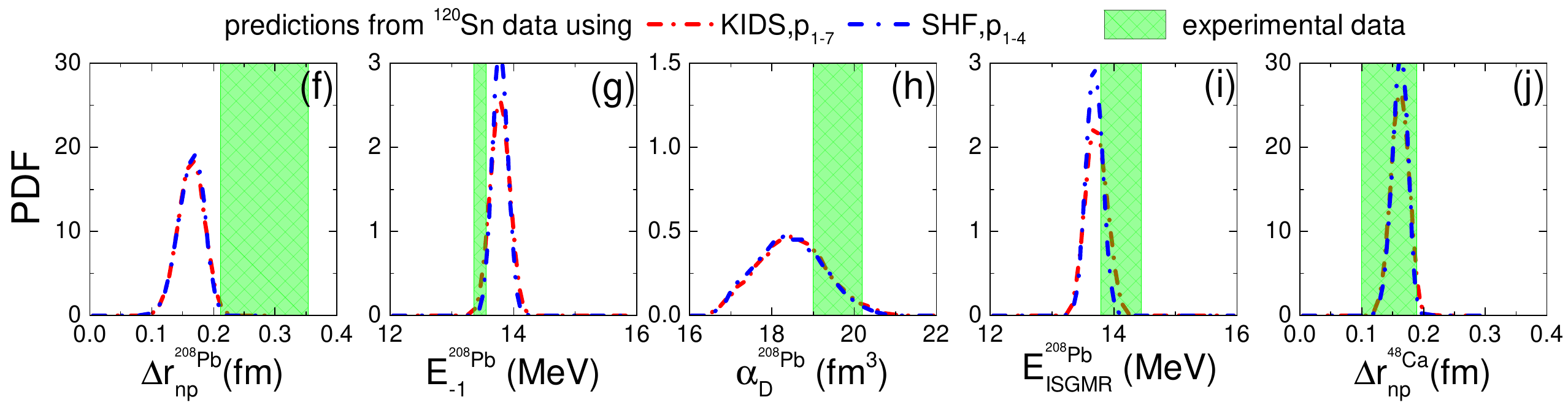}
	\caption{Upper: Predictions of $\Delta r_{np}$, $E_{-1}$, $\alpha_D$, and $E_{ISGMR}$ in $^{120}$Sn and $\Delta r_{np}$ in $^{48}$Ca from the posterior PDFs of seven independent physics variables in the KIDS model as well as those of four independent physics variables in the standard SHF model constrained by the $^{208}$Pb data; Lower: Predictions of $\Delta r_{np}$, $E_{-1}$, $\alpha_D$, and $E_{ISGMR}$ in $^{208}$Pb and $\Delta r_{np}$ in $^{48}$Ca from the posterior PDFs of seven independent physics variables in the KIDS model as well as those of four independent physics variables in the standard SHF model constrained by the $^{120}$Sn data. The corresponding experimental data shown by bands are compared.} \label{fig17}
\end{figure*}

The posterior PDFs of $m_v^\star/m$, $L$, $E_{sym}^0$, $K_{sym}$, $K_\tau$, and $K_0$ from integrating the other variable in the correlated PDFs in Fig.~\ref{fig13} and \ref{fig14} are displayed in Fig.~\ref{fig15}, and results obtained from the Bayesian analysis based on the standard SHF and KIDS model and under the constraints of nuclear structure data of $^{208}$Pb and $^{120}$Sn are compared. The PDFs of $m_v^\star/m$ and $E_{sym}^0$ are similar based on the standard SHF and KIDS model, while those of $L$ and $K_{sym}$ as well as the resulting $K_\tau$ depends on the EDF and the chosen independent variables. The posterior PDFs of these isovector variables, obtained after the ISGMR data are incorporated, are similar to those without incorporating the constraint of ISGMR, as shown in Fig.~\ref{fig12}. For the obtained PDF of $K_0$, both the standard SHF and KIDS model give the overall larger values from the ISGMR data of $^{208}$Pb compared with $^{120}$Sn. This is qualitatively consistent with the "soft Tin puzzle" mentioned in the introduction, while significant overlaps in the PDFs of $K_0$ obtained for $^{208}$Pb and $^{120}$Sn are observed, especially based on the KIDS model that gives a more diffusive PDF.

As shown in Fig.~\ref{fig2}, since the $E_{ISGMR}$ is also sensitive to the skewness EOS parameter $Q_0$ of SNM, $Q_0$ should be varied in the Bayesian analysis based on the KIDS model to get a more reliable posterior PDFs of physics quantities. Again, incorporating $Q_0$ affects mostly the isoscalar EOS parameters and the corresponding correlations. Figure~\ref{fig16} displays the posterior correlated PDFs between $K_0$ and $Q_0$ based on the nuclear structure data of $^{208}$Pb and $^{120}$Sn in this most complete scenario. For the standard SHF model, $Q_0$ is not an independent variable, but can be obtained from other parameters through Eq.~(\ref{Q0shf}), and such intrinsic relation before being confronted with the data is also compared in Fig.~\ref{fig16}. One sees that the correlated PDFs overlap with the curve from Eq.~(\ref{Q0shf}) in a certain range in the standard SHF model, where both $Q_0$ and $K_0$ are significantly constrained mostly from the $E_{ISGMR}$ data. Interesting, without such intrinsic relation as Eq.~(\ref{Q0shf}), the KIDS model gives almost linear positive correlations between $Q_0$ and $K_0$ under the constraint of $E_{ISGMR}$. This can be understood from Fig.~\ref{fig2}, where $E_{ISGMR}$ in both $^{208}$Pb and $^{120}$Sn increases with increasing $K_0$ (decreasing $Q_0$). It is interesting to see that the correlated PDF for $^{120}$Sn is shifted slightly to the lower part but with a similar slope compared with that for $^{208}$Pb.

The final posterior PDFs of $m_v^\star/m$, $L$, $E_{sym}^0$, $K_{sym}$, $K_\tau$, and $K_0$ with seven independent physics variables adjusted within their prior ranges in Table \ref{T1} based on the KIDS model are included in Fig.~\ref{fig15} for both $^{208}$Pb and $^{120}$Sn. Again, the PDFs of isovector parameters $m_v^\star/m$, $L$, $E_{sym}^0$, and $K_{sym}$ are similar to those without incorporating $Q_0$. Although not shown here, the nuclear structure data considered here are unable to constrain $Q_0$, like other higher-order EOS parameters $K_{sym}$ and $Q_{sym}$. On the other hand, incorporating $Q_0$ significantly broadens the PDF of $K_0$, and the PDF of $K_\tau$ is also affected according to Eq.~(\ref{ktau}).

A more complete calculation would involve incorporating at the same time the constraints from both $^{208}$Pb and $^{120}$Sn. However, that would be extremely time-consuming, and not necessarily more illuminating. One expects that the resulting PDFs are roughly close to the average of those from the separate constraints of $^{208}$Pb and $^{120}$Sn, depending on the relative values of $\sigma_i$ in the likelihood function [Eq. (\ref{llh})]. With this limitation in mind, in order to show to what extent the 'soft Tin puzzle' and the 'PREXII puzzle' mentioned in the introduction are resolved, we display in Fig.~\ref{fig17} the predictions of $\Delta r_{np}$, $E_{-1}$, $\alpha_D$, and $E_{ISGMR}$ in $^{120}$Sn from the posterior PDFs of seven independent physics variables in the KIDS model (the scenario of 'KIDS, $p_{1-7}$' in Fig.~\ref{fig15}) as well as those of four independent physics variables in the standard SHF model (the scenario of 'SHF, $p_{1-4}$' in Fig.~\ref{fig15}) constrained by the $^{208}$Pb data, and vice versa. Predictions on the $\Delta r_{np}$ in $^{48}$Ca are displayed in both cases, and the corresponding experimental data for all observables shown by bands are compared. One sees that the posterior PDFs of physics variables from the $^{208}$Pb data predict compatible $\Delta r_{np}$ in $^{120}$Sn, while those from the $^{120}$Sn data underpredict the $\Delta r_{np}$ in $^{208}$Pb. The compatibility in the former case is because the posterior PDFs of isovector model parameters from the Bayesian analysis are dominated by the more accurate IVGDR data of $^{208}$Pb, which favors a softer symmetry energy, rather than the less accurate $\Delta r_{np}$ data of $^{208}$Pb, which favors a stiffer symmetry energy. This shows the effect of incorporating additional constraints compared to Fig.~\ref{fig7}. In addition, the posterior PDFs of physics variables from the $^{208}$Pb data underpredict the $E_{-1}$ in $^{120}$Sn, and those from the $^{120}$Sn data overpredict the $E_{-1}$ in $^{208}$Pb. In other cases, there are appreciable overlaps between the predicted values and the experimental data. The standard SHF model gives very similar predictions with slightly sharper distributions for these observables, compared to those from the KIDS model.

\section{Conclusions}

Based on the KIDS model and using the Bayesian approach, we have obtained the posterior PDFs of physics quantities of interest under the constraints of the neutron-skin thickness, the IVGDR, and the ISGMR data. In the Bayesian analysis, we gradually increase the number of constraints and independent physics variables, to understand where the correlation between physics quantities as well as their PDFs come from. Results are compared with those obtained based on the standard SHF model in order to understand the influence of choosing different independent variables, and those obtained under the different constraints of nuclear structure data of $^{208}$Pb and $^{120}$Sn are also compared.

It is seen that incorporating $K_{sym}$ as an independent variable can significantly broaden the posterior PDFs of $L$ and $K_\tau$. $K_{sym}$ cannot be constrained at all from nuclear data within the KIDS model, although we should note that it can be constrained from astronomical observations, which probe higher densities, with existing data strongly suggesting $-200~\text{MeV} < K_{sym} <0$~\cite{Gil21}. A positive $K_{sym}-L$ correlation based on the KIDS model, instead of the negative $E_{sym}^0-L$ correlation obtained based on the standard SHF model, is observed under the constraint of the neutron-skin thickness. In the isoscalar channel, a positive $K_0-Q_0$ correlation is observed, and incorporating $Q_0$ as an independent variable significantly broadens the PDF of $K_0$ and also affects that of $K_\tau$. In this sense, although the nuclear structure data studied here are good probes of $L$ and $K_0$, the large uncertainty ranges of $K_{sym}$ and $Q_0$ hampers us from constraining accurately the corresponding lower-order EOS parameters. This is different from the standard SHF model, where $K_{sym}$ and $Q_0$ can be well constrained by the same nuclear structure data, mainly because $K_{sym}$ and $Q_0$ can be generally expressed in terms of lower-order EOS parameters. Considering the empirical uncertainty ranges of higher-order EOS parameters, we obtained robust constraints of $L<90$ MeV and $K_0<270$ MeV based on the KIDS model, serving as a baseline to rule out unreasonable parameterizations.

Finally, we have addressed the ``PREXII puzzle" and the ``soft Tin puzzle" quantitatively, by comparing the overlaps of PDFs of $L$ under different constraints as well as those of $K_0$, and predictions of observables using posterior PDFs of physics quantities with the corresponding experimental data. With the posterior PDFs of physics quantities under the constraints of the neutron-skin thickness of $^{208}$Pb from PREXII only, we obtain broad predictions and thus significant overlaps with the data of the neutron-skin thickness of $^{120}$Sn and $^{48}$Ca as well as the IVGDR in $^{208}$Pb. Using the posterior PDFs of physics quantities from more complete nuclear structure data of $^{208}$Pb or $^{120}$Sn, predictions are mostly compatible with the corresponding experimental data though there are exceptions. Predictions from posterior PDFs of physics quantities from all $^{208}$Pb data underestimate the centroid energy of IVGDR in $^{120}$Sn, while predictions from posterior PDFs of physics quantities from all $^{120}$Sn data underestimate the neutron-skin thickness but overestimate the centroid energy of IVGDR in $^{208}$Pb. This shows that the ``PREXII puzzle" remains an issue. On the other hand, the significant overlaps between the PDFs of $K_0$ from the ISGMR in $^{208}$Pb and $^{120}$Sn as well as the compatibility between predictions and ISGMR data indicate that one can find a compromise for the ``soft Tin puzzle". The next challenge would be to make use of the newly revealed correlations and PDFs in order to explore whether indeed a KIDS EDF can simultaneously reproduce the seemingly conflicting data examined in this work. It will also be interesting to incorporate both constraints from nuclear structure data and astrophysical observables in the Bayesian analysis, and thus further constrain model parameters characterizing the nuclear matter EOS from low to high densities.

\appendix

\section{Relation between macroscopic quantities and model parameters in KIDS}
\label{app}

The binding energy per nucleon in isospin asymmetric nuclear matter with nucleon density $\rho=\rho_n+\rho_p$ and isospin asymmetry $\delta = (\rho_n-\rho_p)/\rho$ can be expressed as
\begin{equation}
E(\rho,\delta) = E(\rho,0) + E_{sym}(\rho) \delta^2 + O(\delta^4),
\end{equation}
where the symmetry energy is defined as
\begin{equation}
E_{sym}(\rho) = \frac{1}{2} \left[\frac{\partial^2 E(\rho,\delta)}{\partial \delta^2}\right]_{\delta=0}.
\end{equation}
Around the saturation density $\rho_0$, $E(\rho,0)$ and $E_{sym}(\rho)$ can be expanded in the power of $\chi = \frac{\rho-\rho_0}{3\rho_0}$ as
\begin{eqnarray}
E(\rho,0) &=& E(\rho_0,0) + \frac{K_0}{2!} \chi^2 + \frac{Q_0}{3!} \chi^3 + O(\chi^4), \nonumber \\
E_{sym}(\rho) &=& E_{sym}(\rho_0) + L \chi + \frac{K_{sym}}{2!} \chi^2 + \frac{Q_{sym}}{3!} \chi^3 + O(\chi^4). \nonumber
\end{eqnarray}
In the above, the linear term in the expansion of $E(\rho,0)$ vanishes due to zero pressure of SNM at $\rho_0$. The independent EOS parameters in the KIDS model are the saturation density $\rho_0$, the binding energy $E_0$, the incompressibility $K_0$, and the skewness parameter $Q_0$ of SNM at $\rho_0$, the symmetry energy $E_{sym}^0$ and its slope parameter $L$, curvature parameter $K_{sym}$, and skewness parameter $Q_{sym}$ at $\rho_0$, and they are defined respectively as
\begin{eqnarray}
&&\left[\frac{\partial E(\rho,0)}{\partial \rho}\right]_{\rho=\rho_0} = 0,\\
&&E_0 \equiv E(\rho_0,0),\\
&&K_0 = 9\rho_0^2 \left[\frac{\partial^2 E(\rho,0)}{\partial \rho^2}\right]_{\rho=\rho_0},\\
&&Q_0 = 27\rho_0^3 \left[\frac{\partial^3 E(\rho,0)}{\partial \rho^3}\right]_{\rho=\rho_0},\\
&&E_{sym}^0 \equiv E_{sym}(\rho_0),\\
&&L = 3\rho_0 \left[\frac{\partial E_{sym}(\rho)}{\partial \rho}\right]_{\rho=\rho_0},\\
&&K_{sym} = 9\rho_0^2 \left[\frac{\partial^2 E_{sym}(\rho)}{\partial \rho^2}\right]_{\rho=\rho_0},\\
&&Q_{sym} = 27\rho_0^3 \left[\frac{\partial^3 E_{sym}(\rho)}{\partial \rho^3}\right]_{\rho=\rho_0}.
\end{eqnarray}
Higher-order parameters do not vanish, but they are fully determined by the lower-order ones through the KIDS EOS expansion in terms of the cubic root of the density~\cite{Pap18,Gil19}.

From the Hartree-Fock method, the effective interaction [Eq.~(\ref{SHFI})] leads to the energy per nucleon expressed as
\begin{equation}\label{anmeos}
E(\rho,\delta) = T(\rho,\delta) + \sum_{i=0}^{i_{max}} c_i(\delta) \rho^{1+i/3},
\end{equation}
where the kinetic energy per nucleon is expressed as
\begin{equation}
T(\rho,\delta) = \frac{\hbar^2}{2m\rho}h_k(\rho_n^{5/3}+\rho_p^{5/3}),
\end{equation}
with $h_k\equiv\frac{5}{3}(3\pi^2)^{2/3}$ and $m$ being the bare nucleon mass, and the coefficients in the potential contribution can be parameterized as
\begin{equation}
c_i(\delta) = \alpha_i + \beta_i \delta^2,
\end{equation}
except for $i=2$ for which the corresponding term is related to the momentum-dependent interaction and the coefficient can be written as
\begin{eqnarray}
c_2(\delta) &=& \alpha_2 + \beta_2 \delta^2 \nonumber \\
&+& C_{eff} h_k \left[ \left( \frac{1+\delta}{2} \right)^{5/3} + \left( \frac{1-\delta}{2} \right)^{5/3} \right] \nonumber\\
&+& D_{eff} h_k \delta \left[ \left( \frac{1+\delta}{2} \right)^{5/3} - \left( \frac{1-\delta}{2} \right)^{5/3} \right], \nonumber
\end{eqnarray}
where $\alpha_i$, $\beta_i$, $C_{eff}$, and $D_{eff}$ are constant coefficients. From $i=0$ to $i_{max}=3$ in the present study, they are related to the coefficients in the KIDS model by comparing with Eq.~(\ref{SHFpot}), and their relations can be expressed as
\begin{eqnarray}
&&t_0 = \frac{8}{3} \alpha_0,~~~t_{3i} = 16 \alpha_i~(\text{for}~i=1,2,3), \nonumber\\
&&y_0 = -\frac{1}{2} t_0 - 4\beta_0,~~~y_{3i} = -\frac{1}{2} t_{3i} -24 \beta_i~(\text{for}~i=1,2,3), \nonumber
\end{eqnarray}
and $t_1$, $t_2$, $y_1$, and $y_2$ are related to the nucleon effective mass and the density-gradient coefficients through
\begin{equation}
	        \begin{pmatrix}
		t_{1} \\
		y_{1} \\
		t_{2} \\
        y_{2}
        \end{pmatrix}=	\frac{2}{3}\begin{pmatrix}
		2 & 0 & -8 & 0\\
		-1 & -3 & 4 & 12\\
		6 & -12 & 8 & -16\\
 		-3 & 15 & -4 & 20
		\end{pmatrix}
        \begin{pmatrix}
		C_{eff} \\
		D_{eff} \\
		-G_S/2 \\
        G_V/2
        \end{pmatrix},
\end{equation}
with $G_S$ and $G_V$ being respectively the isoscalar and isovector density gradient coefficients, and
\begin{eqnarray}
C_{eff} &=& \frac{\hbar^2}{2m\rho_0}\left(\frac{m}{m_s^\star}-1\right),\\
D_{eff} &=& \frac{\hbar^2}{2m\rho_0}\left(\frac{m}{m_s^\star}-\frac{m}{m_v^\star}\right),
\end{eqnarray}
where $m_s^\star$ and $m_v^\star$ are the isoscalar and isovector nucleon effective mass, respectively. We note that $G_S$ and $G_V$ are trivially related to the coefficients $C_{12}$ and $D_{12}$ defined in other studies~\cite{Gil21,Pap21}, namely $G_S=-2C_{12}$, $G_V=2D_{12}$. From Eq.~(\ref{anmeos}) the nuclear symmetry energy can be expressed as
\begin{eqnarray}
E_{sym}(\rho) &=& \sigma \left( \frac{\hbar^2}{2m} + C_{eff} \rho \right) h_k \rho^{2/3} + \varphi D_{eff} h_k \rho^{5/3} \nonumber\\
&+& \sum_{i=0}^{i_{max}} \beta_i \rho^{1+i/3},
\end{eqnarray}
where $\sigma \equiv \frac{5}{9} \frac{1}{2^{2/3}} \approx 0.35$ and $\varphi \equiv \frac{5}{3} \frac{1}{2^{2/3}} \approx 1.05$ are constants. Finally, the coefficients $\alpha_i$ for $i=0 \sim 3$ can be expressed as functions of isoscalar EOS parameters through
\begin{widetext}
\begin{equation}
	        \begin{pmatrix}
		\alpha_0 \rho_0 \\
		\alpha_1 \rho_0^{4/3} \\
		\alpha_2 \rho_0^{5/3}  \\
        \alpha_3 \rho_0^2
        \end{pmatrix}=	\begin{pmatrix}
		20 & -19/3 & 1 & -1/6\\
		-45 & 15 & -5/2 & 1/2\\
		36 & -12 & 2 & -1/2\\
 		-10 & 10/3 & -1/2 & 1/6
		\end{pmatrix}
        \begin{pmatrix}
		E_0 - \left( \frac{\hbar^2}{2m} + C_{eff}\rho_0 \right) h_k (\rho_0/2)^{2/3} \\
		- \left( 2\frac{\hbar^2}{2m} + 5C_{eff}\rho_0 \right) h_k (\rho_0/2)^{2/3}  \\
		K_0 + \left(2 \frac{\hbar^2}{2m} - 10 C_{eff}\rho_0 \right) h_k (\rho_0/2)^{2/3}  \\
        Q_0 + \left( -8 \frac{\hbar^2}{2m} + 10 C_{eff}\rho_0 \right) h_k (\rho_0/2)^{2/3}
        \end{pmatrix},
\end{equation}
and the coefficients $\beta_i$ for $i=0 \sim 3$ can be expressed as functions of isovector EOS parameters through
\begin{equation}
	        \begin{pmatrix}
		\beta_0 \rho_0 \\
		\beta_1 \rho_0^{4/3} \\
		\beta_2 \rho_0^{5/3}  \\
        \beta_3 \rho_0^2
        \end{pmatrix}=	\begin{pmatrix}
		20 & -19/3 & 1 & -1/6\\
		-45 & 15 & -5/2 & 1/2\\
		36 & -12 & 2 & -1/2\\
 		-10 & 10/3 & -1/2 & 1/6
		\end{pmatrix}
        \begin{pmatrix}
		E_{sym}^0 - \left[\sigma \left( \frac{\hbar^2}{2m} + C_{eff}\rho_0 \right) + \varphi D_{eff} \rho_0 \right] h_k \rho_0^{2/3} \\
		L - \left[\sigma \left( 2\frac{\hbar^2}{2m} + 5 C_{eff}\rho_0 \right) + 5\varphi D_{eff} \rho_0 \right] h_k \rho_0^{2/3} \\
		K_{sym} + \left[\sigma \left( 2\frac{\hbar^2}{2m} -10 C_{eff}\rho_0 \right) -10\varphi D_{eff} \rho_0 \right] h_k \rho_0^{2/3} \\
        Q_{sym} + \left[\sigma \left( -8\frac{\hbar^2}{2m} +10 C_{eff}\rho_0 \right) +10\varphi D_{eff} \rho_0 \right] h_k \rho_0^{2/3} \\
        \end{pmatrix}.
\end{equation}
\end{widetext}
\begin{acknowledgments}

We thank Bao-An Li for helpful discussions on the incompressibility of nuclear matter, Zhen Zhang for helpful discussions on the random-phase approximation code, and Wen-Jie Xie for helpful discussions on the Bayesian analysis. JX acknowledges the National Natural Science Foundation of China under Grant No. 11922514. The work of PP was supported by the Rare Isotope Science Project of the Institute for Basic Science funded by the Ministry of Science, ICT and Future Planning and the National Research Foundation (NRF) of Korea (2013M7A1A1075764).

\end{acknowledgments}

\end{document}